\documentclass{scrartcl}

\usepackage{
    adjustbox, 
	amsmath, 
	amsthm, 
	booktabs, 
	caption, 
	cite, 
	enumerate,
	float, 
	fullpage, 
    jlcode, 
	listings, 
	longtable, 
	multirow, 
	subcaption, 
	tikz, 
	xcolor 
}

\usepackage[pagebackref]{hyperref}
\usepackage[LGR,T1]{fontenc}
\usepackage[utf8]{inputenc}
\usepackage[english]{babel} 

\usepackage{cleveref} 

\usetikzlibrary{decorations.markings, calc, patterns, arrows}
\tikzstyle{stuff_fill_green}=[circle,draw,fill=green!]
\tikzstyle{stuff_fill_red}=[circle,draw,fill=red!40]
\tikzstyle{stuff_fill_blue}=[circle,draw,fill=cyan!70]
\tikzstyle{stuff_fill_connect}=[circle,draw,fill=orange!70]
\tikzstyle{stuff_fill_high}=[circle,draw,fill=purple!70]

\theoremstyle{plain}
\newtheorem{thrm}{Theorem}[section]
\theoremstyle{definition}
\newtheorem{defn}[thrm]{Definition}
\newtheorem{exmp}[thrm]{Example}
\newtheorem{cor}[thrm]{Corollary}
\newtheorem{lem}{Lemma}
\newtheorem{prop}[thrm]{Proposition}
\newtheorem*{note}{Note}
\newtheorem*{conseq}{Consequence}
\newtheorem{algo}{Algorithm}
\allowdisplaybreaks
\newtheoremstyle{break}  
{\topsep}   
{\topsep}   
{}  
{0pt}       
{\bfseries} 
{:}         
{\newline}  
{}          

\theoremstyle{break}

\makeatletter
\let\@addpunct\@gobble
\makeatother
\newenvironment{myproof}[1][\bfseries \proofname]{%
	\begin{proof}[#1]$ $\par\nobreak\ignorespaces
	}{%
	\end{proof}
}

\DeclareMathOperator{\Pic}{Pic}
\newcommand{\oref}{\Cref}
\newcommand{\textgreek}[1]{\begingroup\fontencoding{LGR}\selectfont#1\endgroup}

\hypersetup{
	pdftitle={Brill-Noether-general Limit Root Bundles: Absence of vector-like Exotics in F-theory Standard Models},
	pdfauthor={Martin Bies, Mirjam Cvetic, Ron Donagi, Marielle Ong},
	pdfsubject={Root bundles, Limit roots, Line bundles, Global sections, F-theory, MSSM, Standard Model, vector-like spectra, Brill-Noether theory}
}

\begin{document}

\vspace*{-2cm}
\begin{flushright}
    {\texttt{UPR-1319-T,  CERN-TH-2022-073}}\\    
\end{flushright}

\vspace*{0.8cm}
\begin{center}
    {\LARGE
    Brill-Noether-general Limit Root Bundles:\\
    Absence of vector-like Exotics in F-theory Standard Models}
    
    \vspace*{1.8cm}
    {Martin Bies$^{1,2}$, Mirjam Cveti{\v c}$^{1,2,3,4}$, Ron Donagi$^{2,1}$, Marielle Ong$^{2}$}
    
    \vspace*{1cm}
    
     \bigskip
    {\textit{$^1$Department of Physics and Astronomy, University of Pennsylvania,  \\Philadelphia, PA 19104-6396, USA}}
    
    \bigskip
    {\textit{$^2$Department of Mathematics, University of Pennsylvania,  \\Philadelphia, PA 19104-6396, USA}}
    
    \bigskip
    {\textit{$^3$Center for Applied Mathematics and Theoretical Physics, University of Maribor, \\Maribor, Slovenia}}
    
    \bigskip
    {\textit{$^4$CERN Theory Department, CH-1211 Geneva, Switzerland}}
    
\vspace*{0.8cm}
\end{center}
\noindent

Root bundles appear prominently in studies of vector-like spectra of 4d F-theory compactifications. Of particular importance to phenomenology are the Quadrillion F-theory Standard Models (F-theory QSMs). In this work, we analyze a superset of the physical root bundles whose cohomologies encode the vector-like spectra for the matter representations $(\mathbf{3}, \mathbf{2})_{1/6}$, $(\mathbf{\overline{3}}, \mathbf{1})_{-2/3}$ and $(\mathbf{1}, \mathbf{1})_{1}$. For the family $B_3( \Delta_4^\circ )$ consisting of $\mathcal{O}(10^{11})$ F-theory QSM geometries, we argue that more than $99.995\%$ of the roots in this superset have no vector-like exotics. This indicates that absence of vector-like exotics in those representations is a very likely scenario.

The QSM geometries come in families of toric 3-folds $B_3( \Delta^\circ )$ obtained from triangulations of certain 3-dimensional polytopes $\Delta^\circ$. The matter curves in $X_\Sigma \in B_3( \Delta^\circ )$ can be deformed to nodal curves which are the same for all spaces in $B_3( \Delta^\circ )$. Therefore, one can probe the vector-like spectra on the entire family $B_3( \Delta^\circ )$ from studies of a few nodal curves. We compute the cohomologies of all limit roots on these nodal curves.

In our applications, for the majority of limit roots the cohomologies are determined by line bundle cohomology on rational tree-like curves. For this, we present a computer algorithm. The remaining limit roots, corresponding to circuit-like graphs, are handled by hand. The cohomologies are independent of the relative position of the nodes, except for a few circuits. On these \emph{jumping circuits}, line bundle cohomologies can jump if nodes are specially aligned. This mirrors classical Brill-Noether jumps. $B_3( \Delta_4^\circ )$ admits a jumping circuit, but the root bundle constraints pick the canonical bundle and no jump happens.

\newpage

\tableofcontents

\newpage

\section{Introduction}

String Theory is a leading candidate for a unified theory of quantum gravity since it elegantly couples gauge dynamics to gravity. In particular, all aspects of our physical reality, including the observed low energy particle physics, must be accounted for by String Theory. Above all, an explicit demonstration of the Standard Model from String Theory is desired.

Enormous efforts have been undertaken in this regard with perturbative String Theory such as the $E_8 \times E_8$ heterotic string \cite{Candelas:1985en,Greene:1986ar,Braun:2005ux,Bouchard:2005ag,Bouchard:2006dn,Anderson:2009mh,Anderson:2011ns,Anderson:2012yf} or intersecting branes models in type II \cite{Berkooz:1996km,Aldazabal:2000dg,Aldazabal:2000cn,Ibanez:2001nd,Blumenhagen:2001te,Cvetic:2001tj,Cvetic:2001nr} (see also \cite{Blumenhagen:2005mu} and references therein). These String Theory compactifications were among the first from which the Standard Model gauge sector emerged with its chiral or, in the case of \cite{Bouchard:2005ag,Bouchard:2006dn}, vector-like spectrum. Typically, these perturbative models suffer from chiral and vector-like exotic matter. The first efforts to overcome this hurdle and provide a globally consistent MSSM construction from String Theory are \cite{Bouchard:2005ag,Bouchard:2006dn} (see \cite{Gomez:2005ii,Bouchard:2008bg} for more details on the subtle global conditions for slope-stability of vector bundles).

A non-perturbative extension of type IIB String Theory must describe the gauge dynamics on 7-branes \emph{including} their back-reactions (to all orders in the string coupling) onto the compactification geometry $B_n$. This is achieved elegantly in F-theory \cite{Vafa:1996xn,oai:arXiv.org:hep-th/9602114,oai:arXiv.org:hep-th/9603161}. Namely, the back-reactions in question are encoded in the geometry of an elliptically fibered Calabi-Yau space $\pi \colon Y_{n + 1} \twoheadrightarrow B_n$. Such a space $Y_{n + 1}$ can be investigated with well-established tools of algebraic geometry. This allows to study and satisfy the global consistency conditions of the physics in $10-2n$ non-compact dimension. For a modern overview of the recent developments towards connecting F-theory and the Standard Model, we point the interested reader to \cite{Cvetic:2022fnv}.

An important signature of the observed particle physics is the chiral fermionic spectrum. A realistic 4d $\mathcal{N} = 1$ F-theory compactifications (i.e., $n = 3$) must therefore reproduce this spectrum. From the F-theory perspective, it is uniquely determined by a background gauge flux. The latter can be conveniently specified by the internal $C_3$ profile in the dual M-theory geometry. Importantly, the chiral spectrum of the F-theory compactification only depends on $G_4 = dC_3 \in H^{(2,2)} (Y_4)$. In recent years, extensive toolkits have been designed to create and enumerate the primary vertical subspace of $G_4$ configurations \cite{Grimm:2011fx,Krause:2012yh,Braun:2013nqa,Cvetic:2013uta,Cvetic:2015txa,Lin:2015qsa,Lin:2016vus}. Applications of these tools to globally consistent chiral F-theory models have been conducted in \cite{Krause:2011xj,Cvetic:2015txa,Lin:2016vus,Cvetic:2018ryq}. Relatively recently, these efforts culminated in the largest currently-known class of explicit string vacua that realize the Standard Model gauge group with their exact chiral spectrum and gauge coupling unification \cite{Cvetic:2019gnh} -- the \emph{Quadrillion F-theory Standard Models} (QSMs).

A 4d $\mathcal{N} = 1$ F-theory compactification contains chiral (super)fields in a representation $\mathbf{R}$ and chiral (super)fields in the charge conjugate representation $\mathbf{\overline{R}}$. The difference between their numbers is the chiral spectrum. Accounting for these fields individually is achieved by the vector-like spectrum. This poses additional challenges as those zero modes depend on the flat directions of the potential $C_3$ and not only on $G_4 = dC_3$. An F-theory ``gauge field'' that uniquely specifies the vector-like spectrum is given by an element of the \mbox{so-called} \emph{Deligne cohomology}. The program initiated in \cite{Bies:2014sra,Bies:2017fam,Bies:2018uzw} rests on the fact that (a subset of) the Deligne cohomology can be parameterized by Chow classes. This parametrization allows one to extract line bundles $L_{\mathbf{R}}$ defined on curves $C_\mathbf{R}$ in $B_3$. In the dual IIB picture, this can be interpreted as the localization of gauge flux on matter curves $C_\mathbf{R}$, which can lift vector-like pairs on these curves. Explicitly, the zero modes are counted by the sheaf cohomology groups of the line bundle $L_{\mathbf{R}}$. There are $h^{0}(C_{\mathbf{R}}, \, {L}_{\mathbf{R}})$ massless chiral superfields in representation $\mathbf{R}$. Similarly, there are $h^1 (C_{\mathbf{R}}, \, {L}_{\mathbf{R}})$ massless chiral superfields in the charge conjugate representation $\mathbf{\overline{R}}$ on $C_{\mathbf{R}}$.

In theory, this procedure always works. In practice however, technical limitations arise since the line bundle cohomologies depend intricately on the complex structure of $Y_4$. As a consequence, even state-of-the-art algorithms, such as \cite{ToricVarietiesProject,M2,sagemath} (see also \cite{Bies:2017fam,Bies:2018uzw}), on supercomputers specifically designed for such computations (such as \texttt{Plesken} at \textit{Siegen University}) can oftentimes not perform the necessary operations in realistic compactification geometries. This is exactly the reason why the works in \cite{Bies:2014sra,Bies:2017fam,Bies:2018uzw} focused on computationally simple geometries. Even then, these models have unrealistically large numbers of chiral fermions. Inspired by the advances in machine learning applications to String Theory \cite{Carifio:2017bov, Halverson:2019tkf, Abel:2021ddu} (see also \cite{Ruehle:2020jrk} and references therein), the complex structure dependency of line bundle cohomologies was investigated more systematically in \cite{Bies:2020gvf}. Inspired by the F-theory GUT models in \cite{Bies:2018uzw} and the help of the algorithms in \cite{ToricVarietiesProject}, it was possible to generate a large data set \cite{Database} of line bundle cohomologies for different complex structure moduli. This data was analyzed with data science techniques. A theoretical understanding was achieved by \mbox{supplementing} these data science insights by Brill-Noether theory \cite{Brill1874} (see \cite{Eisenbud1996} for a modern exposition and \cite{Watari:2016lft} for an earlier application of Brill-Noether theory in F-theory). These insights led to a quantitative study of jumps of charged matter vector pairs as a function of the complex structure moduli of the matter curve $C_\mathbf{R}$.

A $G_4$-flux is subject to the quantization condition $G_4 + \frac{1}{2} c_2( Y_4 ) \in H^4_{\mathbb{Z}}( Y_4 )$. Unfortunately, this condition can often not be checked in a computationally feasible manner. The interested reader is referred to \cite{Jefferson:2021bid} for recent advances in this direction. \emph{Quadrillion F-theory Standard Models}\cite{Cvetic:2019gnh} evaluated necessary conditions for a specific $G_4$-flux to satisfy the quantization condition. In this work, we proceed under the assumption that the quantization condition is satisfied. It is worth noting that the QSM $G_4$-flux candidate also satisfies the D3-tadpole cancelation and masslessness of the $U(1)$-gauge boson \cite{Cvetic:2019gnh}.

Therefore, the vector-like spectra in the QSMs beg to be investigated. Those modes localize on five matter curves and were first studied in \cite{Bies:2021nje}. It was shown that on three of these matter curves, the line bundle ${L}_{\mathbf{R}}$ is necessarily a fractional power $P_{\mathbf{R}}$ of the canonical bundle. On the remaining two matter curves, the line bundles are modified by contributions from the Yukawa points. Such fractional powers of line bundles are known as root bundles, which may be thought of as generalizations of spin bundles. Inspired by \textgreek{Ρίζα}, which is ``root" in Greek, $P$ refers to root bundles throughout this article. Similar to spin bundles, root bundles are far from unique. The mathematics of root bundles indicates that we should think of the different root bundles as being induced from \mbox{inequivalent} gauge potentials for a given $G_4$-flux.

In general, we cannot expect that all physically relevant root bundles on the matter curve ${C}_{\mathbf{R}}$ are induced from F-theory gauge potentials in the Deligne cohomology. This mirrors the expectation that only some of the spin bundles on the matter curves are consistent with the F-theory geometry $Y_4$. Rather, an F-theory gauge potential induces a collection $\{ P_{\mathbf{R}} \}$ consisting of one root bundle on each matter curve $C_{\mathbf{R}}$. By repeating this for all physically relevant F-theory gauge potentials, we should expect that in general only a proper subset of all root bundles on $C_{\mathbf{R}}$ will be reached. Conversely, if we are given a collection of root bundles $\{ P_{\mathbf{R}} \}$ -- one for each matter curve -- then this collection can fail to stem from an F-theory gauge potential. Namely, it is possible for one of the root bundles to not be induced at all, or not in combination with one of the other roots.

While it is of great importance for the physics to determine which roots are physical (i.e. induced from an F-theory gauge potential), this question is also extremely challenging. For this reason, this question was not investigated by \cite{Bies:2021nje} in favor of conducting a systematic study of all root and spin bundles on the individual matter curves. This local bottom-up analysis allowed us to name a root bundle and a spin bundle on all matter curves, except for the Higgs curve in one particular QSM geometry \cite{Cvetic:2019gnh}, such that their tensor product is a line bundle with the physically desired cohomologies. On a technical level, this is achieved by considering a deformation of the smooth, physical matter curves $C_{\mathbf{R}}$ into a nodal curve $C^\bullet_{\mathbf{R}}$. On the latter, root bundles are described in a diagrammatic way by \emph{limit roots} \cite{2004math4078C}. For limit root bundles on the full blowup of $C^\bullet_{\mathbf{R}}$, the global sections can be counted by the techniques put forward in \cite{Bies:2021nje}.

This study was subsequently extended in \cite{Bies:2021xfh} by a computer algorithm which enumerates all root bundles with exactly three global sections on the full blowup of $C^\bullet_{\mathbf{R}}$. This in turn was repeated for $\mathcal{O}( 10^{15} )$ different QSM geometries. The latter rests on the fact that the QSM geometries are defined by families of toric spaces $B_3( \Delta^\circ )$ associated to the (full, regular, star) triangulations of 708 (3-dimensional, reflexive) polytopes $\Delta^\circ$ in the Kreuzer-Skarke list \cite{Kreuzer:1998vb} (see also \cite{Halverson:2016tve}). One may think of these spaces as different desingularizations of toric K3-surfaces \cite{Batyrev:1994hm, Perevalov:1997vw, cox1999mirror, Rohsiepe:2004st,Braun:2017nhi}. These well-known results allowed to argue that the nodal curve $C^\bullet_{\mathbf{R}}$ -- and thus, the number of limit root bundles -- only depends on the polytope $\Delta^\circ$. Hence, by repeating a somewhat involved computer scan for 23 different polytopes, one finds results that apply to a large fraction of the QSM geometries. This allowed, in a first approximation and statistical meaning, to find base spaces for which absence of vector-like exotics in the representation $(\mathbf{3}, \mathbf{2})_{1/6}$ is very likely.

\paragraph{Results}

In this work, we analyze a superset of all physical root bundles for the matter representations $(\mathbf{3}, \mathbf{2})_{1/6}$, $(\mathbf{\overline{3}}, \mathbf{1})_{-2/3}$ and $(\mathbf{1}, \mathbf{1})_{1}$ in $\mathcal{O}(10^{11})$ F-theory QSM geometries $B_3( \Delta_4^\circ )$. We find that more than $99.995\%$ of the roots in this superset have no vector-like exotics. We arrived at this result from a systematic study of a large fraction of QSM geometries. While our results are strongest for the family $B_3( \Delta_4^\circ )$, this study provides statistical evidence that the absence of the vector-like exotics in $(\mathbf{3}, \mathbf{2})_{1/6}$, $(\mathbf{\overline{3}}, \mathbf{1})_{-2/3}$ and $(\mathbf{1}, \mathbf{1})_{1}$ is a very likely scenario for a large fraction of the QSM geometries.

As mentioned above, the vector-like spectra in the QSM geometries are necessarily counted by cohomologies of root bundles\cite{Bies:2021nje}. Such root bundles $P_{\mathbf{R}}$ on the matter curve $C_{\mathbf{R}}$ are by no means unique. Even more, it is not clear exactly which ones are induced top-down from F-theory gauge potentials in the Deligne cohomolgy. First, on one matter curve, only a subset of all mathematically allowed root bundles could be induced from all physically allowed F-theory gauge potentials. Secondly, it is conceivable that fluxes which induce a specific root bundle on matter curve $C_1$, only induce a few selected root bundles on another matter curve $C_2$. In this work, we have not addressed this involved question, but rather conducted a local and bottom-up analysis. Our study is bottom-up in that we study \emph{all} root bundles that could possibly be induced from the $G_4$-flux and all spin bundles on the matter curve in question. Our study is local in that we focus on one matter curve at a time. Correlations among the vector-like spectra of different matter curves are not taken into account.

A crucial step in our study is triangulation independence. Specifically, each family $B_3( \Delta^\circ )$ of toric 3-fold QSM base spaces -- obtained from the (fine, regular, star) triangulations of 708 (3-dimensional, reflexive) polytopes $\Delta^\circ$ in the Kreuzer-Skarke list \cite{Kreuzer:1998vb} admits a triangulation-independent nodal limit \cite{Bies:2021xfh}. This observation allows one to probe the vector-like spectra on the entire family $B_3( \Delta^\circ )$ from the vector-like spectra on just 5 nodal curves. For the family $B_3( \Delta_4^\circ )$ -- associated to the 4-th polytope in the Kreuzer-Skarke list of 3-dimensional polytopes -- which consists of $\mathcal{O}(10^{11})$ different toric spaces, we argue that all limit root bundles on the nodal curves $C_{(\mathbf{3}, \mathbf{2})_{1/6}}^\bullet$, $C_{(\mathbf{\overline{3}}, \mathbf{1})_{-2/3}}^\bullet$ and $C_{(\mathbf{1}, \mathbf{1})_{1}}^\bullet$ have exactly three global sections. By upper semi-continuity, this must hold true on the corresponding smooth, irreducible matter curves in each space of $B_3( \Delta_4^\circ )$. This rules out vector-like exotics.

Apparently, this argument relies on the ability to compute the cohomologies of all limit root bundles. This is exactly the domain of Brill-Noether theory. By following the original work in \cite{2004math4078C}, we find that for some limit roots, this is equivalent to line bundle cohomology on tree-like curves. The remaining limit roots require an understanding of line bundle cohomology on circuit-like curves. In our applications to the F-theory QSMs, the majority of limit roots arise from tree-like curves with rational irreducible components. We present a computer algorithm which counts line bundle cohomology on such curves and handle the circuit-like cases by hand. On tree-like, rational curves, the cohomologies are independent of the relative position of the nodes. While true for most circuit-like cases, special circuits admit jumps provided that the nodes are specially aligned. Such \emph{jumping circuits} mirror classical Brill-Noether jumps. For $B_3( \Delta_4^\circ )$, one jumping circuit exists. The root bundle constraints select the canonical bundle on this jumping circuit so that we find four global sections. This leads to $99.995\%$ instead of $100\%$.

In contrast to earlier works \cite{Bies:2021nje, Bies:2021xfh}, we take the \emph{geometric multiplicity} into account \cite{2004math4078C}. Under a deformation of a smooth to a nodal curve, distinct root bundles on the smooth curve coalesce as we approach the nodal limit. The exact number of distinct roots which coalesce is the geometric multiplicity. Alternatively, this multiplicity ensures that we count as many limit root bundles on nodal curves as root bundles on a smooth curve. This ensures comparability of the limit root counts for different QSM geometries and slightly modifies the conclusions in \cite{Bies:2021xfh}.

\paragraph{Outline}

In \oref{sec:PartialBlowupLimitRootBundles}, we briefly recall the motivation for root bundles in F-theory vacua \cite{Bies:2021nje}. Subsequently, we outline the steps necessary to extend the techniques of \cite{Bies:2021nje, Bies:2021xfh} to limit roots on partial blowup curves, which are compactly summarized in \oref{fig:FullBlowupLimitRoots}, \oref{fig:PartialBlowupLimitRoots} and \oref{fig:CircuitLikeLimitRoots}. This approach makes it natural to think about line bundle cohomology on tree-like nodal curves. We put forward an algorithm that determines these line bundle cohomologies on rational curves in \oref{sec:LineBundleCohomologyOnTreelikeCurves}. This algorithm can be optimized further to detect speciality, i.e. if a line bundle has more sections than on a smooth $\mathbb{P}^1$. While this optimization currently seems not relevant for studying Brill-Noether theory of limit root bundles or for arguing the absence of vector-like exotics in $\mathcal{O}(10^{11})$ QSM geometries, we outline these steps in \oref{sec:Speciality}. With these techniques, we return to the QSM geometries and re-analyze the setups that were originally discussed in \cite{Bies:2021xfh}. In \oref{sec:TowardsBNOfLimitRootBundles}, we first discuss the geometric multiplicity and compare our results to \cite{Bies:2021xfh}. Indeed, we obtain more refined results, which also identify two errors in the earlier work. We summarize our findings in \oref{tab:Results1} and \oref{tab:Results2}. We propose to view these tables as a first approximation of the Brill-Noether theory of the root bundles in question. More details can be found in \oref{appendix:DetailsOnCounts}. We also mention that the necessary computer implementations are available in the \texttt{gap-4}-package \emph{QSMExplorer} as part of the \texttt{ToricVarieties$\_$project} \cite{ToricVarietiesProject}. Many, but definitely not all, of these computations can be conducted on a simple personal computer. Finally in \oref{sec:D4}, we focus on the family of QSM geometries $B_3( \Delta_4^\circ)$ associated to full regular star triangulations of the 4-th polytope $\Delta_4^\circ$ in the Kreuzer-Skarke database \cite{Kreuzer:1998vb}. By hand, we analyze the global sections on a couple of circuit-like nodal curves. For most of them, the line bundle cohomologies are independent of the relative positions of the nodes (cf. \oref{sec:StationaryCircuits}). However, there is one circuit-like nodal curve for which four nodes are located on two irreducible components. By studying the embedding of this nodal curve into each space in $B_3( \Delta_4^\circ )$, we can show that the location of all nodes is determined in terms of the input parameters to the QSMs. In particular, we argue that the root bundle constraints select the canonical bundle on this jump circuit, which leads to exactly four (and not three) global sections. The detailed study of this embedding is conducted in \oref{sec:TriaIndependence}. We summarize and discuss our findings in \oref{sec:ConclusionAndOutlook}.

\section{Partial blowup limit root line bundles} \label{sec:PartialBlowupLimitRootBundles}

\subsection{Vector-like spectra in F-theory}

The story begins with an F-theory compactification to four dimensions given by a singular, elliptically fibered 4-fold $\pi: Y_4 \twoheadrightarrow B_3$. Suppose that $Y_4$ admits a smooth, crepant resolution which is a flat elliptic fibration $\widehat{\pi}: \widehat{Y}_4 \twoheadrightarrow B_3$. A $G_4$-flux in $H^{(2, 2)}(\widehat{Y}_4)$ satisfies the quantization condition precisely if \cite{Witten:1996md}
\begin{align}
G_4 + \frac{1}{2}c_2(T_{\widehat{Y}_4}) \in H^{(2, 2)}_\mathbb{Z}(\widehat{Y}_4):= H^{(2, 2)}(\widehat{Y}_4) \cap H^4(\widehat{Y}_4, \mathbb{Z}) \, .
\end{align}
For F-theory compactifications on an elliptically fibered smooth Calabi-Yau 4-fold with globally defined Weierstrass model, the class $c_2(T_{\widehat{Y}_4})$ is even \cite{Collinucci:2010gz}. Consequently in such geometries, we demand $G_4 \in H^{(2, 2)}_\mathbb{Z}(\widehat{Y}_4)$. Unfortunately, this condition can often not be checked in a computationally feasible manner. The interested reader is referred to \cite{Jefferson:2021bid} for recent advances in this direction. \emph{Quadrillion F-theory Standard Models}\cite{Cvetic:2019gnh} evaluated necessary conditions for a specific $G_4$-flux to satisfy the quantization condition. In this work, we proceed under the assumption that the quantization condition is satisfied. It is worth noting that the QSM $G_4$-flux candidate also satisfies the D3-tadpole cancelation and masslessness of the $U(1)$-gauge boson \cite{Cvetic:2019gnh}.

There is a surjection from the Deligne cohomology group
\begin{align}
\widehat{c}: H^4_D(\widehat{Y}_4, \mathbb{Z}(2)) \twoheadrightarrow  H^{(2, 2)}_\mathbb{Z}(\widehat{Y}_4) \, ,
\end{align}
and a $G_4$-flux lifts to an F-theory gauge potential $A$ in the Deligne cohomology group $H^4_D(\widehat{Y}_4, \mathbb{Z}(2))$. Although this group is intractable when it comes to explicit computations, we can parametrize a subset of it via the map 
\begin{align}
\widehat{\gamma}: \mathrm{CH}^2(\widehat{Y}_4, \mathbb{Z}) \rightarrow H^4_D(\widehat{Y}_4, \mathbb{Z}(2)) \, .
\end{align}
Thus, we turn our attention to studying $\mathcal{A} \in \mathrm{CH}^2(\widehat{Y}_4, \mathbb{Z})$ satisfying $(\widehat{c} \circ \widehat{\gamma})(\mathcal{A})= G_4$. 

For some representation $\textbf{R}$ of the gauge group $G$ of the F-theory compactification in question, let $C_{\textbf{R}} \subseteq B_3$ be the matter curve upon which the massless matter states localize and transform in $\textbf{R}$. The corresponding matter surface $S_{\mathbf{R}} \subseteq \widehat{Y}_4$ -- the mathematicians will notice that $S_{\mathbf{R}}$ is actually a 2-cycle and not a surface -- is a linear combination of $\mathbb{P}^1$-fibration over $C_{\textbf{R}}$ and the coefficients correspond to the weight vector $\mathbf{w}$ of the representation $\mathbf{R}$ (see \cite{Weigand:2018rez} and references therein). By restricting to $S_{\mathbf{R}}$ and integrating over the fibers over $C_{\textbf{R}}$, we can associate to each $\mathcal{A} \in \mathrm{CH}^2(\widehat{Y}_4, \mathbb{Z})$ a divisor via the cylinder map
\begin{align} D_{\textbf{R}}: \mathrm{CH}^2(\widehat{Y}_4, \mathbb{Z}) \rightarrow \mathrm{CH}^1(C_{\textbf{R}}, \mathbb{Z}) \cong \Pic(C_{\textbf{R}}) \, .
\end{align}
Specifically, we consider $\iota_{S_{\mathbf{R}}} \colon S_{\mathbf{R}} \hookrightarrow \widehat{Y}_4$ and $\pi_{S_{\mathbf{R}}} \colon S_{\mathbf{R}} \twoheadrightarrow C_{\mathbf{R}}$. Then the cylinder map is given by
\begin{align}
D_{\mathbf{R}} \left( \mathcal{A} \right) = \pi_{S_{\mathbf{R}}\ast} \left( \iota_{S_{\mathbf{R}}}^\ast \left( \mathcal{A} \right) \right) \in \mathrm{Pic} \left( C_{\mathbf{R}} \right) \, .
\end{align}
As explained in \cite{Bies:2014sra, Bies:2017fam, Bies:2018uzw}, the divisor $D_{\mathbf{R}} \left( \mathcal{A} \right)$ gives rise to the following line bundle
\begin{align}  \label{eq:linebundle}
P_{\mathbf{R}} \left( \mathcal{A} \right) = \mathcal{O}_{C_{\mathbf{R}}} \left( D_{\mathbf{R}} \left( \mathcal{A} \right) \right) \otimes_{\mathcal{O}_{C_{\mathbf{R}}}} \mathcal{O}_{C_{\mathbf{R}}}^{\text{spin}} \, ,
\end{align}
where $\mathcal{O}_{C_{\mathbf{R}}}^{\text{spin}}$ is a spin bundle on $C_{\mathbf{R}}$ compatible with the global structure of the F-theory compactification on $\widehat{Y}_4$, whose sheaf cohomologies encode the vector-like spectrum on $C_{\mathbf{R}}$, i.e. 
\begin{align}
\begin{split}
h^0 \left( C_{\mathbf{R}}, P_{\mathbf{R}} \left( \mathcal{A} \right) \right) & \quad \leftrightarrow \quad \text{chiral zero modes in representation $\mathbf{R}$} \, , \\
h^1 \left( C_{\mathbf{R}}, P_{\mathbf{R}} \left( \mathcal{A} \right) \right) & \quad \leftrightarrow \quad \text{chiral zero modes in charge conjugate representation $\overline{\mathbf{R}}$} \, .
\end{split}
\end{align}
The generic case $h^0 \left( C_{\mathbf{R}}, P_{\mathbf{R}} \left( \mathcal{A} \right) \right) = \chi(P_{\mathbf{R}} \left( \mathcal{A} \right))$ corresponds to the absence of exotic vector-like pairs. The situation $h^0 \left( C_{\mathbf{R}}, P_{\mathbf{R}} \left( \mathcal{A} \right) \right) = \chi(P_{\mathbf{R}} \left( \mathcal{A} \right)) + 1$, which implies the existence of exactly one Higgs pair, features prominently as the desired vector-like spectrum on the Higgs curve of an F-theory MSSM construction.

\subsection{Limit root bundles in the F-theory QSMs}

In \cite{Bies:2021xfh}, we followed this logic to study the vector-like spectra in the largest currently-known class of globally-consistent F-theory Standard Model constructions without chiral exotics and gauge coupling unification \cite{Cvetic:2019gnh}. The geometry of the compactification space of those Quadrillion F-theory Standard Models (F-theory QSMs) determines a class $\mathcal{A}' \in \mathrm{CH}^2(\widehat{Y}_4, \mathbb{Z})$ and an integer $\xi \in \mathbb{Z}_{> 0}$ such that $D_{\textbf{R}}(\mathcal{A}')$ is -- possibly with additional contributions from the Yukawa points -- expressible in terms of the canonical divisor on $C_{\textbf{R}}$ , the integer $\xi$ divides the degree of  $D_{\textbf{R}}(\mathcal{A}')$ and $\mathcal{A}$ is subject to two constraints
\begin{align} \label{eq:condition}
(\widehat{c} \circ \widehat{\gamma})(\mathcal{A})= G_4, \quad \xi \cdot D_{\textbf{R}}(\mathcal{A}) \sim D_{\textbf{R}}(\mathcal{A}')
\end{align}
The condition on the right of \oref{eq:condition} says that $D_{\textbf{R}}(\mathcal{A})$ is a $\xi$-th root divisor of $D_{\textbf{R}}(\mathcal{A}')$ and the associated line bundle $P_{\mathbf{R}} \left( \mathcal{A} \right)$ is a $\xi$-th root bundle on $C_{\textbf{R}}$. To count the zero modes in the F-theory QSMs, we are therefore led to study the sheaf cohomologies of root bundles.

An important task is to find the root bundles that are induced from $F$-theory gauge potentials. First, on one matter curve, only a subset of all mathematically allowed root bundles could be induced from all physically allowed F-theory gauge potentials. Secondly, it is conceivable that fluxes which induce a specific root bundle on matter curve $C_1$, only induce a few selected root bundles on another matter curve $C_2$. The study of these questions is involved and currently beyond our abilities. While we hope to return to these questions in future work, in the current work we opt for a local and bottom-up analysis instead. In this sense, we  follow the spirit of the earlier works \cite{Bies:2021nje, Bies:2021xfh}. Our study is bottom-up in that we study \emph{all} mathematically allowed root bundles. Our study covers all root bundles that could possibly be induced from the $G_4$-flux and all spin bundles on the matter curve in question. Our study is local in that we focus on one matter curve at a time. Correlations among the vector-like spectra of different matter curves are therefore not taken into account.

Consequently, we study all root bundles $P_{\mathbf{R}}$ on $C_{\textbf{R}}$ -- including all spin bundles on $C_{\textbf{R}}$. Since $C_{\textbf{R}}$ is smooth and the integer $\xi$ divides the degree of $D_{\textbf{R}}(\mathcal{A}')$, there are $\xi^{2g}$ root bundles where $g$ is the genus of $C_{\textbf{R}}$. Unfortunately, computing their sheaf cohomology is difficult. In particular, it is not easy to distinguish between trivial and non-trivial root bundles in general. To overcome this challenge, we consider deformations of $C_{\textbf{R}}$ to a reducible, nodal curve $C^\bullet_{\textbf{R}}$ by modifying the defining polynomials of $C_{\textbf{R}}$ in a concrete base geometry $B_3$. This comes at a double cost.

First, the deformation $C^\bullet_{\textbf{R}} \to C_{\textbf{R}}$ is, per se only subject to upper semicontinuity of global sections. This means, that we may lose global sections that exist on $C^\bullet_{\textbf{R}}$ as we deform back to the smooth, irreducible $C_{\textbf{R}}$. However, if the number of global sections match the chiral index, then it is protected by topology and cannot change. We exploited this fact in \cite{Bies:2021nje} and will mimic this strategy in later part of this work.

Second, the $\xi$-th root bundles of a line bundle $L$ on a reducible nodal curve $C$ do not exist if $\xi$ does not divide the degree of the restriction of $L$ to the irreducible components of $C$. Thanks to the work of \cite{2004math4078C}, this is elegantly circumvented by passing to limit root bundles $P^\circ_{\textbf{R}}$ on (partial) blow-ups $C^\circ_{\textbf{R}}$, whose existence only requires for $\xi$ to divide $\deg(L_{\textbf{R}})$. By properly keeping track of multiplicities, there exist $\xi^{2g}$ limit roots, each determined by the combinatorial data associated to the dual graph of $C^\bullet_{\textbf{R}}$. Fortunately, it turns out that there are less than $2 \times 708$ different dual graphs that arise in the context of the QSMs. This observation allows us to gain insights into a large fraction of these geometries.

\subsection{Triangulation independence in the QSM geometries} \label{subsec:Genesis}

To understand why there are only $2 \times 708$ different dual graphs to considered, we briefly recall the genesis of the F-theory QSM geometries, specifically the base spaces in question. These geometries are obtained from desingularizations of toric K3-surfaces. Such desingularizations were first studied in \cite{Batyrev:1994hm}. It is well-established by now that they correspond to three-dimensional, reflexive lattice polytopes $\Delta \subset M_{\mathbb{R}}$ and their polar duals $\Delta^\circ \subset N_{\mathbb{R}}$ defined by $\left\langle \Delta, \Delta^\circ \right\rangle \geq -1$. In \cite{Kreuzer:1998vb}, Kreuzer and Skarke listed all possible 3-dimensional polytopes \cite{Kreuzer:1998vb}. Just as in \cite{Bies:2021xfh}, we consider the i-th polytope in the Kreuzer-Skarke list as subset of $N_{\mathbb{R}}$ and denote it by $\Delta_i^\circ$.

Recall that to each polytope $\Delta \subset M_{\mathbb{R}}$, one can associate a normal fan $\Sigma_\Delta$. The ray generators of this fan are the facet normals of $\Delta$ and the maximal cones correspond to  the vertices of $\Delta$. Even though the toric variety $X_{\Delta} \equiv X_{\Sigma_{\Delta}}$ or the CY-hypersurface may not be smooth, we can resolve these hypersurfaces. Such resolutions were introduced in \cite{Batyrev:1994hm} as \emph{maximal projective crepant partial} (MPCP) desingularizations. Equivalently, \cite{cox1999mirror} refers to them as maximal projective subdivisions of the normal fan. MPCPs are associated to fine, regular, star triangulation (FRST) of the lattice polytope $\Delta^\circ$. We recall that \emph{star} means that every simplex in the triangulation contains the origin, \emph{fine} ensures that every lattice point of $\Delta^\circ$ is used as ray generator\footnote{This property is also referred to as \emph{full}\cite{trias}.} and that \emph{regular} implies that $X_{\Sigma(T)}$ is projective. Together, this implies that $\Sigma(T)$ defines a \emph{maximal} projective refinement of $\Sigma_{\Delta}$. Note that in applications to toric K3-surfaces, $X_{\Sigma(T)}$ is guaranteed to be smooth. Namely, a maximal projective subdivision of $\Sigma_{\Delta}$ then constructs a 3-dimensional Gorenstein orbifold with terminal singularities \cite{cox1999mirror} which must be smooth by proposition 11.4.19 in \cite{cox2011toric}. For more details, we refer the interested reader to \cite{Batyrev:1994hm}.

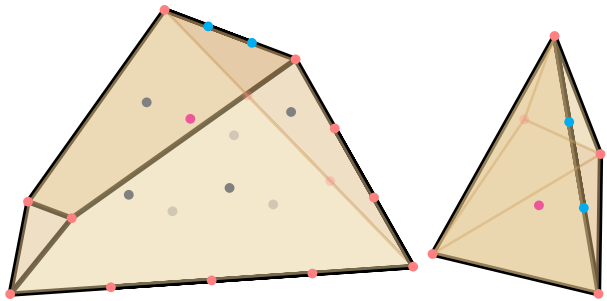
\begin{figure}[ht]
    \centering
    \begin{tikzpicture}
    [x={(0.4497546cm, 0.559773cm)},
    y={(0.1306336cm, 0.681189cm)},
    z={(0.7360895cm, 0.0505690cm)},
    scale=1.80,
    back/.style={line width=1.0pt,opacity=0.50},
    edge/.style={color=brown, line width=2pt},
    facet/.style={fill=yellow!30,fill opacity=0.20},
    vertex/.style={inner sep=1.3pt,circle,fill=red!50,thick,anchor=base}]
    
    \coordinate (-1.0, -1.0, -1.0) at (-1.0, -1.0, -1.0);
    \coordinate (2.0, -1.0, 0.0) at (2.0, -1.0, 0.0);
    \coordinate (0.0, -1.0, -1.0) at (0.0, -1.0, -1.0);
    \coordinate (-1.0, 2.0, 0.0) at (-1.0, 2.0, 0.0);
    \coordinate (-1.0, -1.0, 3.0) at (-1.0, -1.0, 3.0);
    \coordinate (-1.0, 0.0, -1.0) at (-1.0, 0.0, -1.0);
    \coordinate (-1.0, -1.0, 0.0) at (-1.0, -1.0, 0.0);
    \coordinate (-1.0, -1.0, 1.0) at (-1.0, -1.0, 1.0);
    \coordinate (-1.0, -1.0, 2.0) at (-1.0, -1.0, 2.0);
    \coordinate (-1.0, 0.0, 0.0) at (-1.0, 0.0, 0.0);
    \coordinate (-1.0, 0.0, 1.0) at (-1.0, 0.0, 1.0);
    \coordinate (-1.0, 0.0, 2.0) at (-1.0, 0.0, 2.0);
    \coordinate (-1.0, 1.0, 0.0) at (-1.0, 1.0, 0.0);
    \coordinate (-1.0, 1.0, 1.0) at (-1.0, 1.0, 1.0);
    \coordinate (0.0, -1.0, 0.0) at (0.0, -1.0, 0.0);
    \coordinate (0.0, -1.0, 1.0) at (0.0, -1.0, 1.0);
    \coordinate (0.0, -1.0, 2.0) at (0.0, -1.0, 2.0);
    \coordinate (0.0, 0.0, 0.0) at (0.0, 0.0, 0.0);
    \coordinate (0.0, 0.0, 1.0) at (0.0, 0.0, 1.0);
    \coordinate (0.0, 1.0, 0.0) at (0.0, 1.0, 0.0);
    \coordinate (1.0, -1.0, 0.0) at (1.0, -1.0, 0.0);
    \coordinate (1.0, -1.0, 1.0) at (1.0, -1.0, 1.0);
    \coordinate (1.0, 0.0, 0.0) at (1.0, 0.0, 0.0);
    
    \draw[edge, black] (2.0, -1.0, 0.0) -- (0.0, -1.0, -1.0);
    \draw[edge, black] (-1.0, -1.0, -1.0) -- (0.0, -1.0, -1.0);
    \draw[edge, black] (2.0, -1.0, 0.0) -- (-1.0, 2.0, 0.0) -- (1,0,0) -- (0,1,0);
    \draw[edge, back] (-1.0, 2.0, 0.0) -- (-1.0, -1.0, 3.0) -- (-1,1,1) -- (-1,0,2);
    \draw[edge, black] (-1.0, -1.0, -1.0) -- (-1.0, -1.0, 3.0) -- (-1,-1,0) -- (-1,-1,1) -- (-1,-1,2);
    \draw[edge, black] (2.0, -1.0, 0.0) -- (-1.0, -1.0, 3.0) -- (1,-1,1) -- (0,-1,2);
    \draw[edge, black] (-1.0, -1.0, -1.0) -- (-1.0, 0.0, -1.0);
    \draw[edge, black] (0.0, -1.0, -1.0) -- (-1.0, 0.0, -1.0);
    \draw[edge, black] (-1.0, 2.0, 0.0) -- (-1.0, 0.0, -1.0);
    
    \fill[facet, fill opacity=0.30] (2.0, -1.0, 0.0) -- (-1.0, 2.0, 0.0) -- (-1.0, -1.0, 3.0) -- cycle {};
    \fill[facet, brown, fill opacity=0.40] (2.0, -1.0, 0.0) -- (0.0, -1.0, -1.0) -- (-1,0,-1) -- (-1.0, 2.0, 0.0) -- cycle {};
    \fill[facet, brown!80,fill opacity=0.30] (-1,-1,-1) -- (-1.0, -1.0, 3.0) -- (2.0, -1.0, 0.0) -- (0.0, -1.0, -1.0) -- cycle {};
    \fill[facet,  fill opacity=0.30] (-1,-1,-1) -- (-1.0, -1.0, 3.0) -- (-1.0, 2.0, 0.0) -- (-1.0, 0.0, -1.0) -- cycle {};
    \fill[facet, brown!80, fill opacity=0.30] (-1,-1,-1) -- (0.0, -1.0, -1.0) -- (-1.0, 0.0, -1.0) -- cycle {};
    
    \node[vertex] at  (-1.0, -1.0, -1.0) {};
    \node[vertex] at  (2.0, -1.0, 0.0) {};
    \node[vertex] at  (0.0, -1.0, -1.0) {};
    \node[vertex] at  (-1.0, 2.0, 0.0) {};
    \node[vertex] at  (-1.0, -1.0, 3.0) {};
    \node[vertex] at  (-1.0, 0.0, -1.0) {};
    \node[vertex] at  (-1.0, -1.0, 0.0) {};
    \node[vertex] at  (-1.0, -1.0, 1.0) {};
    \node[vertex] at  (-1.0, -1.0, 2.0) {};
    \node[vertex,gray] at  (-1.0, 0.0, 0.0) {};
    \node[vertex,gray] at  (-1.0, 0.0, 1.0) {};
    \node[vertex, fill opacity=0.30] at  (-1.0, 0.0, 2.0) {};
    \node[vertex,gray] at  (-1.0, 1.0, 0.0) {};
    \node[vertex, fill opacity=0.30] at  (-1.0, 1.0, 1.0) {};
    \node[vertex,gray, fill opacity=0.30] at  (0.0, -1.0, 0.0) {};
    \node[vertex,gray, fill opacity=0.30] at  (0.0, -1.0, 1.0) {};
    \node[vertex] at  (0.0, -1.0, 2.0) {};
    \node[vertex, magenta, fill opacity=0.60] at  (0.0, 0.0, 0.0) {};
    \node[vertex,gray] at  (0.0, 0.0, 1.0) {};
    \node[vertex,cyan] at  (0.0, 1.0, 0.0) {};
    \node[vertex, gray, fill opacity=0.30] at  (1.0, -1.0, 0.0) {};
    \node[vertex] at  (1.0, -1.0, 1.0) {};
    \node[vertex, cyan] at  (1.0, 0.0, 0.0) {};
    
    \end{tikzpicture}
    \begin{tikzpicture}
    [x={(0.449656cm, 0.377639cm)},
    y={(-0.77770cm, -0.358578cm)},
    z={(-0.106936cm, 0.633318cm)},
    scale=1.80,
    back/.style={line width=1.0pt,opacity=0.50},
    edge/.style={color=brown, line width=2pt},
    facet/.style={fill=yellow!30,fill opacity=0.20},
    vertex/.style={inner sep=1.3pt,circle,fill=red!50,thick,anchor=base}]
    
    \coordinate (-1.0, -1.0, -1.0) at (-1.0, -1.0, -1.0);
    \coordinate (-1.0, -1.0, 0.0) at (-1.0, -1.0, 0.0);
    \coordinate (-1.0, -1.0, 1.0) at (-1.0, -1.0, 1.0);
    \coordinate (-1.0, -1.0, 2.0) at (-1.0, -1.0, 2.0);
    \coordinate (0.0, 0.0, 1.0) at (0.0, 0.0, 1.0);
    \coordinate (0.0, 1.0, 0.0) at (0.0, 1.0, 0.0);
    \coordinate (1.0, 0.0, 0.0) at (1.0, 0.0, 0.0);
    \coordinate (0.0, 0.0, 0.0) at (0.0, 0.0, 0.0);
    
    \draw[edge, black] (-1.0, -1.0, -1.0) -- (-1.0, -1.0, 2.0) -- (-1,-1,0) -- (-1,-1,1);
    \draw[edge, black] (-1.0, -1.0, 2.0) -- (0.0, 1.0, 0.0);
    \draw[edge, black] (-1.0, -1.0, -1.0) -- (0,1,0);
    \draw[edge, black] (-1.0, -1.0, -1.0) -- (1.0, 0.0, 0.0);
    \draw[edge, black] (-1.0, -1.0, 2.0) -- (1.0, 0.0, 0.0);
    \draw[edge, back] (0.0, 1.0, 0.0) -- (1.0, 0.0, 0.0);
    \draw[edge, back] (-1.0, -1.0, 2.0) -- (0.0, 0.0, 1.0);
    \draw[edge, back] (0.0, 1.0, 0.0) -- (0.0, 0.0, 1.0);
    \draw[edge, back] (1.0, 0.0, 0.0) -- (0.0, 0.0, 1.0);
    
    \fill[facet, fill opacity=0.30] (0.0, 1.0, 0.0) -- (-1.0, -1.0, -1.0) -- (1.0, 0.0, 0.0) -- (0,0,1)-- cycle {};
    \fill[facet, fill opacity=0.30] (-1.0, -1.0, 2.0) -- (0.0, 1.0, 0.0) -- (0,0,1) -- cycle {};
    \fill[facet, brown, fill opacity=0.30] (-1.0, -1.0, -1.0) -- (-1.0, -1.0, 2.0) -- (1,0,0) -- cycle {};
    \fill[facet, fill opacity=0.30] (-1.0, -1.0, 2.0) -- (1.0, 0.0, 0.0) -- (0,0,1) -- cycle {};
    \fill[facet, fill opacity=0.30] (-1,-1,-1) -- (-1.0, -1.0, 2.0) -- (0.0, 1.0, 0.0) -- cycle {};
    \fill[facet, brown, fill opacity=0.30] (-1,-1,-1) -- (-1.0, -1.0, 2.0) -- (0.0, 1.0, 0.0) -- cycle {};
    
    \node[vertex] at (-1.0, -1.0, -1.0) {};
    \node[vertex, cyan] at (-1.0, -1.0, 0.0) {};
    \node[vertex, cyan] at (-1.0, -1.0, 1.0) {};
    \node[vertex] at (-1.0, -1.0, 2.0) {};
    \node[vertex, magenta, fill opacity=0.60] at (0.0, 0.0, 0.0) {};
    \node[vertex, fill opacity=0.30] at (0.0, 0.0, 1.0) {};
    \node[vertex] at (0.0, 1.0, 0.0) {};
    \node[vertex] at (1.0, 0.0, 0.0) {};
    
    \end{tikzpicture}
\caption{$\Delta_{52}^\circ \subset N_{\mathbb{R}}$ on the left and $\Delta_{52} \subseteq M_{\mathbb{R}}$ on the right \cite{Kreuzer:1998vb}. The magenta point is the origin. The generic K3-surface meets trivially with the gray divisors, in an irreducible curve with the pink ones, and in finite families of $\mathbb{P}^1$s with the cyans. This picture is taken from \cite{Bies:2021xfh}.}
\label{fig:Delta13AndDeltaDual13}
\end{figure}

708 of the 4319 polytopes in \cite{Kreuzer:1998vb} satisfy $\overline{K}_{X_\Sigma( T )}^3 \in \{ 6, 10, 18, 30 \}$. These are exactly the polytopes whose geometries define the base spaces for the Quadrillion F-theory Standard Models (QSMs) \cite{Cvetic:2019gnh}. Just as in \cite{Bies:2021xfh}, we use $B_3 ( \Delta^\circ )$ to denote the toric 3-folds obtained from FRSTs of the polytope $\Delta^\circ$. An instructive example is $\Delta_{52}^\circ$, which we display in \oref{fig:Delta13AndDeltaDual13}.

For each QSM geometry, we introduced a nodal quark-doublet curve in \cite{Bies:2021nje}. This curve is canonically associated with the family $B_3( \Delta_4^\circ )$. We can write this curve as
\begin{align}
C_{(\mathbf{3},\mathbf{2})_{1/6}}^\bullet = \bigcup \limits_{a \in A}{V( x_a, s_9 )} \cup \bigcup \limits_{b \in B}{V( y_b, s_9 )} \cup \bigcup \limits_{c \in C}{V( z_c , s_9 )} \, .
\end{align}
We explained the rationale behind this notation in \cite{Bies:2021xfh}. Namely, one can show that $V( x_a, s_9 )$ is irreducible, $V( y_b, s_9 )$ a finite collection of $\mathbb{P}^1$s and $V( z_c, s_9 ) = \emptyset$. This rests on arguments originally presented in \cite{Batyrev:1994hm, Perevalov:1997vw, cox1999mirror, Rohsiepe:2004st, Kreuzer:2006ax}. 

This classification is directly related to the geometry of the polytope $\Delta^\circ$. Namely, the homogeneous coordinates are one-to-one with the lattice points of $\Delta^\circ$ (minus the origin) since we use these as rays of the refined fan. A homogeneous coordinate associated to a facet interior point of $\Delta^\circ$ is denoted by $z_c$. For a lattice point in the interior of an edge $\Theta^\circ \subset \Delta^\circ$, two facets $F_1$, $F_2$ of $\Delta^\circ$ meet at $\Theta^\circ$. In the dual polytope $\Delta$, they correspond to vertices $m_1, m_2$ and the dual edge $\Theta$ is the edge connecting $m_1$ and $m_2$. If $\Theta$ has interior points, we denote the homogeneous coordinate by $y_b$. All remaining coordinates are indicated as $x_a$. In \oref{fig:Delta13AndDeltaDual13}, distinct types of lattice points are marked in different colors. 

Consequently, all irreducible components of $C_{(\mathbf{3},\mathbf{2})_{1/6}}^\bullet$ are smooth, irreducible curves embedded into a smooth and projective K3-surface. This alone implies, and equivalently follows from \cite{Batyrev:1994hm, Perevalov:1997vw, cox1999mirror, Rohsiepe:2004st, Kreuzer:2006ax}, that the pairwise topological intersection numbers are non-negative. Furthermore, we can restrict to open subsets near intersection points. Since an open subset of an irreducible topological space is irreducible, it follows that the multiplicity of intersection at each intersection point is non-negative. Consequently, for generic $s_9$, the number of pairwise intersection points among the irreducible components of $C_{(\mathbf{3},\mathbf{2})_{1/6}}^\bullet$ matches the topological intersection number. This is how we concluded in \cite{Bies:2021xfh} that the dual graph of $C_{(\mathbf{3},\mathbf{2})_{1/6}}^\bullet$ is the same for all FRSTs of $\Delta^\circ$. This leads to at most 708 distinct dual graphs. In fact, one finds that distinct polytopes in the QSM database can lead to identically the same dual graphs. So, there are strictly less than 708 distinct dual graphs for the canonical nodal quark-doublet curve $C_{(\mathbf{3},\mathbf{2})_{1/6}}^\bullet$ in the QSMs. There are two matter curves with distinct topology, namely the Higgs curve $C_{(\mathbf{1},\mathbf{2})_{-1/2}}^\bullet$ and a curve supporting right-handed quarks $C_{(\mathbf{\overline{3}},\mathbf{1})_{1/3}}^\bullet$. However, their topologies are identical, contributing to at most 708 more distinct dual graphs. In this work, we will ignore those two curves and focus on the curve with dual graph identical to that of $C_{(\mathbf{3},\mathbf{2})_{1/6}}^\bullet$.

In \oref{sec:TriaIndependence}, we refine this result for the polytope $\Delta_4^\circ$. We will explain in \oref{subsec:SimplificationForPartialBlowups}, that the counting of limit root bundles on $C_{(\mathbf{3},\mathbf{2})_{1/6}}^\bullet$ only depends on some irreducible components and the nodes on these components. For $\Delta_4^\circ$, we therefore focus on four components $C_0$, $C_1$, $C_2$, $C_3$. Our arguments show that there are homogeneous coordinates for these four irreducible components such that, in these coordinates, the location of the nodes is the same for all FRSTs of $\Delta_4^\circ$. This in turn is critical for our analysis in \oref{subsec:JumpingCircuit}, where we study the one jumping circuit associated to $\Delta_4^\circ$. To the best of our knowledge, this FRSTs independence of the location of the nodes does not follow from earlier works \cite{Batyrev:1994hm, Perevalov:1997vw, cox1999mirror, Rohsiepe:2004st, Kreuzer:2006ax}. At this point, our argument rests on a detailed computational analysis which can, in principle, be extended to all QSM spaces. One can speculate that such an extension would lead to the very same conclusion, namely triangulation independent positions of the relevant nodes. This extension is beyond the scope of the current work and we leave this study for future investigation.

\subsection{The need for partial blowup limit root bundles}

In order to approximate physical root bundle on the smooth matter curve $C_{(\mathbf{3},\mathbf{2})_{1/6}}$, we thus employ limit root bundles on the nodal curve $C_{(\mathbf{3},\mathbf{2})_{1/6}}^\bullet$. As we explained before, the dual graph of $C_{(\mathbf{3},\mathbf{2})_{1/6}}^\bullet$ is independent of the FRSTs. This extends to certain topological intersection numbers and implies that the operations performed on basis of \cite{2004math4078C}, which construct and count limit root line bundles $P^\bullet_{\textbf{R}}$ on (partial) blow-ups of $C^\bullet_{\textbf{R}}$ from combinatorial data, are independent of the FRSTs.

In \cite{Bies:2021nje}, we compute the zeroth sheaf cohomology of all limit root bundles that arise from the full blow-up of $C^\bullet_{\textbf{R}}$. We refer to this work for additional background for the application of limit root bundles towards F-theory QSMs. In particular, we explained that the sheaf cohomology of such limit root bundles is given by the following formula,
\begin{align} h^0(C^\circ_{\textbf{R}}, P^\circ_{\textbf{R}}) = \sum_Z h^0(Z, P^\circ_{\textbf{R}}|_{Z}), 
\end{align}
where $Z$ is an irreducible component of $C^\bullet_{\textbf{R}}$. Since the pushforward of the limit roots $P^\circ_{\textbf{R}}$ along the blow-up map $\pi: C^\circ_{\textbf{R}} \rightarrow C^\bullet_{\mathbf{R}}$ preserves the number of global sections, we also have that
\begin{align} h^0(C^\circ_{\textbf{R}}, P^\circ_{\textbf{R}}) = h^0(C^\bullet_{\textbf{R}}, \pi_*P^\circ_{\textbf{R}}).
\end{align}
Along the deformation of $C^\bullet_{\textbf{R}}$ back to the original curve $C_{\textbf{R}}$, the number of sections either decreases or remains constant as the dimension of the zeroth cohomology is an upper semi-continuous function, i.e. 
\begin{align}
h^0(C_{\textbf{R}}, P_{\textbf{R}}) \leq h^0(C^\bullet_{\textbf{R}}, \pi_*P^\circ_{\textbf{R}}).
\end{align}

In this work, we aim to establish the absence of vector-like exotics for \emph{physical} roots. Since we cannot yet tell which roots are physical, we cover all (mathematical) roots and identify their sheaf cohomologies. With an eye towards the nodal quark-doublet curve $C^\bullet_{(\mathbf{3},\mathbf{2})_{1/6}}$, we would ideally find that all of its roots have exactly three global sections. Then, by upper semicontinuity this remains true as we deform back to the smooth, irreducible (and physical) curve $C_{(\mathbf{3},\mathbf{2})_{1/6}}$. But for this argument to work, we need to extend to all roots, including those on partial blowups. This requires an extension of the techniques in \cite{Bies:2021xfh}. As a side effect of these extensions, we shall present -- what we believe are the very first -- results on Brill-Noether theory of limit root bundles.

\subsection{Beyond full blowup limit roots} \label{subsec:SimplificationForPartialBlowups}

We show the generalizations necessary to go from full to partial blowups in an example that will be relevant in later part of this work. Namely, we focus on the family of F-theory QSMs defined in terms of the fourth polytope $\Delta_4^\circ$ in the Kreuzer-Skarke database \cite{Kreuzer:1998vb}. There are $\mathcal{O}( 10^{11} )$ such geometries \cite{Halverson:2016tve} and by following \cite{Bies:2021nje, Bies:2021xfh}, each contains a canonical nodal curve $C^\bullet_{(\mathbf{3}, \mathbf{2})_{1/6}}$ with dual graph displayed in \oref{equ:Delta4FullDualGraph}. The edges in this graph represent nodal singularities and the vertices are irreducible components of $C^\bullet_{(\mathbf{3}, \mathbf{2})_{1/6}}$. The four curves $C_i$ each are smooth $\mathbb{P}^1$, which are marked in pink in the graph. In addition, $L_{i_1,...,i_k}$ represents a chain of $k$ smooth genus zero curves $C_{i_j}$ such that $C_{i_j}$ and $C_{i_{j+1}}$ intersect each other at a single node. Hence, as dual graph, $L_{i_1, ..., i_k}$ corresponds to a chain or link of $k$ vertices. It is ordered so that $C_{i_1}$ intersects the $\mathbb{P}^1$ on the left and $C_{i_k}$ intersects the $\mathbb{P}^1$ on the right. For $L_{21, 27}$, we have that $C_{21}$ intersects $C_3$ while $C_{27}$ intersects $C_1$.

\paragraph{Simplifying the dual graph}

On this curve, we are looking for $12$-th roots $P^\bullet$ of the $12$-th power of the canonical bundle $K^\bullet_{(\mathbf{3}, \mathbf{2})_{1/6}}$. To this end, we can replace the above dual graph by another graph, for which the combinatorics task of enumerating the limit root bundles is simpler. To see this, we consider the following:
\begin{equation}
\begin{tikzpicture}[scale=0.6, baseline=(current  bounding  box.center)]
      
      \def\s{3.0};
      
      \path[-, black, thick] (0,0) edge (1.5*\s,0);
      \path[-, black, thick] (2.5*\s,0) edge (4*\s,0);
      
      \node at (0,0) [stuff_fill_red, scale=0.7]{$C_0$};
      \node at (1*\s,0) [stuff_fill_red, scale=0.7]{$C_1$};
      \node at (2*\s,0) {$\dots$};
      \node at (3*\s,0) [stuff_fill_red, scale=0.55]{$C_{n-1}$};
      \node at (4*\s,0) [stuff_fill_red, scale=0.7]{$C_n$};
      
\end{tikzpicture}
\end{equation}
In \cite{Bies:2021xfh}, we argued that when constructing limit roots on the full blow-up of a chain of curves, the weights along the chain in the weighted subgraph are uniquely fixed. Thus, we were able to replace these chains by a single edge and we would still count the same number of weighted subgraphs and hence, limit root bundles. Indeed, this is the only weighted subgraph which satisfies the rules (C1) and (C2) in \cite{2004math4078C}. Namely, a curve $C$ with dual graph $\Gamma_C$ has $r^{b_1(\Gamma_C)}$ weighted subgraphs satisfying (C1) and (C2). If $C$ is a chain of rational curves, then
\begin{align}
b_1( \Gamma_C ) = \# \left( \text{edges} \right) + \# \left( \text{connected components} \right) - \# \left( \text{vertices} \right) = 0 \, .
\end{align}
Hence, there is only $1 = r^0$ weighted subgraph, which corresponds to the full blow-up. Consequently, we can transition to the simpler dual graph displayed in \oref{equ:Delta4DualGraph}.

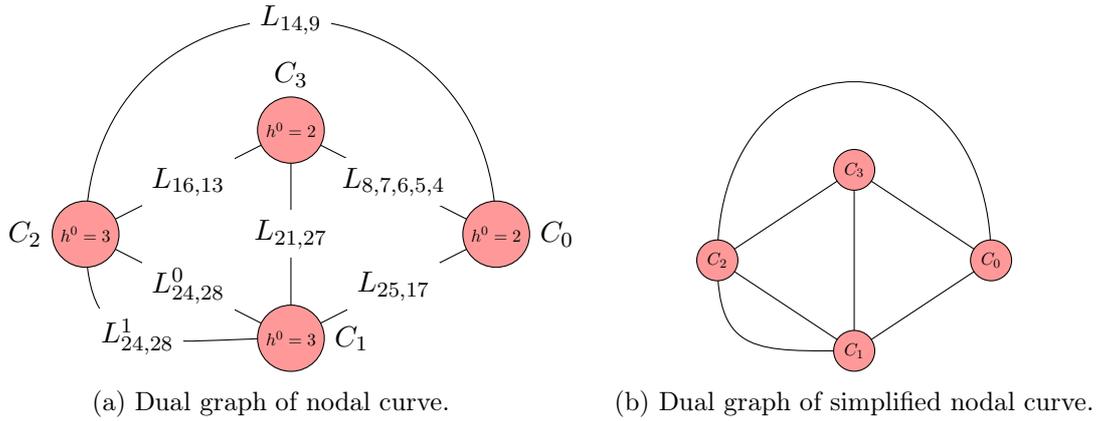
\begin{figure}[tb]
\centering
\begin{subfigure}[b]{0.45\textwidth}
\centering
\begin{tikzpicture}[scale=0.6, baseline=(current  bounding  box.center)]
      
      \def\s{4.5};
      \def\h{2.3};
      
      \path[-] (-\s,0) edge node[fill = white] {$L_{16,13}$} (0,\h);
      \path[-] (-\s,0) edge node[fill = white] {$L^0_{24,28}$} (0,-\h);
      \path[-, out = -90, in = 180, looseness = 1.5] (-\s,0) edge node[fill = white] {$L^1_{24,28}$} (0,-\h);
      \path[-] (\s,0) edge node[fill = white] {$L_{8,7,6,5,4}$} (0,\h);
      \path[-] (\s,0) edge node[fill = white] {$L_{25,17}$} (0,-\h);
      \path[-] (0,\h) edge node[fill = white] {$L_{21,27}$} (0,-\h);
      \path[-, out = 90, in = 90, looseness = 1.8] (-\s,0) edge node[fill = white] {$L_{14,9}$} (\s,0);
      
      \node at (1*\s,0) [stuff_fill_red, scale=0.6, label=right:$C_0$]{$h^0 = 2$};
      \node at (0,-\h) [stuff_fill_red, scale=0.6, label=right:$C_1$]{$h^0 = 3$};
      \node at (-1*\s,0) [stuff_fill_red, scale=0.6, label=left:$C_2$]{$h^0 = 3$};
      \node at (0,\h) [stuff_fill_red, scale=0.6, label=above:$C_3$]{$h^0 = 2$};
      
\end{tikzpicture}
\caption{Dual graph of nodal curve.}
\label{equ:Delta4FullDualGraph}
\end{subfigure}
\quad
\begin{subfigure}[b]{0.45\textwidth}
\centering
\begin{tikzpicture}[scale=0.6, baseline=(current  bounding  box.center)]
      
      \def\s{3.0};
      \def\h{2};
      
      \path[-] (-\s,0) edge (0,\h);
      \path[-] (-\s,0) edge (0,-\h);
      \path[-, out = -90, in = 180, looseness = 1.5] (-\s,0) edge (0,-\h);
      \path[-] (\s,0) edge (0,\h);
      \path[-] (\s,0) edge (0,-\h);
      \path[-] (0,\h) edge (0,-\h);
      \path[-, out = 90, in = 90, looseness = 2.25] (-\s,0) edge (\s,0);
      
      \node at (1*\s,0) [stuff_fill_red, scale=0.6]{$C_0$};
      \node at (0,-\h) [stuff_fill_red, scale=0.6]{$C_1$};
      \node at (-1*\s,0) [stuff_fill_red, scale=0.6]{$C_2$};
      \node at (0,\h) [stuff_fill_red, scale=0.6]{$C_3$};
      
\end{tikzpicture}
\caption{Dual graph of simplified nodal curve.}
\label{equ:Delta4DualGraph}
\end{subfigure}
\caption{Dual graph of (simplified) nodal curve $C^\bullet_{(\mathbf{3}, \mathbf{2})_{1/6}}$ in the QSM spaces $B_3( \Delta_4^\circ )$.}
\label{fig:two graphs}
\end{figure}

\paragraph{Global section counting of partial blowup limit roots}

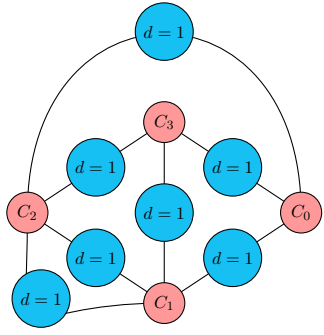
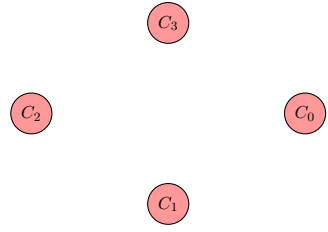
\begin{figure}[tb]
\centering
\begin{subfigure}[b]{0.45\textwidth}
\centering
\begin{tikzpicture}[scale=0.6, baseline=(current  bounding  box.center)]
      
      \def\s{3.0};
      \def\h{2};
      
      \path[-] (-\s,0) edge (0,\h);
      \path[-] (-\s,0) edge (0,-\h);
      \path[-, out = -90, in = 180, looseness = 2.5] (-\s,0) edge (0,-\h);
      \path[-] (\s,0) edge (0,\h);
      \path[-] (\s,0) edge (0,-\h);
      \path[-] (0,\h) edge (0,-\h);
      \path[-, out = 90, in = 90, looseness = 2.25] (-\s,0) edge (\s,0);
      
      \node at (1*\s,0) [stuff_fill_red, scale=0.6]{$C_0$};
      \node at (0,-\h) [stuff_fill_red, scale=0.6]{$C_1$};
      \node at (-1*\s,0) [stuff_fill_red, scale=0.6]{$C_2$};
      \node at (0,\h) [stuff_fill_red, scale=0.6]{$C_3$};
      
      \node at (0.5*\s,0.5*\h) [stuff_fill_blue, scale=0.6]{$d = 1$};
      \node at (0.5*\s,-0.5*\h) [stuff_fill_blue, scale=0.6]{$d = 1$};
      \node at (-0.5*\s,0.5*\h) [stuff_fill_blue, scale=0.6]{$d = 1$};
      \node at (-0.5*\s,-0.5*\h) [stuff_fill_blue, scale=0.6]{$d = 1$};
      \node at (0*\s,0*\h) [stuff_fill_blue, scale=0.6]{$d = 1$};
      \node at (0*\s,2*\h) [stuff_fill_blue, scale=0.6]{$d = 1$};
      \node at (-0.9*\s,-0.95*\h) [stuff_fill_blue, scale=0.6]{$d = 1$};
      
\end{tikzpicture}
\caption{Full blowup of nodal curve.}
\label{equ:Delta4FullBlowup}
\end{subfigure}
\quad
\begin{subfigure}[b]{0.45\textwidth}
\centering
\begin{tikzpicture}[scale=0.6, baseline=(current  bounding  box.center)]
      
      \def\s{3.0};
      \def\h{2};
      
      \node at (1*\s,0) [stuff_fill_red, scale=0.6]{$C_0$};
      \node at (0,-\h) [stuff_fill_red, scale=0.6]{$C_1$};
      \node at (-1*\s,0) [stuff_fill_red, scale=0.6]{$C_2$};
      \node at (0,\h) [stuff_fill_red, scale=0.6]{$C_3$};
      
\end{tikzpicture}
\caption{Disconnected curve from removing nodes.}
\label{equ:Delta4CompletelyDisconnected}
\end{subfigure}
\caption{For full blowup limit roots, each node is replaced by an exceptional $\mathbb{P}^1$. The global sections are isomorphic to those of a line bundle on the disconnected curve.}
\label{fig:FullBlowupLimitRoots}
\end{figure}

In order to find $12$-th roots $P^\bullet$ of the $12$-th power of the canonical bundle $K^\bullet_{(\mathbf{3}, \mathbf{2})_{1/6}}$ and count their global sections, we blew up every node in our earlier works \cite{Bies:2021nje, Bies:2021xfh}. In following \cite{2004math4078C, Bies:2021nje, Bies:2021xfh} we first focus on limit roots $P^\bullet$ that correspond to a certain line bundle on the full blowup curve of $C^\bullet_{(\mathbf{3}, \mathbf{2})_{1/6}}$. This full blowup curve is a curve $C^\circ_{(\mathbf{3}, \mathbf{2})_{1/6}}$ in which each node is replaced by an exceptional $\mathbb{P}^1$. Its dual graph is therefore of the form displayed in \oref{equ:Delta4FullBlowup}. In this diagram, we have marked the exceptional $\mathbb{P}^1$s in blue color. Furthermore, it holds $\left. P \right|_{E_i} \cong \mathcal{O}_{\mathbb{P}^1}(1)$ for each exceptional $\mathbb{P}^1$ by \cite{2004math4078C}. The (local) sections on the exceptional $\mathbb{P}^1$s are uniquely fixed by the demand to bridge between the sections on the components $C_i$ of $C^\bullet_{(\mathbf{3}, \mathbf{2})_{1/6}}$. Therefore, we called these sections \emph{bridging sections} in \cite{Bies:2021nje}. The number of global sections of $P^\bullet$ is therefore identical to those of the restriction of $P^\bullet$ to the completely disconnected curve in \oref{equ:Delta4CompletelyDisconnected}. In particular, it holds
\begin{align}
h^0( C^\bullet, P^\bullet ) = \sum_{i = 0}^{3}{h^0( C_i, \left. P^\bullet \right|_{C_i} )} \, . \label{equ:SimpleSectionCounting}
\end{align}

\begin{figure}[tb]
\centering
\begin{subfigure}[b]{0.3\textwidth}
\centering
\begin{tikzpicture}[scale=0.6, baseline=(current  bounding  box.center)]
      
      \def\s{3.0};
      \def\h{2};
      
      \path[-, thick, red] (-\s,0) edge (0,\h);
      \path[-] (-\s,0) edge (0,-\h);
      \path[-, out = -90, in = 180, looseness = 1.5] (-\s,0) edge (0,-\h);
      \path[-] (\s,0) edge (0,\h);
      \path[-] (\s,0) edge (0,-\h);
      \path[-] (0,\h) edge (0,-\h);
      \path[-, out = 90, in = 90, looseness = 2.5] (-\s,0) edge (\s,0);
      
      \node at (1*\s,0) [stuff_fill_red, scale=0.6]{$C_0$};
      \node at (0,-\h) [stuff_fill_red, scale=0.6]{$C_1$};
      \node at (-1*\s,0) [stuff_fill_red, scale=0.6]{$C_2$};
      \node at (0,\h) [stuff_fill_red, scale=0.6]{$C_3$};
      
\end{tikzpicture}
\caption{No blowup on red edge.}
\label{equ:Delta4MarkedEdge}
\end{subfigure}
\quad
\begin{subfigure}[b]{0.3\textwidth}
\centering
\begin{tikzpicture}[scale=0.6, baseline=(current  bounding  box.center)]
      
      \def\s{3.0};
      \def\h{2};
      
      \path[-, thick, red] (-\s,0) edge (0,\h);
      \path[-] (-\s,0) edge (0,-\h);
      \path[-, out = -90, in = 180, looseness = 2.5] (-\s,0) edge (0,-\h);
      \path[-] (\s,0) edge (0,\h);
      \path[-] (\s,0) edge (0,-\h);
      \path[-] (0,\h) edge (0,-\h);
      \path[-, out = 90, in = 90, looseness = 2.25] (-\s,0) edge (\s,0);
      
      \node at (1*\s,0) [stuff_fill_red, scale=0.6]{$C_0$};
      \node at (0,-\h) [stuff_fill_red, scale=0.6]{$C_1$};
      \node at (-1*\s,0) [stuff_fill_red, scale=0.6]{$C_2$};
      \node at (0,\h) [stuff_fill_red, scale=0.6]{$C_3$};
      
      \node at (0.5*\s,0.5*\h) [stuff_fill_blue, scale=0.6]{$d = 1$};
      \node at (0.5*\s,-0.5*\h) [stuff_fill_blue, scale=0.6]{$d = 1$};
      \node at (-0.5*\s,-0.5*\h) [stuff_fill_blue, scale=0.6]{$d = 1$};
      \node at (0*\s,0*\h) [stuff_fill_blue, scale=0.6]{$d = 1$};
      \node at (0*\s,2*\h) [stuff_fill_blue, scale=0.6]{$d = 1$};
      \node at (-0.9*\s,-0.95*\h) [stuff_fill_blue, scale=0.6]{$d = 1$};
      
\end{tikzpicture}
\caption{Partial blowup.}
\label{equ:Delta4PartialBlowup}
\end{subfigure}
\quad
\begin{subfigure}[b]{0.3\textwidth}
\centering
\begin{tikzpicture}[scale=0.6, baseline=(current  bounding  box.center)]
      
      \def\s{3.0};
      \def\h{2};
      
      \path[-, thick, red] (-\s,0) edge (0,\h);
      
      \node at (1*\s,0) [stuff_fill_red, scale=0.6]{$C_0$};
      \node at (0,-\h) [stuff_fill_red, scale=0.6]{$C_1$};
      \node at (-1*\s,0) [stuff_fill_red, scale=0.6]{$C_2$};
      \node at (0,\h) [stuff_fill_red, scale=0.6]{$C_3$};
      
\end{tikzpicture}
\caption{Tree-like nodal curve.}
\label{equ:Delta4TreeLike}
\end{subfigure}
\caption{For partial blowup limit roots, some nodes are not replaced by an exceptional $\mathbb{P}^1$. In some cases, global sections are then counted on tree-like nodal curves.}
\label{fig:PartialBlowupLimitRoots}
\end{figure}
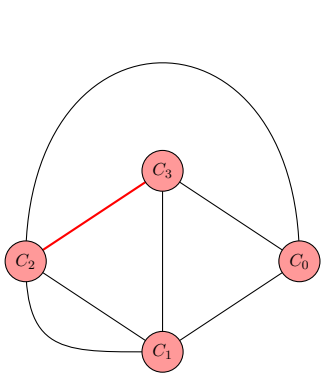
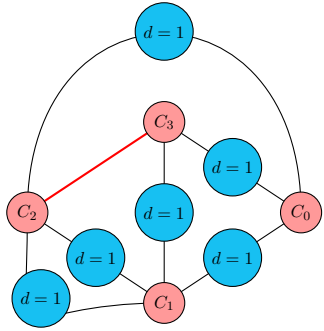
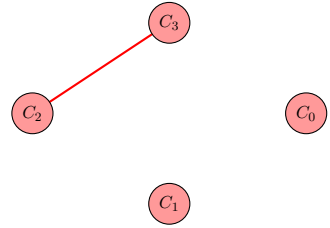

\paragraph{Tree-like and circuit-like graphs}

In order to generalize this procedure to partial blowups, let us first look at the situation in which we decide not to blow up the edge marked in red in \oref{equ:Delta4MarkedEdge}. It is in general not guaranteed that such a partial blowup admits limit roots. However, if limit roots $P^\bullet$ exist, then they correspond to line bundles on the nodal curve in \oref{equ:Delta4PartialBlowup}. Again, the sections on the exceptional $\mathbb{P}^1$s bridge between the sections on neighboring components $C_i$. However, since one node -- namely the one corresponding to the red edge -- has not been blown up, the number of global sections of $P^\bullet$ match that of the restriction of $P^\bullet$ to the nodal curve with dual graph displayed in \oref{equ:Delta4TreeLike}. There, $C_2 \cup C_3$ is a nodal, tree-like curve. If we are able to compute its line bundle cohomology, then we have
\begin{align}
h^0( C^\bullet, P^\bullet) = h^0 \left( C_0, \left. P^\bullet \right|_{C_0} \right) + h^0 \left( C_1, \left. P^\bullet \right|_{C_1} \right) + h^0 \left( C_2 \cup C_3, \left. P^\bullet \right|_{C_2 \cup C_3} \right) \, .
\end{align}
Apparently, this generalizes \oref{equ:SimpleSectionCounting}. 

In our applications, we find that for the majority of limit roots, the number of global sections is the same as the number of global sections of a line bundle on a rational, tree-like nodal curve. For example, 99\% of the limit root bundles on $\Delta_4^\circ$ arise in this way. Consequently, in order to classify the limit root bundles with regard to their number of global sections, it becomes rather important for us to compute line bundle cohomology of line bundles on tree-like nodal curves. We will therefore discuss this at length in \oref{sec:LineBundleCohomologyOnTreelikeCurves}.

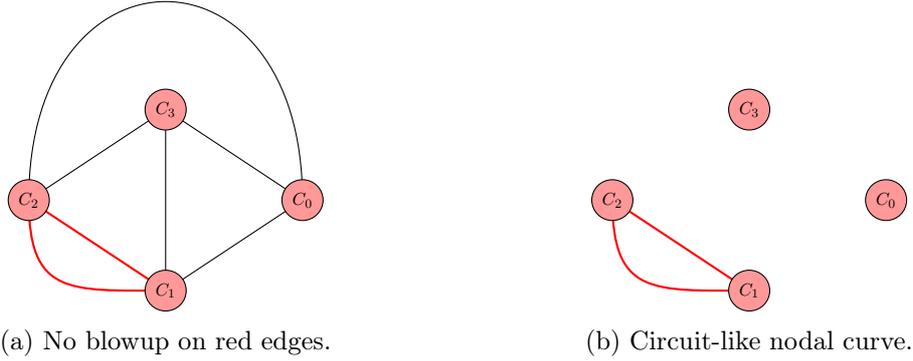
\begin{figure}[tb]
\centering
\begin{subfigure}[b]{0.45\textwidth}
\centering
\begin{tikzpicture}[scale=0.6, baseline=(current  bounding  box.center)]
      
      \def\s{3.0};
      \def\h{2};
      
      \path[-] (-\s,0) edge (0,\h);
      \path[-, red, thick] (-\s,0) edge (0,-\h);
      \path[-, red, thick, out = -90, in = 180, looseness = 1.5] (-\s,0) edge (0,-\h);
      \path[-] (\s,0) edge (0,\h);
      \path[-] (\s,0) edge (0,-\h);
      \path[-] (0,\h) edge (0,-\h);
      \path[-, out = 90, in = 90, looseness = 2.5] (-\s,0) edge (\s,0);
      
      \node at (1*\s,0) [stuff_fill_red, scale=0.6]{$C_0$};
      \node at (0,-\h) [stuff_fill_red, scale=0.6]{$C_1$};
      \node at (-1*\s,0) [stuff_fill_red, scale=0.6]{$C_2$};
      \node at (0,\h) [stuff_fill_red, scale=0.6]{$C_3$};
      
\end{tikzpicture}
\caption{No blowup on red edges.}
\label{equ:Delta4TwiceNoBlowup}
\end{subfigure}
\quad
\begin{subfigure}[b]{0.45\textwidth}
\centering
\begin{tikzpicture}[scale=0.6, baseline=(current  bounding  box.center)]
      
      \def\s{3.0};
      \def\h{2};
      
      \path[-, red, thick] (-\s,0) edge (0,-\h);
      \path[-, red, thick, out = -90, in = 180, looseness = 1.5] (-\s,0) edge (0,-\h);
      
      \node at (1*\s,0) [stuff_fill_red, scale=0.6]{$C_0$};
      \node at (0,-\h) [stuff_fill_red, scale=0.6]{$C_1$};
      \node at (-1*\s,0) [stuff_fill_red, scale=0.6]{$C_2$};
      \node at (0,\h) [stuff_fill_red, scale=0.6]{$C_3$};
      
\end{tikzpicture}
\caption{Circuit-like nodal curve.}
\label{equ:Delta4Circuit}
\end{subfigure}
\caption{Some choices of partial blowups lead to limit roots whose cohomologies are counted by line bundles on circuit-like nodal curves.}
\label{fig:CircuitLikeLimitRoots}
\end{figure}

All remaining root bundles stem from circuit-like graphs. For example, suppose that we do not blow-up the two nodes marked in red in \oref{equ:Delta4TwiceNoBlowup}, then this leads to the computation of line bundle cohomology on the nodal curve with dual graph displayed in \oref{equ:Delta4Circuit}. This curve has a circuit-like component $C_1 \cup C_2$, which we cannot yet treat systematically. 

Fortunately, in our current applications, these circuits account for very few limit root bundles. The reason why seems related to the fact that we are looking for $12$-th and even $20$-th roots of the canonical bundle. If instead, we were looking at second roots, one should expect equal contributions from tree-like and circuit-like graphs. It would be even conceivable that the circuits would dominate. However, in our applications, this is not the case. Therefore, we can conduct a case-by-case study for those few circuit-like curves that we encounter. We will come back to this in \oref{subsec:StationaryCircuits} and \oref{subsec:JumpingCircuit} and distinguish two types of circuits. First, there are circuits for which the number of global sections is the same irrespective of the position of the nodes. We call them \emph{stationary circuits}. On the other hand, circuits for which the number of global sections depends on the (relative) location of the nodes are called \emph{jumping circuits}. For these, it is possible to achieve more than the minimal number of global sections upon rearranging the position of the nodes.

\section{Line bundle cohomology on tree-like, rational curves} \label{sec:LineBundleCohomologyOnTreelikeCurves}

\subsection{Simplifying the tree}

Consider a connected rational nodal curve $C := \bigcup_{i \in I}{C_i}$ and a line bundle $L$ on it. Then, $I$ is a simply connected graph and all components $C_i$ are $\mathbb{P}^1$'s.

\begin{defn}
Let $d_i = \deg(L|_{C_i})$. Write $I = I_+ \cup I_-$, where
\begin{align}
I_- := \left\{ i \in I | d_i < 0 \right\} \, , \qquad I_+ := \left\{ i \in I | d_i \geq 0 \right\} \, .
\end{align}
Let
\begin{align}
C_+ := \bigcup_{i \in I_+}{C_i} \, , \qquad C_- := \bigcup_{i \in I_-}{C_i} \, .
\end{align}
Note that the curves $C_+$, $C_-$ could be disconnected. Let $e_i$ be the intersection number of $C_i$ with $C_-$ and let $L_+$ be the line bundle on $C_+$ with $\deg(L_+|_{C_i}) = d_i - e_i$.
\end{defn}

\begin{prop} \label{prop:simplified}
$h^0( C, L ) = h^0( C_+, L_+ )$.
\end{prop}

\begin{myproof}
Consider a connected component $Y = \cup Y_i$ of $C_+$. Suppose that $e_i$ connected components $Z_j$ of $C_-$ intersect $Y_i$. Since $C$ is connected and rational, $Y_i$ intersects each $Z_j$ at exactly one point $n_j$. Hence, $e_i$ is exactly the number of connected components of $C_-$ that meet $Y_i$. The connected components of $C_-$ support only the trivial section, which automatically glues across the intersections between the components of $Z_j$. Hence, after choosing local coordinates, we can parametrize the local sections as follows: 
\begin{align}
\begin{tabular}{ccc}
\toprule
Curve & Coordinates & Sections \\
\midrule
$Y_i$ & $[x:y]$ & $\mathrm{Span}_{\mathbb{C}} \left( x^{d_i}, x^{d_i-1} y, \dots, y^{d_i} \right)$ \\
$Z_j$ & $[u_j: v_j]$  & 0 \\
\bottomrule
\end{tabular}
\end{align}
Without loss of generality, we can assume that the nodes are at the following positions:
\begin{align}
\begin{tabular}{c|cc}
\toprule
Label & Coordinates in $Y_i$ & Coordinates in $Z_j$ \\
\midrule
$n_j$ & $[x:y] = [a_j: b_j]$ & $[u_j: v_j] = [0:1]$\\
\bottomrule
\end{tabular}
\end{align}
Each intersection between $Y_i$ and $Z_j$ forces the local sections on $Y_i$ to vanish at $n_j$. This amounts to exactly one equation of linear dependence on $h^0(Y_i, L|_{Y_i})$, namely 
\begin{align}
\sum^{d_i}_{k=0} \alpha_k a_j^k b_j^{d_i-k} = 0,
\end{align}
If $d_i \geq e_i$, then 
\begin{align} h^0\Bigl(Y_i \cup \Bigl( \bigcup^{e_i}_{j=1} Z_j \Bigr), L\Bigr) = h^0(Y_i, L|_{Y_i}) - e_i = h^0(\mathbb{P}^1, \mathcal{O}_{\mathbb{P}^1}(d_i - e_i)) = h^0(Y_i, L_+|_{Y_i}).
\end{align}
If $d_i < e_i$, then the number of equations of linear dependence imposed by the $e_i$ intersections exceeds $h^0(Y_i, L|_{Y_i})$. Hence, 
\begin{align} h^0\Bigl(Y_i \cup \Bigl( \bigcup^{e_i}_{j=1} Z_j \Bigr), L\Bigr) = 0,
\end{align}
which coincides with $h^0(Y_i, L_+|_{Y_i}) = h^0(\mathbb{P}^1, \mathcal{O}_{\mathbb{P}^1}(d_i - e_i))$. Counting the sections of $L$ over $C$ is the same as counting the sections over $C_+$ and then, enforcing the gluing conditions at $C_+ \cap C_-$. Hence, $h^0(C, L) = h^0(C_+, L_+)$. 
\end{myproof}
\begin{note}
By iterating this process we have
\begin{align}
h^0 \left( C, L \right) = h^0 \left( C_+, L_+ \right) = \dots = h^0( C_+^{(n)}, L_+^{(n)} ) \, ,
\end{align}
In fact, since $(C_+, L_+)$ has at most as many components as $C$ this process must eventually become ``constant''. Let us illustrate this with two examples.
\end{note}

\begin{exmp}
We first perform the following simplification:
\begin{equation}
\begin{tikzpicture}[scale=0.6, baseline=(current  bounding  box.center)]
      \def\s{3.0};
      \def\h{3.0};
      \path[-] (0,0) edge (2*\s,0);
      \path[-] (1*\s,-\h) edge (2*\s,-\h);
      \node at (0,0) [stuff_fill_red, scale=0.8]{$-2$};
      \node at (1*\s,0) [stuff_fill_red, scale=0.8]{$1$};
      \node at (2*\s,0) [stuff_fill_red, scale=0.8]{$2$};
      \draw[dashed] (-\s,-0.4*\h) -- (3*\s,-0.4*\h);
      \node at (1*\s,-\h) [stuff_fill_red, scale=0.8]{$0$};
      \node at (2*\s,-\h) [stuff_fill_red, scale=0.8]{$2$};
\end{tikzpicture}
\end{equation}
The latter cannot be simplified further.
\end{exmp}

\begin{exmp}
We perform the following simplifications:
\begin{equation}
\begin{tikzpicture}[scale=0.6, baseline=(current  bounding  box.center)]
      
      \def\s{3.0};
      \def\h{2.0};
      \def\t{4.5};
      
      \path[-] (0,0) edge (4*\s,0);
      \path[-] (-0.5*\s,-\h) edge (0,0);
      \path[-] (0.5*\s,-\h) edge (0,0);
      \path[-] (1.5*\s,-\h) edge (2*\s,0);
      \path[-] (2.5*\s,-\h) edge (2*\s,0);
      
      \node at (0,0) [stuff_fill_red, scale=0.8]{$1$};
      \node at (2*\s,0) [stuff_fill_red, scale=0.8]{$1$};
      \node at (4*\s,0) [stuff_fill_red, scale=0.8]{$1$};
      \node at (-0.5*\s,-\h) [stuff_fill_red, scale=0.8]{$-2$};
      \node at (0.5*\s,-\h) [stuff_fill_red, scale=0.8]{$-1$};
      \node at (1.5*\s,-\h) [stuff_fill_red, scale=0.8]{$-1$};
      \node at (2.5*\s,-\h) [stuff_fill_red, scale=0.8]{$-1$};
      
      \draw[dashed] (-\s,-0.68*\t) -- (5*\s,-0.68*\t);
      \path[-] (0,-\t) edge (4*\s,-\t);
      \node at (0,-\t) [stuff_fill_red, scale=0.8]{$-1$};
      \node at (2*\s,-\t) [stuff_fill_red, scale=0.8]{$-1$};
      \node at (4*\s,-\t) [stuff_fill_red, scale=0.8]{$ 1$};

      \draw[dashed] (-\s,-1.3*\t) -- (5*\s,-1.3*\t);
      \node at (4*\s,-1.6*\t) [stuff_fill_red, scale=0.8]{$0$};

\end{tikzpicture}
\end{equation}
The latter cannot be simplified further.
\end{exmp}

\begin{defn}
Let $C$ be a tree-like nodal (but not necessarily connected) curve and $L \in \Pic(C)$. We call a pair $(C,L)$ \emph{terminal} iff $C_+ \cap C_- = \emptyset$.
\end{defn}

\begin{conseq}
Once we know how to compute global sections on a terminal tree-like pair $(C,L)$, we can employ the following algorithm to compute the global sections on a connected nodal pair $(C,L)$:
\begin{enumerate}
 \item By iterating the simplification process, we have:
 \begin{align}
h^0 \left( C, L \right) = h^0 ( C_+, L_+ ) = \dots = h^0( C_+^{(n)}, L_+^{(n)} ) \, ,
\end{align}
with terminal $(C_+^{(n)}, L_+^{(n)} )$.
 \item Compute $h^0( C_+^{(n)}, L_+^{(n)} )$.
\end{enumerate}
\end{conseq}

\subsection{Global sections on terminal trees}

We will now turn to line bundle cohomology on terminal trees. Recall that such terminal trees have $C_+ \cap C_- = \emptyset$. Since $C_-$ does not support global sections, we can focus on just $C_+$. We begin with the following lemma.

\begin{lem} \label{equ:BCLemma}
Consider a curve $C = \mathbb{P}^1_a \cup \mathbb{P}^1_b$ with $\mathbb{P}^1_a \cap \mathbb{P}^1_b = \left\{ n \right\}$ a nodal singularity:
\begin{equation}
\begin{tikzpicture}[scale=0.6, baseline=(current  bounding  box.center)]
      
      \def\s{3.0};
      
      \path[-, black, thick] (-\s,0) edge (\s,0);
      
      \node at (-1*\s,0) [stuff_fill_red, scale=0.8]{$\mathbb{P}^1_a$};
      \node at (1*\s,0) [stuff_fill_red, scale=0.8]{$\mathbb{P}^1_b$};
      
\end{tikzpicture}
\end{equation}
On this curve, consider the line bundle
\begin{align}
\mathcal{L} = \mathcal{O}_C(d_a, d_b) \in \mathrm{Pic}( C ) \, .
\end{align}
That is, we consider $\mathcal{O}_{\mathbb{P}^1_a}(d_a) \in \mathrm{Pic}(\mathbb{P}^1_a)$ and $\mathcal{O}_{\mathbb{P}^1_b}(d_b) \in \mathrm{Pic}(\mathbb{P}^1_b)$ subject to gluing at the node $n$. Assume $d_a, d_b \geq 0$. Then
\begin{align}
    h^0( C, L ) = h^0 \left( \mathbb{P}^1_a, \mathcal{O}_{\mathbb{P}^1_a}(d_a) \right) + h^0 \left( \mathbb{P}^1_b, \mathcal{O}_{\mathbb{P}^1_b}(d_b) \right) - 1 = d_a + d_b + 1 \, .
\end{align}
\end{lem}

\begin{myproof}
We pick the following local coordinates and parametrize the local sections by $\alpha_i, \beta_i \in \mathbb{C}$:
\begin{align}
\begin{tabular}{ccc}
\toprule
Curve & Coordinates & Sections \\
\midrule
$C_1$ & $[x:y]$ & $\sum \limits_{i = 0}^{d_a}{\alpha_i \cdot x^{d_a-i} \cdot y^i}$ \\
$C_2$ & $[w:z]$ & $\sum \limits_{i = 0}^{d_b}{\beta_i \cdot w^{d_a-i} \cdot z^i}$ \\
\bottomrule
\end{tabular}
\end{align}
Without loss of generality, we can assume that the node is at the following position:
\begin{align}
\begin{tabular}{cc|cc}
\toprule
Label & Superset & Coordinates in $C_1$ & Coordinates in $C_2$ \\
\midrule
$n$ & $\mathbb{P}^1_a \cap \mathbb{P}^1_b$ & $[x:y] = [0:1]$ & $[w:z] = [0:1]$\\
\bottomrule
\end{tabular}
\end{align}
Enforcing gluing at $n_1$ is equivalent to $\alpha_{d_a} = \beta_{d_b}$. Hence it follows:
\begin{align}
    h^0( C, L ) = h^0 ( \mathbb{P}^1_a, \mathcal{O}_{\mathbb{P}^1_a}(d_a)) + h^0 ( \mathbb{P}^1_b, \mathcal{O}_{\mathbb{P}^1_b}(d_b) ) - 1 = d_a + d_b + 1 \, .
\end{align}
This completes the argument.
\end{myproof}

It is not too hard to generalize this result.

\begin{lem} \label{claim:NonSpecialIfAllDegreesNonNegative}
Consider a \emph{connected} rational nodal curve $C := \bigcup_{i \in I}{C_i}$ and $L$ a line bundle on it. (This means that $I$ is a simply connected graph and the $C_i$ are $\mathbb{P}^1$'s.)  Let $d_i$ be the degree of the restriction of $L$ to $C_i$. Suppose that $d_i \geq 0$ for all $i \in I$. Then 
\begin{align}
h^0( C, L ) = d + 1 \, , \qquad d = \sum_{i \in I}{d_i} \, .
\end{align}
\end{lem}

\begin{myproof}
We prove this statement by induction on $|I|$. If $|I| = 1$, then the statement is obviously true.

For the induction step, consider a connected, rational curve $C$ with $n$ components. Assume that our statement is true for $C$. By adding one component $C_{n+1}$ to $C$, we obtain a new curve $C^\prime$. Assume that $C^\prime$ is tree-like, i.e. $C_{n+1}$ intersects exactly one component of $C$ in exactly one point. Without loss of generality, let us denote this component as $C_n$ and $p = C_n \cap C_{n+1}$.

The global sections on $C_n$ correspond to a subset $S \subseteq H^0( C_n, \mathcal{O}_{C_n}( d_n))$ on $C_n$. Let us glue these sections to $H^0(C_{n+1}, \mathcal{O}_{C_{n+1}}(d_{n+1}))$ at $p$. Without loss of generality, we make the following choices:
\begin{align}
\begin{tabular}{ccc}
\toprule
Curve & Coordinates & Sections \\
\midrule
$C_n$ & $[x:y]$ & $S \subseteq \mathrm{Span}_{\mathbb{C}} \left( x^{d_n}, x^{d_n-1} y, \dots, y^{d_n} \right)$ \\
$C_{n+1}$ & $[u:v]$ & $\mathrm{Span}_{\mathbb{C}} \left( u^{d_{n+1}}, u^{d_{n+1}-1} v, \dots, v^{d_{n+1} }\right)$ \\
\bottomrule
\end{tabular}
\end{align}
Choose $p = V( x )$ on $C_n$ and $p = V( u )$ on $C_{n+1}$. Let us now assume that no gluing condition in $p$ was required. This would mean that for any $\varphi \in S$ and $\psi \in H^0(C_{n+1}, \mathcal{O}_{C_{n+1}}(d_{n+1}))$, they satisfy
\begin{align}
\varphi( p ) = \psi( p ) \, .
\end{align}
However, since $u^{d_{n+1}}, v^{d_{n+1}} \in H^0( C_{n+1}, \mathcal{O}_{C_{n+1}} (d_{n+1}))$, this would imply that any $\varphi \in S$ satisfies
\begin{align}
\varphi( p ) = u^{d_{n+1}}( p ) = 0 \, , \qquad \varphi( p ) = v^{d_{n+1}}( p ) = 1 \, .
\end{align}
This cannot be satisfied in $\mathbb{C}$. Hence, there is at least one condition to be imposed at $p$. However, since $p$ imposes at most one condition, it imposes exactly one condition. Recall that by assumption, 
\begin{align}
h^0( C, L ) = d + 1 \, , \qquad d = \sum_{i \in I}{d_i} \, .
\end{align}
Consequently, we find for $C^\prime$ that:
\begin{align}
h^0( C^\prime, L ) = h^0( C, L ) + h^0( C_{n+1}, L ) - 1 = d + 1 + ( d_{n+1} + 1 ) - 1 = \left( d + d_{n+1} \right) + 1 \, .
\end{align}
This completes the argument.
\end{myproof}

\begin{cor} \label{cor:H0AllDegreesNonNegative}
Consider a rational nodal curve $C := \bigcup_{i \in I}{C_i}$ and a line bundle $L$ on it. Suppose that $C$ has $k$ connected components and the restrictions of $L$ to $C_i$ have degrees $d_i \geq 0$. Then 
\begin{align}
h^0( C, L ) = \Bigl( \sum_{i \in I}{d_i} \Bigr) + k \, .
\end{align}
\end{cor}

We are now in the position to formulate the result which allows one to count the global sections on any terminal tree.

\begin{prop} \label{cor:H0AllDegrees}
Let $(C,L)$ be terminal and $C_+$ have $k$ connected components. Then, 
\begin{align}
h^0( C, L ) = \Bigl( \sum_{i \in I_+}{d_i} \Bigr) + k \, .
\end{align}
\end{prop}

\begin{myproof}
Since $(C, L)$ is terminal, $C_+ \cap C_- = \emptyset$ and
\begin{align}
h^0(C, L) = h^0(C_+, L|_{C_+}) + h^0(C_-, L|_{C_-}) = h^0(C_+, L|_{C_+}) \, .
\end{align}
Now, let $e$ be the number of intersections between curves in $C_+$. Then, $e = |I_+|-k$ and
\begin{align}
h^0(C, L) = \Bigl( \sum_{i \in I_+}{(d_i + 1)} \Bigr) - e = \Bigl( \sum_{i \in I_+}{d_i} \Bigr) + |I_+| - e = \Bigl( \sum_{i \in I_+}{d_i} \Bigr) + k \, .
\end{align}
This completes the argument.
\end{myproof}

\subsection{Final algorithm}

\begin{algo} \label{algo1}
Consider a connected, rational nodal curve $C := \bigcup_{i \in I}{C_i}$ and a line bundle $L$ on it. Then, the following algorithm computes $h^0( C, L)$.
\begin{enumerate}
 \item Simplify $(C,L) \to (C_+, L_+) \to \dots (C_+^{(n)}, L_+^{(n)} )$ with terminal $(C_+^{(n)}, L_+^{(n)})$.
 \item Then $h^0( C,L ) = d_+\left(L_+^{(n)}\right) + c_+\left( C_+^{(n)} \right)$, where
                  \begin{itemize}
                   \item $d_+(L_+^{(n)}) = \sum_{i \in I_+( C_+^{(n)}}{d_i}$.
                   \item $c_+(C_+^{(n)})$ is the number of connected components of $C_+^{(n)}$.
                  \end{itemize}
\end{enumerate}
\end{algo}

\begin{note}
Since $C_+$ has less or as many components as $C$, this algorithm must terminate after a finite number of iterations. Let us illustrate this algorithm in a couple of examples.
\end{note}

\begin{exmp}
We first perform the following simplification:
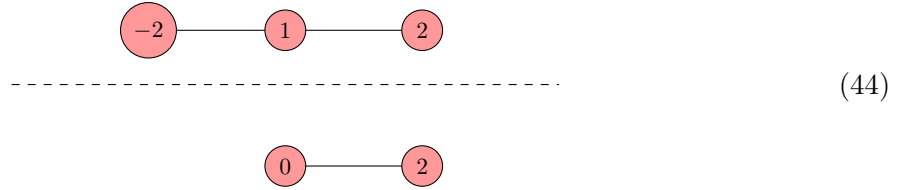
\begin{equation}
\begin{tikzpicture}[scale=0.6, baseline=(current  bounding  box.center)]
      \def\s{3.0};
      \def\h{3.0};
      \path[-] (0,0) edge (2*\s,0);
	  \path[-] (1*\s,-\h) edge (2*\s,-\h);
      \node at (0,0) [stuff_fill_red, scale=0.8]{$-2$};
      \node at (1*\s,0) [stuff_fill_red, scale=0.8]{$1$};
      \node at (2*\s,0) [stuff_fill_red, scale=0.8]{$2$};
      \draw[dashed] (-\s,-0.4*\h) -- (3*\s,-0.4*\h);
      \node at (1*\s,-\h) [stuff_fill_red, scale=0.8]{$0$};
      \node at (2*\s,-\h) [stuff_fill_red, scale=0.8]{$2$};
\end{tikzpicture}
\end{equation}
The latter is terminal and we have
\begin{align}
h^0( C, L ) = (0+2) + 1 = 3 \, .
\end{align}
\end{exmp}

\begin{exmp}
We perform the following simplifications:
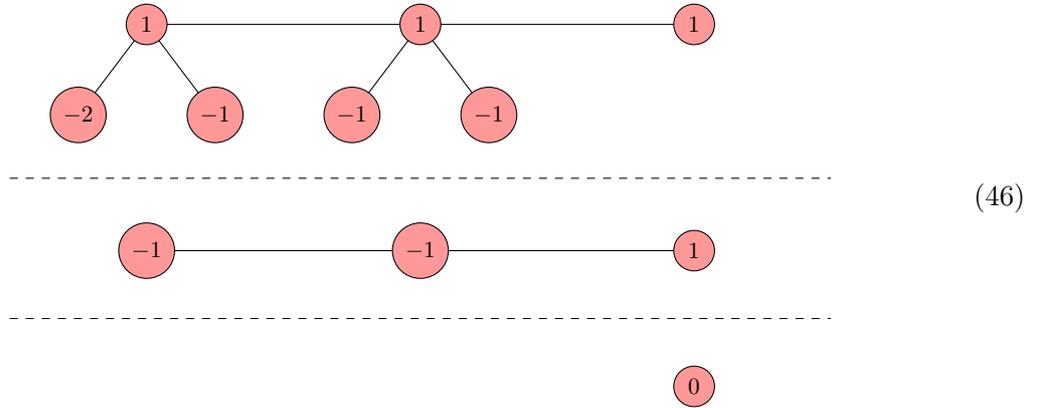
\begin{equation}
\begin{tikzpicture}[scale=0.6, baseline=(current  bounding  box.center)]
      
      \def\s{3.0};
      \def\h{2};
      \def\t{5};
      
      \path[-] (0,0) edge (4*\s,0);
      \path[-] (-0.5*\s,-\h) edge (0,0);
      \path[-] (0.5*\s,-\h) edge (0,0);
      \path[-] (1.5*\s,-\h) edge (2*\s,0);
      \path[-] (2.5*\s,-\h) edge (2*\s,0);
      
      \node at (0,0) [stuff_fill_red, scale=0.8]{$1$};
      \node at (2*\s,0) [stuff_fill_red, scale=0.8]{$ 1$};
      \node at (4*\s,0) [stuff_fill_red, scale=0.8]{$1$};
      \node at (-0.5*\s,-\h) [stuff_fill_red, scale=0.8]{$-2$};
      \node at (0.5*\s,-\h) [stuff_fill_red, scale=0.8]{$-1$};
      \node at (1.5*\s,-\h) [stuff_fill_red, scale=0.8]{$-1$};
      \node at (2.5*\s,-\h) [stuff_fill_red, scale=0.8]{$-1$};
      
      \draw[dashed] (-\s,-0.68*\t) -- (5*\s,-0.68*\t);
      \path[-] (0,-\t) edge (4*\s,-\t);
      \node at (0,-\t) [stuff_fill_red, scale=0.8]{$-1$};
      \node at (2*\s,-\t) [stuff_fill_red, scale=0.8]{$-1$};
      \node at (4*\s,-\t) [stuff_fill_red, scale=0.8]{$1$};

      \draw[dashed] (-\s,-1.3*\t) -- (5*\s,-1.3*\t);
      \node at (4*\s,-1.6*\t) [stuff_fill_red, scale=0.8]{$0$};

\end{tikzpicture}
\end{equation}
The latter is terminal and we have $h^0( C, L) = 0+1 = 1$.
\end{exmp}

\section{Towards Brill-Noether theory of limit root line bundles} \label{sec:TowardsBNOfLimitRootBundles}

We now turn back to the QSM geometries. Our task is to enumerate all physically relevant limit root bundles on the nodal quark-doublet curve of the QSM geometries. Subsequently, we want to identify their number of global sections. To this end, we first briefly explain the geometric multiplicity (see \cite{2004math4078C} for more details) and then turn towards the counts relevant for the QSMs.

\subsection{The geometric multiplicity} \label{sec:GeoMult}

Recall that every line bundle $L$ with degree divisible by $r$ has $r^{2g}$ roots on a smooth curve of genus $g$. For a nodal curve $C^\bullet$, the root bundle construction in \cite{2004math4078C} yields less than $r^{2g}$ limit roots. Indeed, let $\Gamma_C$ be the dual graph of $C^\bullet$ with edge set $E(\Delta)$ and vertex set $V(\Delta)$. Fix a weighted subgraph $\Delta^w$. Its edges are the nodes at which we blow up and so, it determines a partial normalization $\widetilde{\pi}_{\Delta^w}: \widetilde{C}_{\Delta^w} \rightarrow C^\bullet$ with dual graph $\Gamma_C \setminus E(\Delta^w)$ and partial blow-up $C^\circ_{\Delta^w}$. Its weights $u_i, v_i$ determine a line bundle $\widetilde{\pi}^*_{\Delta_w}(L)(-\sum_i (u_ip_i + v_iq_i))$ on $\widetilde{C}_{\Delta^w}$ of which we can take the $r$th root $\widetilde{P}_{\Delta^w} \in \Pic(\widetilde{C}_{\Delta^w})$, i.e. 
\begin{align}
(\widetilde{P}_{\Delta^w})^{\otimes r} \cong \widetilde{\pi}^*_{\Delta_w}(L)\Bigl(-\sum_{n_i \in E(\Delta^w)} (u_ip_i + v_iq_i)\Bigr).
\end{align}
Degree 1 line bundles on the exceptional components are then glued to $\widetilde{P}_{\Delta^w}$ to obtain a limit root $P^\circ_{\Delta^w}$ on $C_{\Delta^w}^\circ$. To count the limit roots associated to $\Delta^w$, consider the tower of normalizations:
\begin{align} \nu: C^\nu \xrightarrow{\nu_{\Delta^w}} \widetilde{C}_{\Delta^w} \xrightarrow{\widetilde{\pi}_{\Delta^w}} C^\bullet,
\end{align}
where $\nu_{\Delta^w}$ is the normalization of $\widetilde{C}_{\Delta^w}$ and $C^\nu$ is the full normalization of $C^\bullet$ with geometric genus $g^{\nu}$ and arithmetic genus $g(C^\nu) = g^\nu + 1 - |V(\Delta)|$. The other arithmetic genera can be expressed in terms of $g^\nu$ as:
\begin{align} 
g(C^\bullet) &= g(C^\nu) + |E(\Delta)| = g^\nu + b_1(\Gamma_C). \\
g(\widetilde{C}_{\Delta^w})  &= g(C^\nu) + |E(\Delta)\setminus E(\Delta^w)| = g^\nu + b_1(\Gamma_C \setminus E(\Delta^w)).
\end{align}
The two genera above are related to each other as
\begin{align} 
g(C^\bullet) &= g(\widetilde{C}_{\Delta^w}) + |E(\Delta^w)| = \sum_{i \in I} g(X_i) + b_1(\Sigma_C),
\end{align}
where $X_i$ is a \emph{connected} component of $\widetilde{C}_{\Delta^w}$ and $\Sigma_C$ is the dual graph whose vertices are $X_i$ and whose edges are the exceptional components. Taking pullbacks induces the composition
\begin{align} \nu^*: \Pic(C^\bullet) \xrightarrow{\widetilde{\pi}^*_{\Delta^w}} \Pic( \widetilde{C}_{\Delta^w}) \xrightarrow{\nu^*_{\Delta^w}}  \Pic(C^\nu).
\end{align}
Each root $\widetilde{P}_{\Delta^w}$ on $\widetilde{C}_{\Delta^w}$ corresponds to a root on the full normalization with a choice of gluings at the nodes that were not blown up. Since there are $r^{2g^\nu}$ roots of $\nu_{\Delta^w}^*\widetilde{\pi}^*(L)(-\sum_i (u_ip_i + v_iq_i))$ on $C^\nu$ and $r^{b_1(\Gamma_C\setminus E(\Delta^w))}$ choices of gluings, every $\Delta^w$ is associated to $r^{2g^\nu + b_1(\Gamma_C\setminus E(\Delta^w))}$ limit $r$th roots. Summing over the $r^{b_1(\Gamma_C)}$ weighted subgraphs gives
\begin{align} \sum_{\Delta^w} r^{2g^{\nu} + b_1(\Gamma_C \setminus E(\Delta^w))} 
\end{align}
limit $r$th roots. To get $r^{2g}$ limit $r$th roots, we need to multiply each summand with a prefactor called the \emph{geometric multiplicity} in \cite{2004math4078C}, namely 
\begin{align*}
\mu_{\Delta^w} = r^{b_1(\Gamma_C)- b_1(\Gamma_C \setminus E(\Delta^w))} = r^{|E(\Delta^w)|} = r^{b_1(\Sigma_C)}.
\end{align*}
One can interpret this geometrically by saying that a single limit root line bundle corresponds to several root bundles on the corresponding smooth, irreducible curve. In \cite{2004math4078C}, $\mu_{\Delta^w}$ is the geometric multiplicity of the connected component of the moduli space of limit roots supported at a given limit root (Lemma 4.1.1). Taking the geometric multiplicity into account, we do indeed get $r^{2g}$ limit roots because
\begin{align} \label{equ:geom}
\sum_{\Delta^w}( r^{2g^{\nu} + b_1(\Gamma_C \setminus E(\Delta^w))} \cdot \mu_{\Delta_w}) = 
 \sum_{\Delta^w} r^{2g^{\nu}+b_1(\Gamma_C)}  = r^{b_1(\Gamma_C)} \cdot r^{2g^\nu + b_1(\Gamma_C)}  = r^{2g}.
\end{align}
In \cite{Bies:2021xfh}, only the blow-ups at the full set of nodes were considered, i.e. $E(\Delta^w) = E(\Delta)$. Hence, $\mu_{\Delta^w} = r^{b_1(\Gamma_C)}$ in that case. The first equality of \oref{equ:geom} shows that each $\Delta^w$ is associated to $r^{2g^\nu + b_1(\Gamma_C)}$ limit roots when the geometric multiplicity is taken into account. Since this quantity does not depend on the weighted subgraph, we will define this specially.
\begin{defn} \label{equ:Multiplicity}
The number of limit roots associated to each weighted subgraph $\Delta^w$ is 
\begin{align}\label{equ:mult}
\mu := r^{2g^\nu + b_1(\Gamma_C)} \,. 
\end{align}
\end{defn} 

We conclude this section by studying a simple example taken from \cite{2004math4078C}. To this end, consider two curves $C_1$, $C_2$ with general genera $g_1$, $g_2$ (marked in dark pink) and assume that they touch in exactly 3 nodes. The goal is to identify all 3rd limit roots of the structure sheaf. Consequently, we look at the following dual graph:
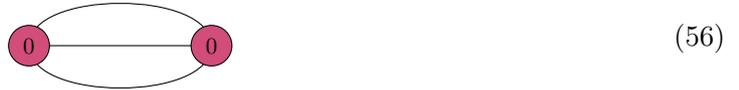
\begin{equation}
\begin{tikzpicture}[scale=0.6, baseline=(current  bounding  box.center)]
      
      \def\s{2.0};
      \def\h{1.3};
      
      \path[-] (-\s,0) edge (\s,0);
      \path[-, out = 90, in = 90, looseness = 0.8] (-\s,0) edge (\s,0);
      \path[-, out = -90, in = -90, looseness = 0.8] (-\s,0) edge (\s,0);
      
      \node at (-1*\s,0) [stuff_fill_high, scale=0.8]{0};
      \node at (1*\s,0) [stuff_fill_high, scale=0.8]{0};
      
\end{tikzpicture}
\end{equation}
This graph has $b_1( \Gamma_C ) = 2$. We list all limit roots in \oref{tab:Example}. In particular, we find a total number of roots:
\begin{align}
\begin{split}
N_{\text{total}} &= 1 \cdot r^{0} \cdot r^{2 g_\nu + 2} + 6 \cdot r^2 \cdot  r^{2 g_\nu} + 2 \cdot r^2 \cdot r^{2 g_\nu} = (1 + 6 + 2) \cdot \mu = 3^{2 \left( 2 + g_{\nu} \right)} \, .
\end{split}
\end{align}
Since $g(C) = 2 + g_{\nu} = 2 + g_1 + g_2$, this is exactly the expected number.
\begin{table}[htb]
\centering{
\begin{adjustbox}{max width=0.9\textwidth}
\begin{tabular}{cc|cc}
\toprule
$N$ & Weighted diagram $\Delta^w$ & $\mu_{\Delta^w}$ & $N_P$\\
\midrule
3 &
\adjustbox{valign=c}{\begin{tikzpicture}[scale=0.6, baseline=(current  bounding  box.center)]
      \def\s{1.0};
      \node at (-1*\s,0) [stuff_fill_high, scale=0.6]{0};
      \node at (1*\s,0) [stuff_fill_high, scale=0.6]{0};
\end{tikzpicture}}
& $r^{2-2}$ & $r^{\left( 2g_1 + 2g_2 \right) + 2}$ \\
\midrule
2 & inconsistent & -- & -- \\
\midrule
1 & 
\adjustbox{valign=c}{\begin{tikzpicture}[scale=0.6, baseline=(current  bounding  box.center)]
      \def\s{1.8};
      \def\h{-1.8};
      \def\d{6.0};
      \path[-, every node/.append style={fill=white},out=90,in =90] (-\s,0) edge node[pos=0.35] {2} node[pos=0.65] {1} (\s,0);
      \path[-, every node/.append style={fill=white},out=0,in =180] (-\s,0) edge node[pos=0.35] {1} node[pos=0.65] {2} (\s,0);
      \node at (-1*\s,0) [stuff_fill_high, scale=0.6]{0};
      \node at (1*\s,0) [stuff_fill_high, scale=0.6]{0};
      \path[-, every node/.append style={fill=white},out=90,in =90] (-\s+\d,0) edge node[pos=0.35] {1} node[pos=0.65] {2} (\s+\d,0);
      \path[-, every node/.append style={fill=white},out=0,in =180] (-\s+\d,0) edge node[pos=0.35] {2} node[pos=0.65] {1} (\s+\d,0);
      \node at (-1*\s+\d,0) [stuff_fill_high, scale=0.6]{0};
      \node at (1*\s+\d,0) [stuff_fill_high, scale=0.6]{0};
      \path[-, every node/.append style={fill=white},out=90,in =90] (-\s,1.0*\h) edge node[pos=0.35] {2} node[pos=0.65] {1} (\s,1.0*\h);
      \path[-, every node/.append style={fill=white},out=-90,in =-90] (-\s,1.0*\h) edge node[pos=0.35] {1} node[pos=0.65] {2} (\s,1.0*\h);
      \node at (-1*\s,1.0*\h) [stuff_fill_high, scale=0.6]{0};
      \node at (1*\s,1.0*\h) [stuff_fill_high, scale=0.6]{0};
      \path[-, every node/.append style={fill=white},out=90,in =90] (-\s+\d,1.0*\h) edge node[pos=0.35] {1} node[pos=0.65] {2} (\s+\d,1.0*\h);
      \path[-, every node/.append style={fill=white},out=-90,in =-90] (-\s+\d,1.0*\h) edge node[pos=0.35] {2} node[pos=0.65] {1} (\s+\d,1.0*\h);
      \node at (-1*\s+\d,1.0*\h) [stuff_fill_high, scale=0.6]{0};
      \node at (1*\s+\d,1.0*\h) [stuff_fill_high, scale=0.6]{0};
      \path[-, every node/.append style={fill=white},out=0,in =180] (-\s,2.0*\h) edge node[pos=0.35] {2} node[pos=0.65] {1} (\s,2.0*\h);
      \path[-, every node/.append style={fill=white},out=-90,in =-90] (-\s,2.0*\h) edge node[pos=0.35] {1} node[pos=0.65] {2} (\s,2.0*\h);
      \node at (-1*\s,2.0*\h) [stuff_fill_high, scale=0.6]{0};
      \node at (1*\s,2.0*\h) [stuff_fill_high, scale=0.6]{0};
      \path[-, every node/.append style={fill=white},out=0,in =180] (-\s+\d,2.0*\h) edge node[pos=0.35] {1} node[pos=0.65] {2} (\s+\d,2.0*\h);
      \path[-, every node/.append style={fill=white},out=-90,in =-90] (-\s+\d,2.0*\h) edge node[pos=0.35] {2} node[pos=0.65] {1} (\s+\d,2.0*\h);
      \node at (-1*\s+\d,2.0*\h) [stuff_fill_high, scale=0.6]{0};
      \node at (1*\s+\d,2.0*\h) [stuff_fill_high, scale=0.6]{0};
\end{tikzpicture}} & $r^{2-0}$ & $r^{\left( 2g_1 + 2g_2 \right) + 0}$ \\
\midrule
0 & \adjustbox{valign=c}{\begin{tikzpicture}[scale=0.6, baseline=(current  bounding  box.center)]
      \def\s{2.0};
      \def\d{6.0};
      \path[-, every node/.append style={fill=white},out=90,in =90] (-\s,0) edge node[pos=0.35] {2} node[pos=0.65] {1} (\s,0);
      \path[-, every node/.append style={fill=white},out=0,in =180] (-\s,0) edge node[pos=0.35] {2} node[pos=0.65] {1} (\s,0);
      \path[-, every node/.append style={fill=white},out=-90,in =-90] (-\s,0) edge node[pos=0.35] {2} node[pos=0.65] {1} (\s,0);
      \node at (-1*\s,0) [stuff_fill_high, scale=0.6]{0};
      \node at (1*\s,0) [stuff_fill_high, scale=0.6]{0};
      \path[-, every node/.append style={fill=white},out=90,in =90] (-\s+\d,0) edge node[pos=0.35] {1} node[pos=0.65] {2} (\s+\d,0);
      \path[-, every node/.append style={fill=white},out=0,in =180] (-\s+\d,0) edge node[pos=0.35] {1} node[pos=0.65] {2} (\s+\d,0);
      \path[-, every node/.append style={fill=white},out=-90,in =-90] (-\s+\d,0) edge node[pos=0.35] {1} node[pos=0.65] {2} (\s+\d,0);
      \node at (-1*\s+\d,0) [stuff_fill_high, scale=0.6]{0};
      \node at (1*\s+\d,0) [stuff_fill_high, scale=0.6]{0};      
\end{tikzpicture}} & $r^{2-0}$ & $r^{\left( 2g_1 + 2g_2 \right) + 0}$ \\
\bottomrule
\end{tabular}
\end{adjustbox}}
\caption{All limit roots on a nodal curve example taken from \cite{2004math4078C}. The geometric multiplicity is given by $\mu_{\Delta^w} = r^{b_1(\Gamma_C) - b_1(\Gamma_C \setminus E(\Delta^w))}$. The number of nodes left from the blow-up is denoted by $N$. The number of limit roots is denoted by $N_P = r^{2 g_\nu + b_1(\Gamma_C \setminus E(\Delta^w))}$.}
\label{tab:Example}
\end{table}

\subsection{Errata, corrections and immediate extensions of earlier work}

By taking the geometric multiplicity into account, we can refine the statistics derived in the earlier publication \cite{Bies:2021xfh}. There, lower bounds $\check{N}^{(3)}_P$ on the number of limit root bundles with $h^0 = 3$ were derived. By taking the geometric multiplicity into account, we can refine these results. Our findings are listed in \oref{tab:ImprovedResultsPreviousWork}. Note that we have grouped the spaces $\Delta^\circ_{128}$, $\Delta^\circ_{130}$, $\Delta^\circ_{136}$, $\Delta^\circ_{236}$ together. This is because the dual graphs are identical for these spaces. We point out a typo in \cite{Bies:2021xfh} for the result of $\Delta^\circ_{387}$. The correct result is highlighted in green color in \oref{tab:ImprovedResultsPreviousWork}. There is also an incorrect result for $\Delta^\circ_{856}$ in this earlier publication. The dual graph of $\Delta^\circ_{856}$ is identical to that of $\Delta^\circ_{882}$, so the lower bounds $\check{N}^{(3)}_P$ must agree for both spaces. A detailed check shows that the results for $\Delta^\circ_{856}$ coincide with the numbers listed for $\Delta^\circ_{882}$ in \cite{Bies:2021xfh}.

This brief study shows that roughly 50\% of all limit root bundles are obtained from a full blowup and have exactly three global sections. The largest percentage is found for $\Delta_{88}$ (61.1\%).
In \cite{Bies:2021xfh}, the geometric multiplicity was not taken into account. These earlier counts falsely led us to the conclusion that for the polytope $\Delta_8$, the largest fraction of full blowup roots with three global sections were found. However, as listed above, the actual percentage is only 57.3\% which is strictly smaller than the result for $\Delta_{88}$.

Before we continue, let us point out that the observations from this initial study are subject to the following subtleties. First, there are more limit root bundles, resulting from partial blowups. Second, for physics applications, we are eventually interested in physical roots. We will discuss the first point momentarily and turn to the second in \oref{sec:D4}.

\begin{table}[tb]
\begin{center}
\begin{tabular}{cc|ccc|c}
\toprule
& & $\check{N}^{(3)}_P$ & $\mu$ & $N_{\text{total}}$ & $\check{N}^{(3)}_P \cdot \mu /N_{\text{total}}$ [\%]\\
\midrule
\multirow{4}{*}{$\overline{K}_{B_3}^3 = 6$} & $\Delta^\circ_{8}$ & $142560$ & $12^3$ & $12^8$ & $57.3$ \\
& $\Delta^\circ_{4}$ & $11110$ & $12^4$ & $12^8$ & $53.6$ \\
& $\Delta^\circ_{134}$ & $10100$ & $12^4$ & $12^8$ & $48.7$ \\
& $\Delta^\circ_{128}$, $\Delta^\circ_{130}$, $\Delta^\circ_{136}$, $\Delta^\circ_{236}$ & $8910$ & $12^4$ & $12^8$ & $42.0$ \\
\midrule
\multirow{21}{*}{$\overline{K}_{B_3}^3 = 10$} & $\Delta^\circ_{88}$ & $781.680.888$ & $20^5$ & $20^{12}$ & $61.1$ \\ 
& $\Delta^\circ_{110}$ & $738.662.983$ & $20^5$ & $20^{12}$ & $57.8$ \\
& $\Delta^\circ_{272}$, $\Delta^\circ_{274}$ & $736.011.640$ & $20^5$ & $20^{12}$ & $57.5$ \\
& $\Delta^\circ_{387}$ & \textcolor{green}{$733.798.300$} & $20^5$ & $20^{12}$ & $57.3$ \\
& $\Delta^\circ_{798}$, $\Delta^\circ_{808}$, $\Delta^\circ_{810}$, $\Delta^\circ_{812}$ & $690.950.608$ & $20^5$ & $20^{12}$ & $54.0$ \\
\cmidrule{2-6}
& $\Delta^\circ_{254}$ & $35.004.914$ & $20^6$ & $20^{12}$ & $54.7$ \\
& $\Delta^\circ_{302}$ & $34.908.682$ & $20^6$ & $20^{12}$ & $54.7$ \\
& $\Delta^\circ_{52}$ & $34.980.351$ & $20^6$ & $20^{12}$ & $54.7$ \\
& $\Delta^\circ_{786}$ & $32.860.461$ & $20^6$ & $20^{12}$ & $51.3$ \\
& $\Delta^\circ_{762}$ & $32.858.151$ & $20^6$ & $20^{12}$ & $51.3$ \\
\cmidrule{2-6}
& $\Delta^\circ_{417}$ & $32.857.596$ & $20^6$ & $20^{12}$ & $51.3$ \\
& $\Delta^\circ_{838}$ & $32.845.047$ & $20^6$ & $20^{12}$ & $51.3$ \\
& $\Delta^\circ_{782}$ & $32.844.379$ & $20^6$ & $20^{12}$ & $51.3$ \\
& $\Delta^\circ_{377}$, $\Delta^\circ_{499}$, $\Delta^\circ_{503}$ & $30.846.440$ & $20^6$ & $20^{12}$ & $48.2$ \\
& $\Delta^\circ_{1348}$ & $30.845.702$ & $20^6$ & $20^{12}$ & $48.2$ \\
\cmidrule{2-6}
& $\Delta^\circ_{882}$, \textcolor{red}{$\Delta^\circ_{856}$} & $30.840.098$ & $20^6$ & $20^{12}$ & $48.2$ \\
& $\Delta^\circ_{1340}$ & $28.954.543$ & $20^6$ & $20^{12}$ & $45.2$ \\
& $\Delta^\circ_{1879}$ & $28.950.852$ & $20^6$ & $20^{12}$ & $45.2$ \\
& $\Delta^\circ_{1384}$ & $27.178.020$ & $20^6$ & $20^{12}$ & $42.5$ \\
\bottomrule
\end{tabular}
\end{center}
\caption{The results of \cite{Bies:2021xfh} listed an incorrect result for $\Delta^\circ_{856}$ and displayed a typo for $\Delta^\circ_{387}$. We fix both and refine the earlier findings  by use of the geometric multiplicity.}
\label{tab:ImprovedResultsPreviousWork}
\end{table}

\subsection{Statistics of partial blowup limit roots in QSM geometries}

Let us employ these techniques to the nodal curves relevant for the QSMs. To this end we employ the algorithms available in \texttt{gap-4}-package \emph{QSMExplorer}. The latter is part of the \texttt{ToricVarieties$\_$project} \cite{ToricVarietiesProject}. Before we list the results, let us discuss the limitations:
\begin{enumerate}
 \item Not all partial blowups come from tree-like subgraphs. Whenever this is the case, the technology of \oref{sec:LineBundleCohomologyOnTreelikeCurves} does not apply. Indeed, such configurations matter for the   QSM geometries but their number is typically small compared to the tree-like cases. This includes circuits which admit a jump provided that the nodes are moved in special alignment. We will analyze curves of this type in more detail in \oref{sec:D4}. At this point, we cannot treat such curves systematically, but rather rely on a case-by-case study. Therefore, our algorithm merely provides a lower bound in such instances. 
 \item If the nodal quark-doublet curve contains at least one elliptic curve, then we face the following limitations for partial blowup root bundles. First, if the elliptic curve contains one node that is not blown-up, then the technology of \oref{sec:LineBundleCohomologyOnTreelikeCurves} does not apply. Otherwise, if there is a $d = 0$ line bundle on an elliptic curve and we are looking for $r$-th roots, then we know that $r^2-1$ of the $r^2$ roots are guaranteed to have no global sections. The number of global sections on the remaining root depends on whether the original bundle was trivial or not. Distinguishing this is highly non-trivial and definitely goes beyond the topological data considered in our computer scan. Thus, in those cases we provide a lower bound.
\end{enumerate}
If we face one of these two cases, then we can only provide a lower bound to the number of global sections. In particular, in non-generic setups, the actual number of global sections could be strictly larger than this lower bound. Therefore, we will list the roots for which we determine the exact number of global sections and the roots for which we can currently only provide a lower bound separately. For $\Delta_4^\circ$ with $\mu = 12^{4}$, our counter finds the following:
\begin{align}
\begin{tabular}{c|cc|cc}
\toprule
 & \multicolumn{2}{c|}{Percentages} & \multicolumn{2}{c}{Absolute numbers} \\
$N$ & $h^0 = 3$ & $h^0 \geq 3$ & $h^0 = 4$ & $h^0 \geq 4$ \\
\midrule
0 & 53.6 & & 11110 & \\
1 & 36.7 & & 7601 & \\
2 & 7.5 & 0.5 & 1562 & 110 \\
3 & 1.3 & 0.1 & 264 & 11 \\
4 & & 0.3 & & 66 \\
5 & & 0.1 & & 11 \\
7 & & 0.0 & & 1 \\
\midrule
$\Sigma$ & 99.0 & 1.0 & 20537 & 199 \\
\bottomrule
\end{tabular} \label{equ:DetailedCount}
\end{align}
Note that we omit trivial results. Also, the percentages are rounded to one decimal place. Hence, the number 0.0 indicates a non-zero result smaller than $0.045$. We observe that $12^4 \cdot \left( 20537 + 199 \right) = 12^8$. Hence, our algorithm indeed enumerated all limit root bundles, in particular partial blowup limit roots. Therefore, it could tell that at least 99\% of all limit roots have exactly three sections on $C_{(\mathbf{3},\mathbf{2})_{1/6}}^\bullet$ irrespective of the complex structure! The remaining percentage of limit roots begs to be investigated in more detail. We will return to this task momentarily in \oref{sec:D4}. Counts similar to \oref{equ:DetailedCount} are listed in \oref{appendix:DetailsOnCounts}. For an overview, we will list the percentages rather than the absolute numbers. For the families of QSM spaces with $\overline{K}_{B_3}^3 = 6$, we summarize our findings in \oref{tab:Results1}. Note that $E$ denotes the number of edges and that the nodal quark-doublet curve in $\Delta_8^\circ$ contains an elliptic curve. This is the reason why our ignorance is much larger for this curve. The results in \oref{tab:Results1} can be computed in a few minutes on a personal computer. However, for the spaces with $\overline{K}_{B_3}^3 = 10$, the computational challenge increases a lot:
\begin{enumerate}
 \item The number of ways to conduct partial blowups grows as $2^N$ with $E$ the number of edges. For $\overline{K}_{B_3}^3 = 6$ spaces, there are at most edges 9 in the dual graph of $C_{(\mathbf{3},\mathbf{2})_{1/6}}^\bullet$. However, for spaces with $\overline{K}_{B_3}^3 = 10$, we can have up to 15 edges.  Thus, the number of combinatorial possibilities increases from at most 512 for $\overline{K}_{B_3}^3 = 6$ to up to 32768 for $\overline{K}_{B_3}^3 = 10$.
 \item For each of these partial blowups, determining the number of limit root line bundles takes longer since the graph is more complicated.
 \item We repeat such a scan for each value of $h^0$ that we want to find the limit root bundles for. For the spaces with $\overline{K}_{B_3}^3 = 6$, we find all roots after running our algorithm for $h^0 = 3$. However, for $\overline{K}_{B_3}^3 = 10$, we find limit root bundles with $h^0 \in \{ 3,4,5,6 \}$, so that we have to repeat this step up to four times.
\end{enumerate}
In consequence, even on the super-computer \texttt{plesken.mathematik.uni-siegen.de}, the computations for the most complicated setups take almost four days. Eventually, this leads to the results summarized in \oref{tab:Results2}.

As we can see from \oref{tab:Results1} and \oref{tab:Results2}, the biggest limitation for our algorithm is the handling of elliptic curve components. Indeed, whenever at least one $g = 1$ component is present, we can only guarantee that roughly at least $75\%$ of the limit root bundles have $h^0 = 3$. However, whenever no such elliptic curve is present, our ignorance drops significantly and our algorithm finds that at least about $95\%$ have $h^0 = 3$. With an eye towards the F-theory QSMs, it is interesting to wonder if we can lower this ignorance even further. Our immediate focus turns to the polytopes $\Delta_4^\circ$, $\Delta_{134}^\circ$, $\Delta_{128}^\circ$, $\Delta_{130}^\circ$, $\Delta_{136}^\circ$ and $\Delta_{236}^\circ$ in which we find that at least $99\%$ of all limit root line bundles have $h^0 = 3$. We will study $\Delta_4$ momentarily in \oref{sec:D4}.

Another interesting aspect is to compare our findings to the classical Brill-Noether theory in which one has a continuous space of line bundles and estimates the dimension of the variety of all line bundles with a certain $h^0$. This concept does not apply for root bundles since there are only finitely many of them in the first place. To the best of our knowledge, the above tables \oref{tab:Results1} and \oref{tab:Results2} (more details in \oref{appendix:DetailsOnCounts}) are the first counts/estimates for such bundles. In this sense, we propose to consider them a first response to the question ``what is Brill-Noether theory for limit roots?''. In particular notice from \oref{tab:Results2}, that for the QSM setups with $\overline{K}_{B_3}^3 = 10$ we find limit root bundles which always have more than minimal number of global sections.

In the next section, we will study limit roots on circuit-like subgraphs. Since these setups lead to the computation of line bundle cohomology on circuit-like nodal curves, we cannot handle them systematically but rather conduct a case-by-case study. By focusing on $\Delta_4^\circ$, we then find limit roots for which the number of global sections depends on the relative position of the nodes, i.e. depends on the complex structure of this nodal curve. For this reason, we term such limit roots \emph{jumping-circuits}. In particular, this phenomenon seems to resemble the classical Brill-Noether jump on a smooth, irreducible curve. In the latter case, the points on the curve corresponding to the line bundle divisor in question align such that the Serre dual bundle admits a global section.

\begin{table}[tb]
\begin{center}
\begin{tabular}{cc|cc|cc}
\toprule
Polytope & $E$ & $h^0 = 3$ & $h^0 \geq 3$ & $h^0 = 4$ & $h^0 \geq 4$ \\
\midrule
$\Delta_8^\circ$ & 6 & 76.4 & 23.6 & & \\
$\Delta_4^\circ$ & 7 & 99.0 & 1.0 & & \\
$\Delta_{134}^\circ$ & 8 & 99.8 & 0.2 & & \\
$\Delta_{128}^\circ$, $\Delta_{130}^\circ$, $\Delta_{136}^\circ$, $\Delta_{236}^\circ$ & 9 & 99.9 & 0.1 & & \\
\bottomrule
\end{tabular}
\end{center}
\caption{Statistics of limit roots on $C_{(\mathbf{3}, \mathbf{2})_{1/6}}^\bullet$ in all QSM spaces with $\overline{K}_{B_3}^3 = 6$.}
\label{tab:Results1}
\end{table}

\begin{table}[tb]
\begin{center}
\begin{adjustbox}{max width=0.9\textwidth}
\begin{tabular}{c|cc|cc|cc|cc}
\toprule
Polytope & $h^0 = 3$ & $h^0 \geq 3$ & $h^0 = 4$ & $h^0 \geq 4$ & $h^0 = 5$ & $h^0 \geq 5$ & $h^0 = 6$ & $h^0 \geq 6$ \\
\midrule
$\Delta_{88}^\circ$ & 74.9&22.1&2.5&0.5&0.0&0.0 \\
$\Delta_{110}^\circ$&82.4&14.1&3.1&0.4&0.0 & \\
$\Delta_{272}^\circ$, $\Delta_{274}^\circ$&78.1&18.0&3.4&0.5&0.0&0.0 \\
$\Delta_{387}^\circ$ &73.8&21.9&3.5&0.7&0.0&0.0 \\
$\Delta_{798}^\circ$, $\Delta_{808}^\circ$, $\Delta_{810}^\circ$, $\Delta_{812}^\circ$ &77.0&17.9&4.4&0.7&0.0&0.0 \\
\midrule
$\Delta_{254}^\circ$ &95.9&0.5&3.5&0.0&0.0&0.0 \\
$\Delta_{52}^\circ$ &95.3&0.7&3.9&0.0&0.0&0.0 \\
$\Delta_{302}^\circ$ &95.9&0.5&3.5&0.0&0.0 &\\
$\Delta_{786}^\circ$ &94.8&0.3&4.8&0.0&0.0&0.0 \\
$\Delta_{762}^\circ$ &94.8&0.3&4.9&0.0&0.0&0.0 \\
\midrule
$\Delta_{417}^\circ$ & 94.8&0.3&4.8&0.0&0.0&0.0&0.0 \\
$\Delta_{838}^\circ$ &94.7&0.3&5.0&0.0&0.0&0.0 \\
$\Delta_{782}^\circ$ &94.6&0.3&5.0&0.0&0.0&0.0 \\
$\Delta_{377}^\circ$, $\Delta_{499}^\circ$, $\Delta_{503}^\circ$&93.4&0.2&6.2&0.0&0.1&0.0 \\
$\Delta_{1348}^\circ$ &93.7&0.0&6.2&0.0&0.1&&0.0 \\
\midrule
$\Delta_{882}^\circ$, $\Delta_{856}^\circ$&93.4&0.3&6.2&0.0&0.1&0.0&0.0 \\
$\Delta_{1340}^\circ$ & 92.3&0.0&7.6&0.0&0.1&&0.0 \\
$\Delta_{1879}^\circ$ & 92.3&0.0&7.5&0.0&0.1&&0.0 \\
$\Delta_{1384}^\circ$ &90.9&0.0&8.9&0.0&0.2&&0.0 \\
\bottomrule
\end{tabular}
\end{adjustbox}
\end{center}
\caption{Statistics of limit roots on $C_{(\mathbf{3}, \mathbf{2})_{1/6}}^\bullet$ in selected QSM spaces with $\overline{K}_{B_3}^3 = 10$.}
\label{tab:Results2}
\end{table}

\section{\texorpdfstring{$\mathcal{O}(10^{11})$}{10 to 11} F-theory QSM without vector-like quark-doublets} \label{sec:D4}

In this section, we focus on the F-theory QSMs associated to the polytope $\Delta_4^\circ$ in the Kreuzer-Skarke database \cite{Kreuzer:1998vb}. Specifically, we wish to identify the number of global sections for all limit roots for which our algorithm could only provide a lower bound in \oref{tab:Results1}. The upshot of this analysis is that we can identify a condition on the complex structure moduli of those QSM geometries, such that all limit roots on $C^\bullet_{(\mathbf{3}, \mathbf{2})_{1/6}}$ have exactly three global sections. This then implies that all physical roots on the smooth, irreducible curve $C_{(\mathbf{3}, \mathbf{2})_{1/6}}$ have exactly three global sections. In other words, we can then guarantee that there are no vector-like exotics present. To this end, we now study all setups in \oref{equ:DetailedCount} for which only a lower bound is listed.

\subsection{Stationary circuits} \label{subsec:StationaryCircuits}

We begin by looking at two non-resolved nodes. It is not too hard to modify our algorithm to print out the 110 limit roots, for which it can only provided the lower bound $h^0( C^\bullet, P^\bullet ) \geq 3$. It turns out that all of them are associated to the following setup:
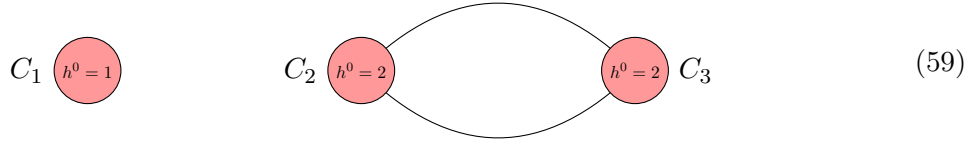
\begin{equation}
\begin{tikzpicture}[scale=0.6, baseline=(current  bounding  box.center)]
      
      \def\s{3.0};
      
      \path[-,out = 45, in = 135, looseness = 1.2] (0,0) edge (2*\s,0);
      \path[-,out = -45, in = -135, looseness = 1.2] (0,0) edge (2*\s,0);
      
      \node at (-2*\s,0) [stuff_fill_red, scale=0.6, label=left:$C_1$]{$h^0 = 1$};
      \node at (0,0) [stuff_fill_red, scale=0.6, label=left:$C_2$]{$h^0 = 2$};
      \node at (2*\s,0) [stuff_fill_red, scale=0.6, label=right:$C_3$]{$h^0 = 2$};
      
\end{tikzpicture} \label{equ:StationaryCircuit1}
\end{equation}
To find out if the lower bound is saturated, we recall that a line bundle on a nodal curve whose irreducible components are all isomorphic to $\mathbb{P}^1$ is uniquely specified by two pieces of information \cite{harris2006moduli}:
\begin{itemize}
 \item The degree of the line bundle on each irreducible component.
 \item Descent data which specifies the gluing conditions at the nodes. The latter corresponds to a number $\lambda \in \mathbb{C}^\ast$ for each node. In particular, the value of the global sections have to agree at each node up to this factor $\lambda$.
\end{itemize}
So in an ideal world, we would specify the descent data of a line bundle $\mathcal{L}^\bullet$ on a nodal curve $C^\bullet$, feed this information into the root bundle description of \cite{2004math4078C} and would end up with the multidegree \emph{and} descent data for each root. In this sense, the information in \oref{equ:StationaryCircuit1} is incomplete, as it does not detail the descent data. Indeed, we will try simpler. Instead of following the descent data through \cite{2004math4078C}, we will now study all different descent data that we can put on \oref{equ:StationaryCircuit1}. Equivalently, we discuss all line bundles  on this nodal curve with the indicated multidegree. The rational behind this is that we can easily see that all of them admit exactly two global sections on $C_2 \cup C_3$, which is sufficient for our study. The name \emph{stationary circuit} derives from this very observation that \emph{all} of these line bundles have the same number of global sections.

To see that indeed all these line bundles have exactly two global sectiosn on $C_2 \cup C_3$, we pick homogeneous coordinates for $C_2$, $C_3$ and parametrize the sections on the curves $C_2$ and $C_3$ by $\left( \alpha_1, \alpha_2, \alpha_3, \alpha_4 \right) \in \mathbb{C}^4$:
\begin{align}
\begin{tabular}{ccc}
\toprule
Curve & Coordinates & Sections \\
\midrule
$C_2$ & $[a:b]$ & $\alpha_1 a + \alpha_2 b$ \\
$C_3$ & $[c:d]$ & $\alpha_3 c + \alpha_4 d$ \\
\bottomrule
\end{tabular}
\end{align}
By use of a M\"obius transformation, we can assume that the two nodes $C_2 \cap C_3$ are at the following positions:
\begin{align}
\begin{tabular}{c|cc}
\toprule
Label & Coordinates in $C_2$ & Coordinates in $C_3$ \\
\midrule
$n_1$ & $[a:b] = [1:0]$ & $[c:d] = [1:0]$\\
$n_2$ & $[a:b] = [0:1]$ & $[c:d] = [0:1]$\\
\bottomrule
\end{tabular}
\end{align}
We now enforce the gluing conditions. At $n_1$ we demand
\begin{align}
\left( \alpha_1 a + \alpha_2 b \right)([1:0]) = \lambda_1 \cdot \left(\alpha_3 c + \alpha_4 d\right)([1:0]) \, ,
\end{align}
where $\lambda_1 \in \mathbb{C}^\ast$. Similarly, we enforce at $n_2$ the gluing by demanding
\begin{align}
\left( \alpha_1 a + \alpha_2 b \right)([0:1]) = \lambda_2 \cdot \left(\alpha_3 c + \alpha_4 d\right)([0:1]) \, ,
\end{align}
where $\lambda_2 \in \mathbb{C}^\ast$. It is not too hard to see that this is equivalent to
\begin{align}
\alpha_1 = \lambda_1 \alpha_3 \, , \qquad \alpha_2 = \lambda_2 \alpha_4 \, .
\end{align}
Hence, the global sections on $C_2 \cup C_3$ are parametrized by $(\alpha_3, \alpha_4) \in \mathbb{C}^2$ via
\begin{align}
\begin{tabular}{ccc}
\toprule
Curve & Coordinates & Sections \\
\midrule
$C_2$ & $[a:b]$ & $\lambda_1 \alpha_3 a + \lambda_2 \alpha_4 b$ \\
$C_3$ & $[c:d]$ & $\alpha_3 c + \alpha_4 d$ \\
\bottomrule
\end{tabular}
\end{align}
Hence, irrespective of $\lambda_1, \lambda_2 \in \mathbb{C}^\ast$, we find exactly two global sections on $C_2 \cup C_3$. Note that we could have rescaled the sections on $C_2$ such that $\lambda_2 \to 1$, so that only $\lambda_1 \in \mathbb{C}^\ast$ appeared in the analysis. We will make use of this simplification in the next section.

In going back to the setup in \oref{equ:StationaryCircuit1} we also take $C_1$ into account. Thereby we find $h^0( C^\bullet, P^\bullet ) = 3$. Hence, the lower bound is indeed saturated for this circuit. It is not too hard to repeat this analysis for all circuits with up to five non-resolved nodes. In all of these cases, we find that the lower bound is saturated. We list details in \oref{sec:StationaryCircuits} and suffice it to note that this improves \oref{equ:DetailedCount} as follows:
\begin{align}
\begin{tabular}{c|cc|cc}
\toprule
& \multicolumn{2}{c|}{Percentages} & \multicolumn{2}{c}{Absolute numbers} \\
$N$ & $h^0 = 3$ & $h^0 \geq 3$ & $h^0 = 3$ & $h^0 \geq 3$ \\
\midrule
0 & 53.6 &  & 11110 &  \\
1 & 36.7 &  & 7601 &  \\
2 & 8.0 &  & 1672 &  \\
3 & 1.3 &  & 275 &  \\
4 & 0.3 &  & 66 &  \\
5 & 0.1 &  & 11 &  \\
7 &  & 0.0 &  & 1 \\
\midrule
$\Sigma$ & 100.0 & 0.0 & 20735 & 1 \\
\bottomrule
\end{tabular}
\end{align}

\subsection{Jumping circuit} \label{subsec:JumpingCircuit}

\paragraph{Jump from special node alignment}

It remains to study the one circuit corresponding to seven remaining nodes:
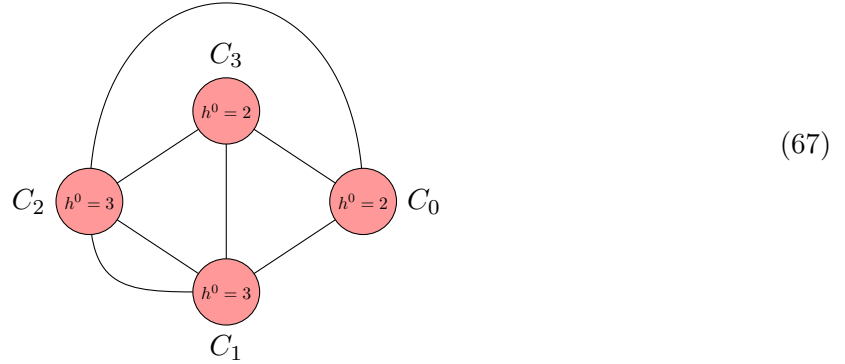
\begin{equation}
\begin{tikzpicture}[scale=0.6, baseline=(current  bounding  box.center)]
      
      \def\s{3.0};
      \def\h{2};
      
      \path[-] (-\s,0) edge (0,\h);
      \path[-] (-\s,0) edge (0,-\h);
      \path[-, out = -90, in = 180, looseness = 1.5] (-\s,0) edge (0,-\h);
      \path[-] (\s,0) edge (0,\h);
      \path[-] (\s,0) edge (0,-\h);
      \path[-] (0,\h) edge (0,-\h);
      \path[-, out = 90, in = 90, looseness = 2.5] (-\s,0) edge (\s,0);
      
      \node at (1*\s,0) [stuff_fill_red, scale=0.6, label=right:$C_0$]{$h^0 = 2$};
      \node at (0,-\h) [stuff_fill_red, scale=0.6, label=below:$C_1$]{$h^0 = 3$};
      \node at (-1*\s,0) [stuff_fill_red, scale=0.6, label=left:$C_2$]{$h^0 = 3$};
      \node at (0,\h) [stuff_fill_red, scale=0.6, label=above:$C_3$]{$h^0 = 2$};
      
\end{tikzpicture} \label{equ:JumpingCircuit2}
\end{equation}
We parametrize the local sections by $\beta_i \in \mathbb{C}$ and pick homogeneous coordinates of the $\mathbb{P}^1$s:
\begin{align}
\begin{tabular}{ccc}
\toprule
Curve & Coordinates & Sections \\
\midrule
$C_0$ & $[u_0 \colon v_0]$ & $\beta_1 u_0 + \beta_2 v_0$ \\
$C_1$ & $[u_1 \colon v_1]$ & $\beta_3 u_1^2 + \beta_4 u_1 v_1 + \beta_5 v_1^2$ \\
$C_2$ & $[u_2 \colon v_2]$ & $\beta_6 u_2^2 + \beta_7 u_2 v_2 + \beta_8 v_2^2$ \\
$C_3$ & $[u_3 \colon v_3]$ & $\beta_9 u_3 + \beta_{10} v_3$ \\
\bottomrule
\end{tabular}
\end{align}
In the same spirit as in \oref{subsec:StationaryCircuits}, it is in theory possible to follow the descent data through \cite{2004math4078C}. Again, this is not necessary for this work and we reserve this task for future investgation. For the time being, we thus again study all descent data for \oref{equ:JumpingCircuit2}, respectively all line bundles on this nodal curves with the displayed multidegree.

Recall that a M\"obius transformation allows us to map any three distinct points on a $\mathbb{P}^1$ to $0$, $1$ and $\infty$. Since there are four nodes on $C_1$ and $C_2$ each, a M\"obius transformation cannot fix the location of that fourth node. This adds more parameters and allows to tune the location of these nodes. This leads to jumping phenomenon -- there is descent data such that the corresponding line bundle jumps from having exactly three global sections to four global sections once the nodes are specially aligned. This mirrors a classical Brill-Noether jump.

While we will not use this effect in the current work, it might be of use in future F-theory MSSM constructions. Let us therefore briefly illustrate these jumps. To this end, we use M\"obius transformations and rescalings of the sections on the irreducible components $C_i$ such that the locations of the nodes and the descent data are as follows:
\begin{align}
\begin{tabular}{c|cccc|c}
\toprule
Node & $C_0$ & $C_1$ & $C_2$ & $C_3$ & Descent data \\
\midrule
$n_1$ & $[1:0]$ & $[1:0]$ & & & $\lambda_1$ \\
$n_2$ & $[1:1]$ & & $[1:0]$ & & 1\\
$n_3$ & $[0:1]$ & & & $[1:0]$ & 1 \\
\midrule
$n_4$ & & $[1:1]$ & $[1:1]$ & & $\lambda_4$ \\
$n_5$ & & $[0:1]$ & $[0:1]$ & & $\lambda_5$ \\
$n_6$ & & $[1:a]$ & & $[1:1]$ & $1$ \\
\midrule
$n_7$ & & & $[1:b]$ & $[0:1]$ & $\lambda_7$ \\
\bottomrule
\end{tabular}
\label{tab:NodePositionsByMoebius}
\end{align}
Recall that $\lambda_i \in \mathbb{C}^\ast$ are employed for the gluings of the global sections at the nodes and that the position of the fourth node on $C_1$ and $C_2$ is not uniquely specified by a M\"obius transformation. The positions of the latter are parametrized by $a,b \in \mathbb{C} \setminus \{0,1\}$. It then follows that the null space of the following matrix $A \in \mathbb{M}(7 \times 10, \mathbb{C})$ describes the global sections on this circuit:
\begin{align}
A = \left[ \begin{array}{cccccccccc} \lambda_1 & & -1 \\ 1 & 1 & & & & -1 \\ 0 & 1 & & & & & & & -1 \\ & & 1 & 1 & 1 & -\lambda_4 & -\lambda_4 & -\lambda_4 \\ & & & & 1 & & & -\lambda_5 \\ & & 1 & a & a^2 & & & & -1 & -1 \\ & & & & & 1 & b & b^2 & & - \lambda_7 \end{array} \right] \in \mathbb{M}( 7 \times 10, \mathbb{C} )\, . \label{equ:BottomUpMatrix}
\end{align}
For generic values of $\lambda_1, \lambda_4, \lambda_5, \lambda_7 \in \mathbb{C}^\ast$ and $a,b \in \mathbb{C} \setminus \{0,1\}$ it is not too hard to see that $N(A) \cong \mathbb{C}^3$. The key observation is that all $7 \times 7$-minors of this matrix vanish iff
\begin{align}
\lambda_1 = \frac{1}{a-1} \, , \qquad \lambda_4 = \frac{b}{a \left( b-1 \right)} \, , \qquad \lambda_5 = \frac{b}{a \left( a-1 \right)} \, , \qquad \lambda_7 = b-1 \, .
\label{equ:DescentDataCanonicalBundle}
\end{align}
Hence, there is a unique line bundle with four global sections and multidegree matching that of the canonical bundle on this curve. Therefore, \oref{equ:DescentDataCanonicalBundle} list nothing other than the descent data of the canonical bundle on this curve. To see the jumping phenomenon mentioend above, focus on a line bundles with descent data
\begin{align}
\left( \lambda_1, \lambda_4, \lambda_5, \lambda_7 \right) = \left( \lambda_1, \frac{\lambda_1 \left(\lambda_7+1 \right)}{\lambda_7 \left(\lambda_1+1\right)}, \frac{\lambda_1^2 \cdot \left( \lambda_7+1 \right)}{\lambda_1+1}, \lambda_7 \right) \, ,
\end{align}
where $\lambda_1, \lambda_7 \in \mathbb{C} \setminus \{ -1, 0 \}$. For generic $\lambda_1, \lambda_7 \in \mathbb{C} \setminus \{ -1, 0 \}$, this line bundle has exactly three global sections. However, once the nodes $n_6$, $n_7$ are positioned according to
\begin{align}
a = \frac{\lambda_1+1}{\lambda_1} \, , \qquad b = \lambda_7 + 1 \, ,
\end{align}
the number of global sections jumps to four and the line bundle in question turns into the canonical bundle. Such a jump in the number of global sections resembles a classical Brill-Noether jump. This is the reason why we propose to call such a pair of a nodal curve and a multidegree a \emph{jumping circuit}. While we will not make use of such jumps in this article, it can be speculated that they may be relevant to string model building. We hope to investigate such applications, in particular to F-theory MSSMs, in future works.

\paragraph{Sections of the physically relevant root}

Let us return to the F-theory QSMs. In this context, the parameters $a$ and $b$ are not free, but dictated by the embedding of the nodal curve into the QSM geometries $B_3( \Delta_4^\circ )$. Hence, in order to analyze the global number of sections, we wish to parametrize the location of the nodes with the input parameters of the QSM. Recall that $C_{(\mathbf{3}, \mathbf{2})_{1/6}} = V( s_3, s_9 )$ with $s_3, s_9 \in H^0( B_3, \overline{K}_{B_3} )$. In the QSM geometries associated to $\Delta_4^\circ$, we can write (cf. \oref{sec:TriaIndependence})
\begin{align}
\begin{split}
s_9 & = \alpha_1 \cdot x_{0}^6 x_{4}^5 x_{5}^4 x_{6}^3 x_{7}^2 x_{8} x_{9}^4 x_{10}^3 x_{11}^2 x_{12} x_{14}^2 x_{15} x_{17}^4 x_{18}^3 x_{19}^2 x_{20} x_{22}^2 x_{25}^2 x_{26} \\
   &\qquad + \alpha_2 \cdot x_{0}^3 x_{3}^3 x_{4}^3 x_{5}^3 x_{6}^3 x_{7}^3 x_{8}^3 x_{9}^2 x_{10}^2 x_{11}^2 x_{12}^2 x_{13}^2 x_{14} x_{15} x_{16} x_{17}^2 x_{18}^2 x_{19}^2 x_{20}^2 x_{21}^2 x_{22} x_{23} x_{25} x_{26} x_{27} \\
   &\qquad + \alpha_3 \cdot x_{0} x_{1} x_{2} x_{3} x_{4} x_{5} x_{6} x_{7} x_{8} x_{9} x_{10} x_{11} x_{12} x_{13} x_{14} x_{15} x_{16} x_{17} x_{18} \\
   & \hspace{20em} \times x_{19} x_{20} x_{21} x_{22} x_{23} x_{24} x_{25} x_{26} x_{27} x_{28} \\
   &\qquad + \alpha_4 \cdot x_{3}^6 x_{4} x_{5}^2 x_{6}^3 x_{7}^4 x_{8}^5 x_{10} x_{11}^2 x_{12}^3 x_{13}^4 x_{15} x_{16}^2 x_{18} x_{19}^2 x_{20}^3 x_{21}^4 x_{23}^2 x_{26} x_{27}^2 \\
   &\qquad + \alpha_5 \cdot x_{2}^3 x_{9} x_{10} x_{11} x_{12} x_{13} x_{14}^2 x_{15}^2 x_{16}^2 x_{22} x_{23} x_{24}^2 x_{28}\\
   &\qquad + \alpha_6 \cdot x_{1}^3 x_{17} x_{18} x_{19} x_{20} x_{21} x_{22} x_{23} x_{24} x_{25}^2 x_{26}^2 x_{27}^2 x_{28}^2 \, .
\end{split}
\end{align}
We proceed under the assumption that the coefficients $\alpha_i \in \mathbb{C}$ satisfy
\begin{align}
\alpha_1, \alpha_4, \alpha_5 \neq 0 \, , \qquad \alpha_2^2 \neq 4 \alpha_1 \alpha_4 \, . \label{equ:Assumption}
\end{align}
We have worked out the embedding of the nodal curve
\begin{align}
C^\bullet_{(\mathbf{3}, \mathbf{2})_{1/6}} = V \left( \prod_{i = 0}^{28}{x_i}, s_9 \right) = \bigcup_{i = 0}^{28}{V \left( x_i, s_9 \right)}\ \, ,
\end{align}
including the location of the nodes of the circuit in \oref{equ:JumpingCircuit2}. The details are listed in \oref{sec:TriaIndependence}. Suffice it to state that this leads to the following parametrization of the nodes:
\begin{align}
\begin{adjustbox}{max width=0.9\textwidth}
\begin{tabular}{c|cccc}
\toprule
Node & $C_0$ & $C_1$ & $C_2$ & $C_3$ \\
\midrule
$n_1$ & $[1:-\frac{\alpha_4}{\alpha_5}]$ & $[0:1]$\\
$n_2$ & $[1:0]$ & & $[0:1]$\\
$n_3$ & $[0:1]$ & & & $[0:1]$\\
\midrule
$n_4$ & & $\left[ 1 : \frac{- \alpha_2 -  \sqrt{\alpha_2^2 - 4 \alpha_1 \alpha_4}}{2 \alpha_4} \right]$ & $\left[ 1 : \frac{- \alpha_2 -  \sqrt{\alpha_2^2 - 4 \alpha_1 \alpha_4}}{2 \alpha_4} \right]$ \\
$n_5$ & & $\left[ 1 : \frac{\alpha_2 - \sqrt{\alpha_2^2 - 4 \alpha_1 \alpha_4}}{2 \alpha_4} \right]$ & $\left[ 1 : \frac{\alpha_2 - \sqrt{\alpha_2^2 - 4 \alpha_1 \alpha_4}}{2 \alpha_4} \right]$ \\
$n_6$ & & $[1:0]$ & & $[1:- \frac{\alpha_1}{\alpha_5}]$\\
\midrule
$n_7$ & & & $[1:0]$ & $[1:0]$\\
\bottomrule
\end{tabular}
\end{adjustbox}
\label{tab:NodePositionsEmbedded}
\end{align}
Note that these positions are rather different than those in \oref{tab:NodePositionsByMoebius}. Hence, we look at a different matrix $A \in \mathbb{M}( 7 \times 10, \mathbb{C} )$, whose right null space is the vector space of global sections:
\begin{footnotesize}
\begin{align}
A = \left[
\begin{array}{cccccccccc}
\lambda_1 & \lambda_1 \cdot \left( -\frac{\alpha_4}{\alpha_5} \right) & 0 & 0 & -1 & 0 & 0 & 0 & 0 & 0 \\
1 & 0 & 0 & 0 & 0 & 0 & 0 & -1 & 0 & 0 \\
0 & 1 & 0 & 0 & 0 & 0 & 0 & 0 & 0 & -1 \\
0 & 0 & 1 & \frac{-\alpha_2 - R}{2 \alpha_4} & \left( \frac{\alpha_2 + R}{2 \alpha_4} \right)^2 & - \lambda_4 & - \lambda_4 \left( \frac{-\alpha_2 - R}{2 \alpha_4} \right) & - \lambda_4 \left( \frac{\alpha_2 + R}{2 \alpha_4} \right)^2 & 0 & 0 \\
0 & 0 & 1 & \frac{-\alpha_2 + R}{2 \alpha_4} & \left(\frac{-\alpha_2 + R}{2 \alpha_4} \right)^2 & - \lambda_5 & - \lambda_5 \left( \frac{-\alpha_2 + R}{2 \alpha_4} \right) & - \lambda_5 \left( \frac{-\alpha_2 + R}{2 \alpha_4} \right)^2 & 0 & 0 \\
0 & 0 & 1 & 0 & 0 & 0 & 0 & 0 & -1 & \frac{\alpha_1}{\alpha_5} \\
0 & 0 & 0 & 0 & 0 & 1 & 0 & 0 & -\lambda_7 & 0
\end{array}
\right]
\label{equ:MatrixA}
\end{align}
\end{footnotesize}
In this matrix we use $R = \sqrt{\alpha_2^2 - 4 \alpha_1 \alpha_4}$. For generic parameters $\lambda_1, \lambda_4, \lambda_5, \lambda_7 \in \mathbb{C}^\ast$, this matrix has $N(A) \cong \mathbb{C}^3$. All $7 \times 7$-minors vanish iff $\lambda_1 = \lambda_4 = \lambda_5 = \lambda_7 = 1$ and
\begin{align}
N(A) = \mathrm{Span}_{\mathbb{C}} \left( \left[ \begin{array}{c} 0\\0\\0\\1\\0\\0\\1\\0\\0\\0 \end{array} \right] \, , \left[ \begin{array}{c} 1\\0\\0\\0\\1\\0\\0\\1\\0\\0 \end{array} \right] \, , \left[ \begin{array}{c} 0\\0\\1\\0\\0\\1\\0\\0\\1\\0 \end{array} \right] \, , \left[ \begin{array}{c} 0\\1\\-\frac{\alpha_1}{\alpha_5}\\- \frac{\alpha_2}{\alpha_5}\\-\frac{\alpha_4}{\alpha_5}\\0\\0\\0\\0\\1 \end{array} \right] \right) \, . \label{equ:ExplicitNullspace}
\end{align}
Hence, $\lambda_1 = \lambda_4 = \lambda_5 = \lambda_7 = 1$ is nothing but the descent data of the canonical bundle. Note that this parametrization differs from \oref{equ:DescentDataCanonicalBundle} since we are now employing the node positions in \oref{tab:NodePositionsEmbedded}. As a consistency check, recall that we are looking at the nodal curve
\begin{align}
C^\bullet = V \left(\prod_{i = 1}^{28}{x_i},s_9 \right) \, .
\end{align}
This curve, and also the line bundle in question, remain completely unchanged as we replace $s_9$ by $\mu \cdot s_9$, where $\mu \in \mathbb{C}^\ast$ is arbitrary but fixed. Therefore, the null space computed above must be invariant upon rescaling the coefficients of $s_9$ via $\alpha_i \to \mu \cdot \alpha_i$. Indeed, this is the case for the null space in \oref{equ:ExplicitNullspace}.

We must now wonder which descent data encodes the physically relevant root. To this end, recall that in following \cite{Bies:2021nje}, we are solving the root bundle constraint $P_\bullet^{12} = K_{C^\bullet}^{12}$. We found in \oref{subsec:StationaryCircuits} that all solutions but the jumping circuit have exactly three global sections. Apparently, $P_\bullet = K_{C^\bullet}$ is a solution and by theory it must always hold $h^0( C^{\bullet}, K_{C^\bullet} ) = 4$. Hence, this solution was not found before and must correspond to the jumping circuit. Therefore, the physically relevant root bundle on this jumping circuit is nothing but the canonical bundle.

We recall that the root bundle conditions are necessary conditions, in that it is not yet clear which of these solutions are obtained top-down from an F-theory gauge potential. Our analysis shows that more than $99.995\%$ of the solutions to these necessary conditions have exactly three global sections. If realized top-down from an F-theory gauge potential, this means that this solution has no vector-like exotics in the representation $C^\bullet_{(\mathbf{3}, \mathbf{2})_{1/6}}$.

\subsection{Towards absence of vector-like exotics} \label{subsec:AbsenceOfVectorlikeExotics}

Recall that in every QSM geometry, there are three curves with identical topology:
\begin{align}
C_{(\mathbf{\overline{3}}, \mathbf{2})_{1/6}} = V( s_3, s_9 ) \, , \qquad C_{(\mathbf{\overline{3}}, \mathbf{1})_{-2/3}} = V( s_5, s_9 ) \, , \qquad C_{(\mathbf{1}, \mathbf{1})_{1}} = V( s_1, s_5 ) \, .
\end{align}
Our arguments apply to each of these curves. In focusing on the spaces in $B_3( \Delta_4^\circ )$, let us explicitly parametrize $s_1$ and $s_9$:
\begin{align}
\begin{split}
s_1 & = \alpha_1 \cdot x_{0}^6 x_{4}^5 x_{5}^4 x_{6}^3 x_{7}^2 x_{8} x_{9}^4 x_{10}^3 x_{11}^2 x_{12} x_{14}^2 x_{15} x_{17}^4 x_{18}^3 x_{19}^2 x_{20} x_{22}^2 x_{25}^2 x_{26} \\
   &\qquad + \alpha_2 \cdot x_{0}^3 x_{3}^3 x_{4}^3 x_{5}^3 x_{6}^3 x_{7}^3 x_{8}^3 x_{9}^2 x_{10}^2 x_{11}^2 x_{12}^2 x_{13}^2 x_{14} x_{15} x_{16} x_{17}^2 x_{18}^2 x_{19}^2 x_{20}^2 x_{21}^2 x_{22} x_{23} x_{25} x_{26} x_{27} \\
   &\qquad + \alpha_3 \cdot x_{0} x_{1} x_{2} x_{3} x_{4} x_{5} x_{6} x_{7} x_{8} x_{9} x_{10} x_{11} x_{12} x_{13} x_{14} x_{15} x_{16} x_{17} x_{18} \\
   & \hspace{20em} \times x_{19} x_{20} x_{21} x_{22} x_{23} x_{24} x_{25} x_{26} x_{27} x_{28} \\
   &\qquad + \alpha_4 \cdot x_{3}^6 x_{4} x_{5}^2 x_{6}^3 x_{7}^4 x_{8}^5 x_{10} x_{11}^2 x_{12}^3 x_{13}^4 x_{15} x_{16}^2 x_{18} x_{19}^2 x_{20}^3 x_{21}^4 x_{23}^2 x_{26} x_{27}^2 \\
   &\qquad + \alpha_5 \cdot x_{2}^3 x_{9} x_{10} x_{11} x_{12} x_{13} x_{14}^2 x_{15}^2 x_{16}^2 x_{22} x_{23} x_{24}^2 x_{28}\\
   &\qquad + \alpha_6 \cdot x_{1}^3 x_{17} x_{18} x_{19} x_{20} x_{21} x_{22} x_{23} x_{24} x_{25}^2 x_{26}^2 x_{27}^2 x_{28}^2 \, , \\
s_9 & = \beta_1 \cdot x_{0}^6 x_{4}^5 x_{5}^4 x_{6}^3 x_{7}^2 x_{8} x_{9}^4 x_{10}^3 x_{11}^2 x_{12} x_{14}^2 x_{15} x_{17}^4 x_{18}^3 x_{19}^2 x_{20} x_{22}^2 x_{25}^2 x_{26} \\
   &\qquad + \beta_2 \cdot x_{0}^3 x_{3}^3 x_{4}^3 x_{5}^3 x_{6}^3 x_{7}^3 x_{8}^3 x_{9}^2 x_{10}^2 x_{11}^2 x_{12}^2 x_{13}^2 x_{14} x_{15} x_{16} x_{17}^2 x_{18}^2 x_{19}^2 x_{20}^2 x_{21}^2 x_{22} x_{23} x_{25} x_{26} x_{27} \\
   &\qquad + \beta_3 \cdot x_{0} x_{1} x_{2} x_{3} x_{4} x_{5} x_{6} x_{7} x_{8} x_{9} x_{10} x_{11} x_{12} x_{13} x_{14} x_{15} x_{16} x_{17} x_{18} \\
   & \hspace{20em} \times x_{19} x_{20} x_{21} x_{22} x_{23} x_{24} x_{25} x_{26} x_{27} x_{28} \\
   &\qquad + \beta_4 \cdot x_{3}^6 x_{4} x_{5}^2 x_{6}^3 x_{7}^4 x_{8}^5 x_{10} x_{11}^2 x_{12}^3 x_{13}^4 x_{15} x_{16}^2 x_{18} x_{19}^2 x_{20}^3 x_{21}^4 x_{23}^2 x_{26} x_{27}^2 \\
   &\qquad + \beta_5 \cdot x_{2}^3 x_{9} x_{10} x_{11} x_{12} x_{13} x_{14}^2 x_{15}^2 x_{16}^2 x_{22} x_{23} x_{24}^2 x_{28}\\
   &\qquad + \beta_6 \cdot x_{1}^3 x_{17} x_{18} x_{19} x_{20} x_{21} x_{22} x_{23} x_{24} x_{25}^2 x_{26}^2 x_{27}^2 x_{28}^2 \, .
\end{split}
\end{align}
Now, we demand that
\begin{align}
\alpha_1, \alpha_4, \alpha_5 \neq 0 \, , \qquad \alpha_2^2 \neq 4 \alpha_1 \alpha_4 \, , \qquad \beta_1, \beta_4, \beta_5 \neq 0 \, , \qquad \beta_2^2 \neq 4 \beta_1 \beta_4 \, . \label{label:FinalCondition}
\end{align}
Then, by the above arguments, we are guaranteed that for each of the $\mathcal{O}( 10^{11} )$ toric spaces in $B_3( \Delta_4^\circ )$, more than $99.995\%$ of the limit root bundle on the nodal curves $C^\bullet_{(\mathbf{\overline{3}}, \mathbf{2})_{1/6}}$, $C^\bullet_{(\mathbf{\overline{3}}, \mathbf{1})_{-2/3}}$ and $C^\bullet_{(\mathbf{1}, \mathbf{1})_{1}}$ which satisfies the necessary condition identified in \cite{Bies:2021nje} have exactly three global sections. Let us emphasize that these root bundle conditions are necessary but not sufficient for a physical (i.e. induced from an F-theory gauge potential) root bundle. Hence, the physical roots are special among the ones covered in this article. Since we have found that more than $99.995\%$ of the limit roots in this superset have exactly three sections, this indicates that the chances are good that this carries over to the physical roots in the representations $(\mathbf{3}, \mathbf{2})_{1/6}$, $(\mathbf{\overline{3}}, \mathbf{1})_{-2/3}$ and $(\mathbf{1}, \mathbf{1})_{1}$.

\section{Conclusion and Outlook} \label{sec:ConclusionAndOutlook}

Root bundles appear prominently in studies of vector-like spectra of 4d F-theory compactifications. Of particular importance to phenomenology are the Quadrillion F-theory Standard Models (F-theory QSMs) \cite{Cvetic:2019gnh}. In this work, we have analyzed a superset of all physical root bundles for the matter representations $(\mathbf{3}, \mathbf{2})_{1/6}$, $(\mathbf{\overline{3}}, \mathbf{1})_{-2/3}$ and $(\mathbf{1}, \mathbf{1})_{1}$ in $\mathcal{O}(10^{11})$ F-theory QSM geometries. We have found that more than $99.995\%$ of the roots in this superset have exatly three global sections.

We arrived at this finding from a systematic study of a large fraction of QSM geometries. While our results are strongest for the family $B_3( \Delta_4^\circ )$ of toric 3-folds obtained from full, regular, star triangulations of the 4-th polytope in the Kreuzer-Skarke list of 3-dimensional, reflexive poytopes \cite{Kreuzer:1998vb}, the results for this large fraction of QSM geometries provide statistical evidence that the absence of the vector-like exotics in the representations $(\mathbf{3}, \mathbf{2})_{1/6}$, $(\mathbf{\overline{3}}, \mathbf{1})_{-2/3}$ and $(\mathbf{1}, \mathbf{1})_{1}$ is a very likely scenario within the QSMs.

In the QSM geometries, vector-like spectra are counted by the cohomologies of line bundles which are necessarily roots of (twists of certain powers of) the canonical bundle \cite{Bies:2021nje}. Such root bundles $P_{\mathbf{R}}$ on the matter curve $C_{\mathbf{R}}$ are by no means unique. Even more, it is not clear exactly which ones are induced top-down from F-theory gauge potentials in the Deligne cohomolgy. First, on one matter curve, only a subset of all mathematically allowed root bundles could be induced from all physically allowed F-theory gauge potentials. Secondly, it is conceivable that fluxes which induce a specific root bundle on matter curve $C_1$, only induce a few selected root bundles on another matter curve $C_2$. The study of these questions is involved and currently beyond our abilities. While we hope to return to these questions in future work, in the current work we opted for a local and bottom-up analysis instead. In this sense, we followed the philosophy of the earlier works \cite{Bies:2021nje, Bies:2021xfh}. The presented study is bottom-up in that we studied \emph{all} mathematically allowed root bundles. In particular, our study covers all root bundles that could possibly be induced from the $G_4$-flux and all spin bundles on the matter curve in question. Our study is local in that we focused on one matter curve at a time. Correlations among the vector-like spectra of different matter curves were not taken into account. Hence, we study a set of root bundles on each matter curve, which is a (proper) superset to all physically allowed root bundles. For one family of QSM geometries, we argue that all those roots in this superset have exactly three sections. Consequently, since the physical roots are a subset of the set of roots that we study, the absence of vector-like exotics follows.

A crucial step towards this goal was triangulation independence. Specifically, each family $B_3( \Delta^\circ )$ of toric 3-fold QSM base spaces admits a triangulation independent nodal limit \cite{Bies:2021xfh}. This observation allows us to probe the vector-like spectra on the entire family $B_3( \Delta^\circ )$ from vector-like spectra on just 5 nodal curves. Another motivation for this deformation is that on smooth, irreducible curves, it is hard to construct root bundles, not to mention count their global sections. On nodal curves however, the construction of all limit root bundles has been detailed in \cite{2004math4078C}. For the family $B_3( \Delta_4^\circ )$ -- associated to the 4-th polytope in the Kreuzer-Skarke list of 3-dimensional polytopes -- which consists of $\mathcal{O}(10^{11})$ different toric spaces, we showed that all limit root bundles on the nodal curves $C_{(\mathbf{3}, \mathbf{2})_{1/6}}^\bullet$, $C_{(\mathbf{\overline{3}}, \mathbf{1})_{-2/3}}^\bullet$ and $C_{(\mathbf{1}, \mathbf{1})_{1}}^\bullet$ have exactly three global sections. By upper semi-continuity, this must hold true on the corresponding smooth, irreducible matter curves in each space of $B_3( \Delta_4^\circ )$. This then implies the absence of vector-like exotics.

Apparently, this argument relies on the ability to compute the cohomologies of \emph{all} limit root bundles. This is exactly the domain of Brill-Noether theory. By following the original work in \cite{2004math4078C}, we find that for some limit roots, this is equivalent to line bundle cohomology on tree-like curves. The remaining limit roots require an understanding of line bundle cohomology on circuit-like curves. We explained this in \oref{sec:PartialBlowupLimitRootBundles}. In our applications to the F-theory QSMs, the majority of limit roots arise from tree-like curves with rational irreducible components. Therefore, we presented a computer algorithm that counts line bundle cohomology on such curves in \oref{sec:LineBundleCohomologyOnTreelikeCurves}. The few circuit-like cases are handled by hand.

The reason why there are so few circuits seems related to the fact that we are looking for $12$-th and even $20$-th roots of the canonical bundle on nodal curves with relatively small Betti numbers, namely 4 and 6. In fact, we observe that lower root indices shift the ratio towards having more circuits. For instance, if instead we were looking at second roots, one should expect equal contributions from tree-like and circuit-like graphs. It would even be conceivable that the circuits would dominate. We reserve a systematic study of line bundle cohomology on circuit-like curves for future work.

Our algorithm simplifies the pair $(C^\bullet, L)$ of a tree-like, rational, nodal curve and a line bundle $L$ on it by removing components of $C^\bullet$ and adjusting $L$ such that the number of global sections remains unchanged. This simplification eventually terminates and leaves us with a final configuration for which the number of global sections can be read off easily. If we only want to tell if the number of sections is larger than that on a smooth $\mathbb{P}^1$, then this algorithm can be optimized further. While this optimization is not directly relevant in order to establish the absence of vector-like exotics in $\mathcal{O}(10^{11})$ QSM geometries, we outlined these steps in \oref{sec:Speciality}.

In addition to integrating this algorithm into the algorithm used for \cite{Bies:2021xfh}, we have also taken a combinatorial factor into account. In \cite{2004math4078C}, this factor was introduced as the \emph{geometric multiplicity}. In layman's terms, this multiplicity ensures that we count as many limit root bundles as root bundles on a smooth curve. Alternatively, under a deformation of a smooth to a nodal curve, distinct root bundles on the smooth curve coalesce as we approach the nodal curve. The exact number of distinct roots which coalesce is the geometric multiplicity. In \oref{sec:TowardsBNOfLimitRootBundles}, we have elaborated in more detail on this factor before pointing out two errors in \cite{Bies:2021xfh}. The new and refined results were listed in \oref{tab:Results1} and \oref{tab:Results2}. The relevant computer algorithms are available in the \texttt{gap-4}-package \emph{QSMExplorer} as part of the \texttt{ToricVarieties$\_$project} \cite{ToricVarietiesProject}. For convenience of the reader, let us reproduce \oref{tab:Results1} (cf. \oref{appendix:DetailsOnCounts} for more details):
\begin{align}
\begin{tabular}{cc|cc|cc}
\toprule
Polytope & $E$ & $h^0 = 3$ & $h^0 \geq 3$ & $h^0 = 4$ & $h^0 \geq 4$ \\
\midrule
$\Delta_8^\circ$ & 6 & 76.4 & 23.6 & & \\
$\Delta_4^\circ$ & 7 & 99.04 & 0.96 & & \\
$\Delta_{134}^\circ$ & 8 & 99.8 & 0.2 & & \\
$\Delta_{128}^\circ$, $\Delta_{130}^\circ$, $\Delta_{136}^\circ$, $\Delta_{236}^\circ$ & 9 & 99.9 & 0.1 & & \\
\bottomrule
\end{tabular}
\end{align}
Hence, by focusing on limit roots on tree-like partial blowups of $C^\bullet_{(\mathbf{3}, \mathbf{2})_{1/6}}$, we have established for the QSM spaces with $\overline{K}_{B_3}^3 = 6$ that are not obtained from FRSTs of $\Delta_8^\circ$, that at least 99\% of the limit roots have exactly three global sections. This is a massive improvement compared to \cite{Bies:2021xfh} in which we merely reached about $60\%$. Even more, these setups beg for a further investigation of the remaining 1\% of the limit roots, which are associated to circuit-like partial blowups.

This is exactly what we have done in \oref{sec:D4} for the family $B_3( \Delta_4^\circ )$ of QSM spaces obtained from triangulations of the polytope $\Delta_4^\circ$. It turned out that, up to symmetries, only a few line bundles on circuit-like, nodal curves had to be investigated. We found that for some of these circuits, the line bundle cohomology is independent of the relative position of the nodes. We call such circuit-like curves \emph{stationary circuits} (cf. \oref{sec:StationaryCircuits} for details). These results mirror our finding that for the tree-like blow-up limit roots, the cohomologies are independent of the relative position of the nodes. However, there are special circuits for which such a dependence does exist. In particular, there exists one jumping circuit for the limit roots on $C^\bullet_{(\mathbf{3}, \mathbf{2})_{1/6}}$ in $B_3( \Delta_4^\circ )$. In this case, the number of global sections of line bundles can jump from 3 to 4 as we move the nodes in special alignment. We propose to call such curves \emph{jumping circuits}. It can be speculated that such jumps may be relevant to string model building. We hope to investigate such applications, in particular to F-theory MSSMs, in future works.

To study the jumping circuit for $B_3( \Delta_4^\circ )$ in full detail, we investigated its embedding into all spaces of $B_3( \Delta_4^\circ )$ in \oref{sec:TriaIndependence}. This embedding depends on the complex structure moduli chosen for $C^\bullet_{(\mathbf{3}, \mathbf{2})_{1/6}}$. For the spaces in $B_3( \Delta_4^\circ )$, it holds that
\begin{align}
\begin{split}
C^\bullet_{(\mathbf{3}, \mathbf{2})_{1/6}} &= V \left( \prod_{i = 0}^{28}{x_i}, s_9 \right) \,,  \\
s_9 & = \alpha_1 \cdot x_{0}^6 x_{4}^5 x_{5}^4 x_{6}^3 x_{7}^2 x_{8} x_{9}^4 x_{10}^3 x_{11}^2 x_{12} x_{14}^2 x_{15} x_{17}^4 x_{18}^3 x_{19}^2 x_{20} x_{22}^2 x_{25}^2 x_{26} \\
&\quad + \alpha_2 \cdot x_{0}^3 x_{3}^3 x_{4}^3 x_{5}^3 x_{6}^3 x_{7}^3 x_{8}^3 x_{9}^2 x_{10}^2 x_{11}^2 x_{12}^2 x_{13}^2 x_{14} x_{15} x_{16} x_{17}^2 x_{18}^2 x_{19}^2 \\
& \hspace{15em} \times x_{20}^2 x_{21}^2 x_{22} x_{23} x_{25} x_{26} x_{27} \\
&\quad + \alpha_3 \cdot x_{0} x_{1} x_{2} x_{3} x_{4} x_{5} x_{6} x_{7} x_{8} x_{9} x_{10} x_{11} x_{12} x_{13} x_{14} x_{15} x_{16} x_{17} x_{18} \\
& \hspace{15em} \times x_{19} x_{20} x_{21} x_{22} x_{23} x_{24} x_{25} x_{26} x_{27} x_{28} \\
&\quad + \alpha_4 \cdot x_{3}^6 x_{4} x_{5}^2 x_{6}^3 x_{7}^4 x_{8}^5 x_{10} x_{11}^2 x_{12}^3 x_{13}^4 x_{15} x_{16}^2 x_{18} x_{19}^2 x_{20}^3 x_{21}^4 x_{23}^2 x_{26} x_{27}^2 \\
&\quad + \alpha_5 \cdot x_{2}^3 x_{9} x_{10} x_{11} x_{12} x_{13} x_{14}^2 x_{15}^2 x_{16}^2 x_{22} x_{23} x_{24}^2 x_{28}\\
&\quad + \alpha_6 \cdot x_{1}^3 x_{17} x_{18} x_{19} x_{20} x_{21} x_{22} x_{23} x_{24} x_{25}^2 x_{26}^2 x_{27}^2 x_{28}^2 \, .
\end{split}
\end{align}
For simplicity, we assumed $\alpha_1, \alpha_4, \alpha_5 \neq 0$ and $\alpha_2^2 \neq 4 \alpha_1 \alpha_4$. The embedding was then worked out and we found that the global sections on this circuit are encoded by the right null space of
\begin{footnotesize}
\begin{align}
A = \left[
\begin{array}{cccccccccc}
\lambda_1 & -\lambda_1 \cdot \frac{\alpha_4}{\alpha_5} & 0 & 0 & -1 & 0 & 0 & 0 & 0 & 0 \\
1 & 0 & 0 & 0 & 0 & 0 & 0 & -1 & 0 & 0 \\
0 & 1 & 0 & 0 & 0 & 0 & 0 & 0 & 0 & -1 \\
0 & 0 & 1 & \frac{-\alpha_2 - R}{2 \alpha_4} & \left( \frac{\alpha_2 + R}{2 \alpha_4} \right)^2 & -\lambda_4 & - \lambda_4 \left( \frac{-\alpha_2 - R}{2 \alpha_4} \right) & - \lambda_4 \left( \frac{\alpha_2 + R}{2 \alpha_4} \right)^2 & 0 & 0 \\
0 & 0 & 1 & \frac{-\alpha_2 + R}{2 \alpha_4} & \left(\frac{-\alpha_2 + R}{2 \alpha_4} \right)^2 & -\lambda_5 & - \lambda_5 \left( \frac{-\alpha_2 + R}{2 \alpha_4} \right) & - \lambda_5 \left( \frac{-\alpha_2 + R}{2 \alpha_4} \right)^2 & 0 & 0 \\
0 & 0 & 1 & 0 & 0 & 0 & 0 & 0 & -1 & \frac{\alpha_1}{\alpha_5} \\
0 & 0 & 0 & 0 & 0 & 1 & 0 & 0 & -\lambda_7 & 0
\end{array}
\right]
\end{align}
\end{footnotesize}
In this matrix we use $R = \sqrt{\alpha_2^2 - 4 \alpha_1 \alpha_4}$. An explicit computation reveals that this right null space is generically of dimension $3$. All $7 \times 7$ minors of this matrix vanish iff $\lambda_1 = \lambda_4 = \lambda_5 = \lambda_7 = 1$ and the corresponding line bundle then has exactly four sections. Since the multidegrees match those of the canoical bundle on the jumping circuit in \oref{equ:JumpingCircuit2}, we conclude that this very line bundle is nothing but the canonical bundle on this nodal curve. Indeed, we are solving $P_\bullet^{12} = K_{C^\bullet}^{12}$. One solution is apparently $P_\bullet = K_{C^\bullet}$ and by theory $h^0( C_\bullet, K_{C^\bullet} ) = 4$. We argued that all other solutions other than the jumping circuit have exactly three global sections. Hence, in this jumping circuit we are interested in $K_{C^\bullet}$ which by theory has to have four global sections. Therefore, we conclude that for the family of toric 3-folds $B_3( \Delta_4^\circ )$, more than 99.995\% of the solutions to the necessary root bundle condition -- originally derived in \cite{Bies:2021nje} -- have exactly three global sections. We recall that it is not clear at this point which of these solutions are realized top-down from an F-theory gauge potential. This will likely select a subset of the set of root bundles analyzed in this work. The fact, that more than 99.995\% of the roots in this superset have exactly three global sections, indicates that absence of vector-like exotics in the representation $(\mathbf{3}, \mathbf{2})_{1/6}$, $(\mathbf{\overline{3}}, \mathbf{1})_{-2/3}$ and $(\mathbf{1}, \mathbf{1})_{1}$ is a very likely scenario in the $\mathcal{O}( 10^{11} )$ QSM geometries $B_3( \Delta_4^\circ )$.

It is, at least in principle, possible to extend this analysis to other QSM spaces. Based on \oref{tab:Results1}, one should focus on $B_3( \Delta_{134}^\circ)$, $B_3(\Delta_{128}^\circ)$, $B_3(\Delta_{130}^\circ)$, $B_3(\Delta_{136}^\circ)$, $B_3(\Delta_{236}^\circ)$ next. These geometries and our limit root counts are very similar to those of $B_3( \Delta_4^\circ )$. We should therefore expect similar findings, possibly establishing an even larger class of F-theory Standard Models without vector-like exotics. However, once we turn to spaces with $\overline{K}_{B_3}^3 = 10$, then our argument breaks down. Namely, we can read off from \oref{tab:Results2} that there are root bundles on the canonical nodal matter curves that admit more than three global sections. To understand such setups, one has to formulate conditions that guarantee that the number of sections remains constant/drops down as one deforms back to a smooth curve. Such conditions will also be vital to guarantee exactly one Higgs pair. It can be speculated that the physics of Yukawa interactions, which informs us how, when and which fields acquire masses upon deformations, will be a good guide towards such conditions. A quest for F-theory MSSM must therefore address this question eventually. We hope to investigate this fascinating and challenging interplay in future works.

As far as the techniques in the current work are concerned, the $\mathcal{O}( 10^{11} )$ QSM geometries $B_3( \Delta_4^\circ )$ behaved identically. To some extent, this must be seen as very fortunate circumstances. However, many of the intersection numbers -- and this applies more generally for all QSM geometries -- are triangulation-dependent. This affects the cosmological aspects of these models, including inflation. Namely, as already mentioned in \cite{Cvetic:2019gnh}, the classical K\"{a}hler potential depends on exactly these triangulation-dependent intersection numbers.

While our analysis is streamlined towards the particle phenomenology, specifically the vector-like spectra, of the F-theory QSMs, we were led to establish -- what we believe to be the first -- computational approaches towards Brill-Noether theory of limit root bundles. The classical Brill-Noether theory concerns a classification of a continuous space of line bundles by their global sections. However, there always exists a finite number $N_{\text{total}}$ of root bundles. In this regard, their Brill-Noether theory becomes a question of finding the partition $N_{\text{total}} = N_0 + N_1 + \dots$ in which $N_i$ is the number of root bundles with $i$ global sections. This is exactly how we propose to read our central results in \oref{tab:Results1} and \oref{tab:Results2}. Also, we propose to consider a \emph{jumping circuit} as the analogue of a classical Brill-Noether jump.

The ignorances in \oref{tab:Results1} and \oref{tab:Results2} stem from two sources. First, we only provide a lower bound if a node remains on an elliptic curve. For this, one could use theta characteristics to identify a basis of the sections on the elliptic curve in question and then work out the conditions imposed by the gluings at the node. The second limitation arises due to the lack of a systematic treatment of line bundle cohomology on circuit-like nodal curves (with rational or even elliptic) components. In this work, we have treated the relevant cases by hand and proposed to distinguish stationary and jumping circuits. In particular, one could speculate that a circuit of rational curves is jumping if there are at least four nodes on one curve component, owing to the fact that the action of $SL(2,\mathbb{Z})$ can be employed to fix the location of three points. Such a study is reserved for future works and should help to further sharpen our understanding of Brill-Noether theory of limit roots.

Our approach towards Brill-Noether theory of limit root bundles also opens an avenue for a potential machine learning application. Indeed when focusing on (powers of) the canonical bundle, a major part of the information is contained in the dual graph of the nodal curve. For sufficiently simple graphs, our algorithm can -- modulo the above-mentioned ignorances -- compute an approximation of the Brill-Noether theory of the corresponding limit roots. It would be fascinating to seek a pattern that relates those graphs and the corresponding Brill-Noether approximations (such as \oref{tab:Results1} and \oref{tab:Results2}). For a mathematical perspective, this should ideally lead to insights that establish a connection among dual graphs and counts of limit root bundles in a provable manner. With a phenomenological application in mind, one could instead hope to acquire a well-trained algorithm that can estimate the limit root bundles on the Higgs curve, which is currently in most setups beyond our computational abilities. Of course, such counts are then best repeated with conditions that preserve (or explicitly enumerate the defect of) the number of global sections upon deformations, as elaborated upon above.

Finally, it must be mentioned that we have avoided to answer a key question in this program, namely which mathematical roots are induced from an F-theory gauge field in the Deligne cohomology. We anticipate that the answer to that question is involved and have therefore, decided to analyze the full set of mathematically admissible root bundles instead. In this current work, we have found that in the family $B_3( \Delta_4^\circ )$ of $\mathcal{O}( 10^{11} )$ different QSM geometries, more than 99.995\% of the root bundles in this superset of mathematically admissible root bundles do not have vector-like exotics. This is indicative that absence of vector-like exotics in the representations $(\mathbf{3}, \mathbf{2})_{1/6}$, $(\mathbf{\overline{3}}, \mathbf{1})_{-2/3}$ and $(\mathbf{1}, \mathbf{1})_{1}$ is a very likely scenario. We reserve a detailed study of the top-down conditions for future work.

\paragraph{Acknowledgement}
We are grateful to Muyang Liu for past collaboration, insightful discussions and ongoing collaboration on limit roots on the Higgs curve. The \texttt{OSCAR} computer algebra system was used for key computations in this work, and we thank Lars Kastner and Benjamin Lorenz for valuable discussions. The reliable computations conducted by the supercomputer \texttt{plesken.mathematik.uni-siegen.de} allowed us to count limit root bundles. This we truly appreciate. M.B.~and M.C.~thank the \emph{Ludwig-Maximilians-Universitaet Muenchen} for hospitality during early stages of this project. M.B., R.D. and M.O.~are partially supported by the NSF grant DMS~2001673 and by the Simons Foundation Collaboration grant \#390287 on ``Homological Mirror Symmetry''. M.O. is grateful for the support by the Ph.D. Presidential Fellowship research fund. The work of M.C.~is supported by DOE Award DE-SC0013528Y. M.B.~ and M.C.~further acknowledge support by the Simons Foundation Collaboration grant \#724069 on ``Special Holonomy in Geometry, Analysis and Physics''. M.C.~thanks the Slovenian Research Agency \mbox{No. P1-0306} and the Fay R.~and Eugene L.~Langberg Chair for support.

\newpage

\appendix

\section{Speciality} \label{sec:Speciality}

Consider a rational, connected, nodal curve $C := \bigcup_{i \in I}{C_i}$ and $L \in \Pic(C)$.
\begin{defn}
We say that $(C, L)$ with $\deg(L) \geq -1$ is \emph{special} if the following equivalent conditions hold.
\begin{enumerate}[(i)]
 \item $h^0(C,L)$ jumps (down) under deformation of $C$,
 \item $h^1(C,L) \neq 0$,
 \item $h^0(C,L) \neq \chi(C,L) = \deg(L) + 1$.
\end{enumerate}
\end{defn} 

\begin{defn}
Let $d_i = \deg(L|_{C_i})$. Write $I = I_+ \cup I_-$, where
\begin{align}
I_- := \left\{ i \in I | d_i < 0 \right\} \, , \qquad I_+ := \left\{ i \in I | d_i \geq 0 \right\} \, .
\end{align}
Let
\begin{align}
C_+ := \bigcup_{i \in I_+}{C_i} \, , \qquad C_- := \bigcup_{i \in I_-}{C_i} \, .
\end{align}
Note that the curves $C_+$, $C_-$ could be disconnected. So, let $k_+$ and $k_-$ be the numbers of connected components of $C_+$ and $C_-$. Let $e_i$ be the intersection number of $C_i$ with $C_-$ and let $L_+$ be the line bundle on $C_+$ with $\deg(L_+|_{C_i}) = d_i - e_i$.
\end{defn}

\begin{prop} \label{equ:RonsCriterion1}
The pair $(C,L)$ with $\deg(L) \geq -1$ is non-special iff the following conditions hold:
\begin{enumerate}[(i)]
 \item $h^0(C_+,L_+) = \chi \left( C_+, L_+ \right)$,
 \item $d_i \geq -1$ for all $i \in I$ and $C_i \cap C_j = \emptyset$ for all distinct $i, j \in I_-$.
\end{enumerate}
\end{prop}

\begin{myproof}
Since $C$ is a rational nodal curve, we have the identity
\begin{align}
\sum_{i \in I_+}{e_i} = k_+ + k_- - 1 \, .
\end{align}
Since Euler characteristics are additive under disjoint union, applying Riemann-Roch to components $C_i$ of $C_+$ where $g(C_i) = 0$ yields
\begin{align} \label{eq:above}
\begin{split}
\chi( C_+,L_+) &=  \deg(L_+) + k_+ = \sum_{i \in I_+}{ d_i } -  \sum_{i \in I_+}{ e_i }+ k_+ \, .
\end{split}
\end{align}
Substituting \eqref{eq:above} into the above gives
\begin{align}
\begin{split}
\chi( C_+,L_+) = \sum_{i \in I_+}{d_i} - k_- +1  = \mathrm{deg}(L) +1 - \Bigl( k_- +\sum_{i \in I_-}{d_i} \Bigr) \geq \chi(C,L) \, ,
\end{split}
\end{align}
The last inequality follows because $d_i \leq -1$ for all $i \in I_-$ and so,
\begin{equation} \label{eq:inequality}
 k_- +\sum_{i \in I_-}{d_i}  \leq k_- - |I_-| \leq 0  \, . 
\end{equation}
By \oref{prop:simplified}, $h^0(C,L) = h^0(C_+, L_+)$. Therefore, 
\begin{align}
h^0(C,L) = h^0(C_+,L_+) \geq \chi(C_+,L_+) \geq \chi(C,L) \, .
\end{align}
Condition (i) ensures that $h^0(C_+, L_+) = \chi(C_+, L_+)$. Also, $\chi(C_+, L_+) = \chi(C,L)$ iff condition (ii) holds. Indeed, if two curves in $C_-$ intersect, then $k_-$ is strictly less than $|I_-|$. Therefore, the right inequality in \oref{eq:inequality} is strict. On the other hand, if $d_i < -1$ for some $i \in I_-$, then the left inequality in \oref{eq:inequality} is strict. Conversely, if condition (ii) holds, then \oref{eq:inequality} is an equality and $\chi(C_+, L_+) = \chi(C, L)$.
\end{myproof}

\begin{algo}[$\deg(L) \geq -1$] Start with $(C, L)$, where $C$ is a rational, connected, nodal curve and $L \in \Pic(C)$ with $\deg(L) \geq -1$
\begin{enumerate}
\item Check condition (ii) of \oref{equ:RonsCriterion1}, which involves the following.
\begin{enumerate}[(i)]
\item Are there any degrees smaller than $-1$? If yes, then $(C, L)$ is special. Otherwise, proceed.
\item Is $I_-$ trivial? If yes, then $(C, L)$ is not special. Otherwise, proceed.
\item Do curves with negative degrees intersect? If yes, then $(C, L)$ is special. Otherwise, proceed.
\end{enumerate}
\item Form $I^{(1)} = I^{(1)}_+ \cup I^{(1)}_-$, where $d_i = \deg(L|_{C_i})$, 
\begin{align} I^{(1)}_+ = \{i \in I : d_i \geq 0 \}  \, , \quad I^{(1)}_- = \{i \in I : d_i < 0\}  \, .
\end{align}
If $C_i$ is a curve component of $C$, then set
\begin{align} C^{(1)}_+ = \bigcup_{i \in I^{(1)}_+} C_i  \, , \quad C^{(1)}_- = \bigcup_{i \in I^{(1)}_-} C_i  \, .
\end{align}
Define a line bundle $L^{(1)}_+$ over $C^{(1)}_+$ such that $\deg(L^{(1)}_+|_{C_i}) = d_i - e_i$ where $e_i$ is the intersection number between $C_i$ and $C^{(1)}_-$. 
\item Is $ C^{(1)}_+ \cap C^{(1)}_- = \emptyset$? If not, feed $(C^{(1)}_+, L^{(1)}_+)$ back into step 2 and iterate. Otherwise, proceed. 
\item Stop iterating when $C^{(n)}_+ \cap C^{(n)}_- = \emptyset$, which means that  $(C^{(n)}_+, L^{(n)}_+)$ is terminal. Since $C$ has finitely many components, the algorithm will eventually terminate. 
\item Check condition (i) of \oref{equ:RonsCriterion1} by computing $\chi(C, L) = \deg(L) + 1$ and 
\begin{align}h^0(C_+^{(n)}, L_+^{(n)}) = \sum_{i \in I^{(n)}_+} d_i + k_+(C^{(n)}_+).\end{align}
If $h^0(C_+^{(n)}, L_+^{(n)}) \neq \chi(C,L)$, then $(C, L)$ is special. Otherwise, $(C, L)$ is non-special.
\end{enumerate}
\end{algo}
This algorithm ultimately seeks to determine if $h^0(C, L) = \chi(C, L)$ by simplifying the curve and checking the conditions of \oref{equ:RonsCriterion1}. Whenever condition (ii) is satisfied in step 1, $\chi (C^{(i)}_+, L^{(i)}_+ ) = \chi (C^{(i+1)}_+, L^{(i+1)}_+ )$ by \oref{equ:RonsCriterion1}. In particular, if the algorithm reaches a terminal curve and condition (ii) is satisfied all throughout, then  
\begin{align} \chi(C, L) = \cdots = \chi (C_+^{(n)}, L_+^{(n)} ).\end{align}
Thanks to \oref{prop:simplified}, we can use the terminal curve to compute $h^0(C, L) = h^0(C_+^{(n)}, L_+^{(n)} )$ in step 5. We check if condition (i) is satisfied by comparing $h^0(C_+^{(n)}, L_+^{(n)})$ with $ \chi (C_+^{(n)}, L_+^{(n)})$.
\begin{exmp}
Consider the following pair $(C, L)$ with $\deg(L) = 0$.
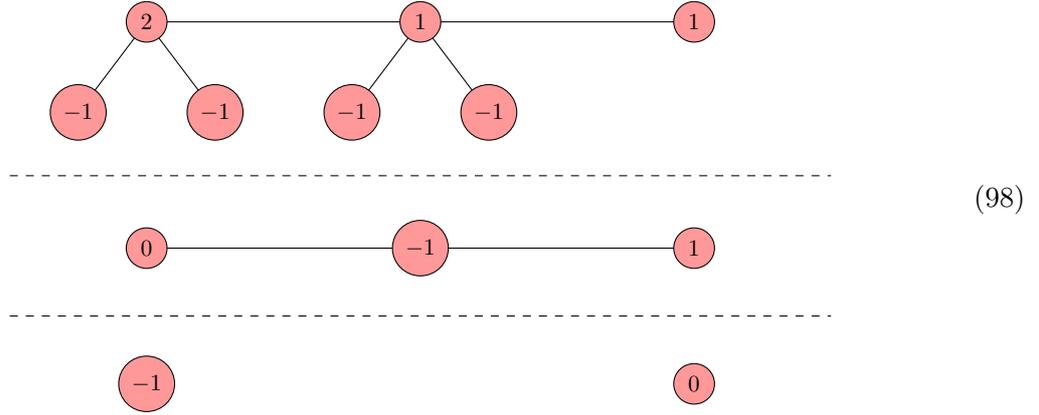
\begin{equation}
\begin{tikzpicture}[scale=0.6, baseline=(current  bounding  box.center)]
      
      \def\s{3.0};
      \def\h{2};
      \def\t{5};
      
      \path[-] (0,0) edge (4*\s,0);
      \path[-] (-0.5*\s,-\h) edge (0,0);
      \path[-] (0.5*\s,-\h) edge (0,0);
      \path[-] (1.5*\s,-\h) edge (2*\s,0);
      \path[-] (2.5*\s,-\h) edge (2*\s,0);
      
      \node at (0,0) [stuff_fill_red, scale=0.8]{$2$};
      \node at (2*\s,0) [stuff_fill_red, scale=0.8]{$1$};
      \node at (4*\s,0) [stuff_fill_red, scale=0.8]{$1$};
      \node at (-0.5*\s,-\h) [stuff_fill_red, scale=0.8]{$-1$};
      \node at (0.5*\s,-\h) [stuff_fill_red, scale=0.8]{$-1$};
      \node at (1.5*\s,-\h) [stuff_fill_red, scale=0.8]{$-1$};
      \node at (2.5*\s,-\h) [stuff_fill_red, scale=0.8]{$-1$};
      
      \draw[dashed] (-\s,-0.68*\t) -- (5*\s,-0.68*\t);
      \path[-] (0,-\t) edge (4*\s,-\t);
      \node at (0,-\t) [stuff_fill_red, scale=0.8]{$0$};
      \node at (2*\s,-\t) [stuff_fill_red, scale=0.8]{$-1$};
      \node at (4*\s,-\t) [stuff_fill_red, scale=0.8]{$1$};

      \draw[dashed] (-\s,-1.3*\t) -- (5*\s,-1.3*\t);
      \node at (0*\s,-1.6*\t) [stuff_fill_red, scale=0.8]{$-1$};
      \node at (4*\s,-1.6*\t) [stuff_fill_red, scale=0.8]{$0$};

\end{tikzpicture}
\end{equation}
Since $I^{(2)}_+ \cap I^{(2)}_- = \emptyset$, the algorithm terminates. Since $h^0(C^{(2)}_+, L^{(2)}_+) =  1 = \chi(C, L)$, the curve is non-special. 
\end{exmp}
\begin{defn}
The pair $(C, L)$ with $\deg(L) < -1$ is \emph{special} iff $h^0(C, L) >0$.

This is because $\chi(C, L) = \deg(L) + 1 < 0$. 
\end{defn}
\begin{algo}[$\deg(L) < -1$] Start with $(C, L)$, where $C$ is a rational, connected, nodal curve and $L \in \Pic(C)$ with $\deg(L) < -1$. 
\begin{enumerate} 
\item Form $I^{(1)} = I^{(1)}_+ \cup I^{(1)}_-$, where $d_i = \deg(L|_{C_i})$, 
\begin{align} I^{(1)}_+ = \{i \in I : d_i \geq 0 \} \, , \quad I^{(1)}_- = \{i \in I: d_i < 0\}  \, . 
\end{align}
If $C_i$ is a curve component of $C$, then set
\begin{align} C^{(1)}_+ = \bigcup_{i \in I^{(1)}_+} C_i  \, , \quad C^{(1)}_- = \bigcup_{i \in I^{(1)}_-} C_i  \, . 
\end{align}
Define a line bundle $L^{(1)}_+$ over $C^{(1)}_+$ such that $\deg(L^{(1)}_+|_{C_i}) = d_i - e_i$ where $e_i$ is the intersection number between $C_i$ and $C^{(1)}_-$.
\item After forming $(C^{(1)}_+, L^{(1)}_+)$, we perform the following checks:
\begin{enumerate}[(i)]
\item Is $I^{(1)}_+ = \emptyset$? If yes, then $(C, L)$ is non-special. Otherwise, proceed.
\item Is $I^{(1)}_+ \cap I^{(1)}_- = \emptyset$? If not, then feed $(C^{(1)}_+, L^{(1)}_+)$ into back into Step 2 and iterate. Otherwise proceed.
\end{enumerate}
\item  Stop iterating when $C^{(n)}_+ \cap C^{(n)}_- = \emptyset$, which means that  $(C^{(n)}_+, L^{(n)}_+)$ is terminal. Since $C$ has finitely many components, the algorithm will eventually terminate. Compute 
\begin{align}
h^0(C^{(n)}_+, I^{(n)}_+) = \sum_{i \in I_+} d_i + k_+(C^{(n)}_+)  \, .
\end{align}
\item If $h^0(C^{(n)}_+, I^{(n)}_+) > 0$, then $(C, L)$ is special. If not, then $(C, L)$ is non-special.
\end{enumerate}
\end{algo}
This algorithm simplifies $(C, L)$ and allows us to compute $h^0(C, L) = h^0(C_+^{(n)}, L_+^{(n)})$ using the terminal curve thanks to \oref{prop:simplified}. 

\begin{exmp}
Consider the following pair $(C, L)$ with $N = 3$ and $\deg(L) = -3$.
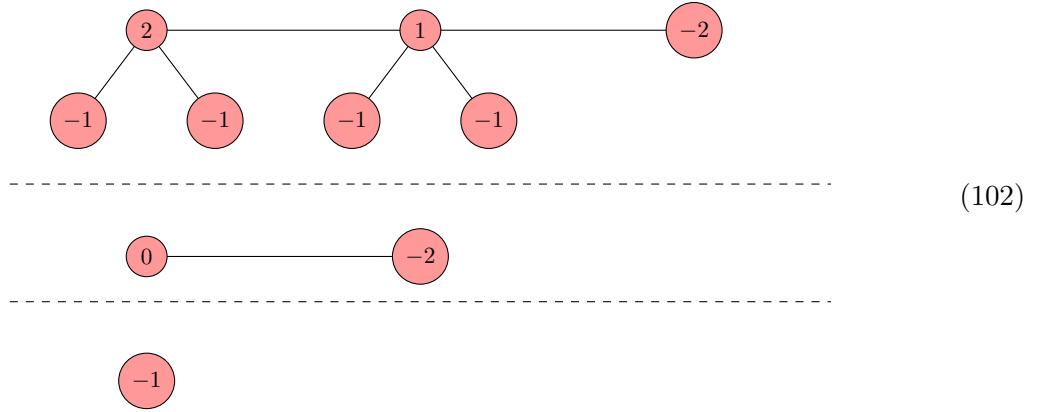
\begin{equation}
\begin{tikzpicture}[scale=0.6, baseline=(current  bounding  box.center)]
      
  	  \def\s{3.0};
      \def\h{2};
      \def\t{5};
      
      \path[-] (-0.5*\s,-\h) edge (0,0);
      \path[-] (0.5*\s,-\h) edge (0,0);
      \path[-] (1.5*\s,-\h) edge (2*\s,0);
      \path[-] (2.5*\s,-\h) edge (2*\s,0);
 	  \path[-] (4*\s,0) edge (0,0);
      
      \node at (0,0) [stuff_fill_red, scale=0.8]{$2$};
      \node at (2*\s,0) [stuff_fill_red, scale=0.8]{$1$};
      \node at (4*\s,0) [stuff_fill_red, scale=0.8]{$-2$};
      \node at (-0.5*\s,-\h) [stuff_fill_red, scale=0.8]{$-1$};
      \node at (0.5*\s,-\h) [stuff_fill_red, scale=0.8]{$-1$};
      \node at (1.5*\s,-\h) [stuff_fill_red, scale=0.8]{$-1$};
      \node at (2.5*\s,-\h) [stuff_fill_red, scale=0.8]{$-1$};
      
      \draw[dashed] (-\s,-0.68*\t) -- (5*\s,-0.68*\t);
      \path[-] (2*\s,-\t) edge (0,-\t);
      \node at (0,-\t) [stuff_fill_red, scale=0.8]{$0$};
      \node at (2*\s,-\t) [stuff_fill_red, scale=0.8]{$-2$};	
      \draw[dashed] (-\s,-1.2*\t) -- (5*\s,-1.2*\t);
      \node at (0,-1.55*\t) [stuff_fill_red, scale=0.8]{$-1$};	
\end{tikzpicture}
\end{equation}
Since $I^{(2)}_+ \cap I^{(2)}_- = \emptyset$, the algorithm terminates. Since $h^0(C_+^{(2)}, L_+^{(2)}) = 0$, the pair $(C, L)$ is non-special.
\end{exmp}
We can also formulate an equivalent criterion for determining speciality when $\deg(L) \geq -1$.

\begin{prop}
Let $\mathrm{deg}(L) \geq -1$. Then the following are equivalent: 
\begin{enumerate}
 \item $(C,L)$ is non-special,
 \item the restriction of $L$ to every connected subcurve of $C$ has degree $\geq -1$.
\end{enumerate}
\end{prop}

\begin{myproof}
\item $(1) \Rightarrow (2)$:\\
Suppose that $(C, L)$ is non-special and there exists a connected subcurve $C_{\text{sub}}$ of $C$ such that $\deg(L|_{C_{\text{sub}}}) < -1.$ Let $I_{\text{sub}} \subseteq I$ parametrize the components of $C_{\text{sub}}$. Since $(C, L)$ is non-special, $(C_{\text{sub}}, L|_{C_{\text{sub}}})$ satisfies condition (ii) of \oref{equ:RonsCriterion1} since components of $C_{\text{sub}}$ are components of $C$. So, we have 
\begin{align}
-1 > \deg(L|_{C_{\text{sub}}}) =\sum_{i \in I_{\text{sub}}} d_i \geq  \sum_{i \in (I_{\text{sub}})_+} (d_i-e_i) = \deg(L_+|_{(C_{\text{sub}})_+})  \, .
\end{align}
The inequality above is strict whenever a component of $(I_{\text{sub}})_-$ intersects more than one component of $(I_{\text{sub}})_+$. Since $(C, L)$ is non-special, $(C^{(i)}_+, L^{(i)}_+t)$ is non-special for all $i \in \mathbb{N}$ by condition (i) of \oref{equ:RonsCriterion1}. By iterating the argument, we have that $\left((C^{(i)}_{\text{sub}})_+, (L^{(i)}_{\text{sub}})_+\right)$ satisfies condition (ii) and 
\begin{align} \label{eq:negative} 
-1 > \deg(L|_{C_{\text{sub}}}) \geq  \deg(L^{(i)}_+|_{(C^{(i)}_{\text{sub}})_+})
\end{align}
for all $i \in \mathbb{N}$. This includes the terminal pair $\left((C^{(n)}_{\text{sub}})_+, (L^{(n)}_{\text{sub}})_+\right)$ for which $(C^{(n)}_{\text{sub}})_+ \cap (C^{(n)}_{\text{sub}})_- = \emptyset$. However, for \oref{eq:negative} to hold, the negative degrees must be concentrated in $(I^{(n)}_{\text{sub}})_-$, which violates condition (ii) and contradicts the non-speciality of $(C, L)$. 

\item $(2) \Rightarrow (1)$:\\
We first make two simple observations. First, by assumption, $d_i = \mathrm{deg}( \left. L \right|_{C_i} ) \geq -1$. Second, if two components $C_i$ and $C_j$ with degree $-1$ intersect, then the restriction of $L$ to $C_i \cup C_j$ would have degree $-2 < -1$, which contradicts (2). Hence, no curves with negative degrees intersect.
        
Recall from \oref{equ:RonsCriterion1} that $(C,L)$ is non-special iff $d_i \geq -1$, no two components with negative degrees intersect and $(C_+, L_+)$ is non-special. Since we just argued that the first two are satisfied, it remains to show that $(C_+, L_+)$ is non-special.
        
Consider a connected subcurve $C_{\text{sub}}$ of $C_+$. Then,
\begin{align}
\mathrm{deg} ( \left. L_+ \right|_{C_{\text{sub}}} ) = \sum_{i \in C_{\text{sub}}}{(d_i - e_i)} = \sum_{i \in \widetilde{C}_{\text{sub}}}{d_i} = \mathrm{deg} ( \left. L \right|_{\widetilde{C}_{\text{sub}}} ) \, ,
\end{align}
where $\widetilde{C}_{\text{sub}}$ is the extension of $C_{\text{sub}}$ by all curves in $C$ adjacent to $C_{\text{sub}}$ with $d_i = -1$. In particular, $\widetilde{C}_{\text{sub}}$ is connected. Hence, $\mathrm{deg} ( \left. L_+ \right|_{C_{\text{sub}}})  \geq -1$ for all connected subcurves $C_{\text{sub}}$ of $C_+$ by assumption.
        
By iterating our argument, we conclude that the degree of $L_+$ on all components in $C_+$ is greater or equal to $-1$. Also, no two components of $C_+$ with negative degree of $L_+$ do intersect. Therefore by \oref{equ:RonsCriterion1}, it remains to show that $(C_{++}, L_{++})$ is non-special.
        
After sufficiently many iterations, we achieve one of the following configurations $C_+^{(n)}$:
\begin{itemize}
\item $( C_+^{(n)} )_-$ is disconnected from $( C_+^{(n)} )_+$: No components with negative degrees intersect. Hence, $( C_+^{(n)} )_-$ is a finite, disjoint union of components, each with $d_i = -1$. So, $h^1( C_+^{(n)}, L_+^{(n)} ) = 0$ and $( C^{(n)}_+, L^{(n)}_+ )$ is non-special.
\item $( C_+^{(n)} )_- = \emptyset$: Then, $h^1( C_+^{(n)}, L_+^{(n)} ) = 0$ and $( C^{(n)}_+, L^{(n)}_+ )$ is non-special. \qedhere
\end{itemize}
\end{myproof}

\section{Counts of partial blowup limit roots} \label{appendix:DetailsOnCounts}

In this section we present the detailed counts of limit roots for a number $N$ of remaining nodes. For $\Delta_{8}^\circ$, there are no limit roots if $N = 4$ or $N = 5$. This repeats for all other spaces. For ease of presentation we omit trivial cases and list absolute numbers of limit roots as multiples of the multiplicity $\mu$ introduced in \oref{equ:Multiplicity}. Percentages are always rounded to one decimal place. Hence, in the following tables 0.0 indicates a non-zero result smaller than $0.045$.

\subsection{Spaces with \texorpdfstring{$\overline{K}_{B_3}^3 = 6$}{KB3=6}}

\begin{align}
\begin{tabular}{cc|cc|cc}
\toprule
& \multirow{2}{*}{$N$} & \multicolumn{2}{c|}{Percentages} & \multicolumn{2}{c}{Absolute numbers} \\
& & $h^0 = 3$ & $h^0 \geq 3$ & $h^0 = 3$ & $h^0 \geq 3$ \\
\midrule
\multirow{6}{*}{$\Delta_8^\circ$ ($\mu = 12^3$)} & 0 & 57.3 & & 142560 & \\ 
& 1 & 19.1 & 19.1 & 47520 & 47520 \\
& 2 & & 1.9 & & 4752\\
& 3 & & 2.5 & & 6336 \\
& 6 & & 0.1 & & 144 \\
& $\Sigma$ & 76.4 & 23.6 & 190080 & 58752 \\
\midrule
\multirow{8}{*}{$\Delta_4^\circ$ ($\mu = 12^4$)} & 0 & 53.6 & & 11110 & \\
& 1 & 36.7 & & 7601 & \\
& 2 & 7.5 & 0.5 & 1562 & 110 \\
& 3 & 1.3 & 0.1 & 264 & 11 \\
& 4 & & 0.3 & & 66 \\
& 5 & & 0.1 & & 11 \\
& 7 & & 0.0 & & 1 \\
& $\Sigma$ & 99.0 & 1.0 & 20537 & 199 \\
\midrule
\multirow{8}{*}{$\Delta_{134}^\circ$ ($\mu = 12^4$)} & 0 & 48.3 & & 10010 & \\
& 1 & 40.3 & & 8360 & \\
& 2 & 8.6 & & 1782 & \\ 
& 3 & 2.3 & & 484 & \\ 
& 4 & 0.3 & & 55 & \\ 
& 5 & & 0.2 & & 44 \\ 
& 8 & & 0.0 & & 1 \\
& $\Sigma$ & 99.8 & 0.2 & 20691 & 45 \\
\midrule
\multirow{9}{*}{$\Delta_{128}^\circ$, $\Delta_{130}^\circ$, $\Delta_{136}^\circ$, $\Delta_{236}^\circ$ ($\mu = 12^4$)} & 0 & 43.0 & & 8910 & \\
& 1 & 44.6 & & 9240 & \\ 
& 2 & 8.0 & & 1650 & \\ 
& 3 & 3.9 & & 814 & \\
& 4 & 0.3 & & 66 & \\ 
& 5 & 0.2 & & 33 & \\ 
& 6 & & 0.1 & & 22 \\
& 9 & & 0.0 & & 1 \\
& $\Sigma$ & 99.9 & 0.1 & 20713 & 23 \\
\bottomrule
\end{tabular}
\end{align}

\newpage

\subsection{Spaces with \texorpdfstring{$\overline{K}_{B_3}^3 = 10$}{KB3=10}}

\begin{footnotesize}
\begin{longtable}{cc|cc|cc|cc|ccc}
\caption{Percentages}\\
\toprule
& $N$ & $h^0 = 3$ & $h^0 \geq 3$ & $h^0 = 4$ & $h^0 \geq 4$ & $h^0 = 5$ & $h^0 \geq 5$ & $h^0 = 6$ & $h^0 \geq 6$ \\
\midrule
\endfirsthead
\multicolumn{8}{c}
{\tablename\ \thetable\ -- \textit{Percentages -- continued from previous page}} \\
\midrule
& $N$ & $h^0 = 3$ & $h^0 \geq 3$ & $h^0 = 4$ & $h^0 \geq 4$ & $h^0 = 5$ & $h^0 \geq 5$ & $h^0 = 6$ & $h^0 \geq 6$ \\
\midrule
\endhead
\midrule \multicolumn{8}{r}{\textit{Percentages -- continued on next page}} \\
\endfoot
\endlastfoot
\multirow{8}{*}{$\Delta_{88}^\circ$}
&0&61.1&0.0&2.0&&0.0&&& \\*
&1&12.8&16.2&0.5&0.4&&0.0&& \\*
&2&1.0&5.2&0.0&0.1&&&& \\*
&3&0.0&0.7&&0.0&&&& \\*
&4&&0.1&&&&&& \\*
&5&&0.0&&&&&& \\*
&6&&0.0&&&&&& \\*
&$\Sigma$&74.9&22.1&2.5&0.5&0.0&0.0&& \\
 \midrule
 \multirow{8}{*}{$\Delta_{110}^\circ$}
&0&57.7&0.0&2.2&0.0&0.0&&& \\*
&1&21.1&9.1&0.9&0.3&&&& \\*
&2&3.3&4.1&0.1&0.1&&&& \\*
&3&0.3&0.8&&0.0&&&& \\*
&4&0.0&0.1&&&&&& \\*
&5&&0.0&&&&&& \\*
&6&&0.0&&&&&& \\*
&$\Sigma$&82.4&14.1&3.1&0.4&0.0&&& \\
 \midrule
 \multirow{4}{*}{$\Delta_{272}^\circ$}
&0&57.5&0.0&2.4&0.0&0.0&&& \\*
&1&17.9&12.2&0.9&0.4&0.0&0.0&& \\*
&2&2.4&4.9&0.1&0.1&&&& \\*
&3&0.2&0.9&&0.0&&&& \\*
\multirow{4}{*}{$\Delta_{274}^\circ$}&4&0.0&0.1&&&&&& \\*
&5&&0.0&&&&&& \\*
&6&&0.0&&&&&& \\*
&$\Sigma$&78.1&18.0&3.4&0.5&0.0&0.0&& \\
 \midrule
 \multirow{8}{*}{$\Delta_{387}^\circ$}
&0& \textcolor{green}{57.3}&0.0&2.6&&0.0&&& \\*
&1&14.8&15.2&0.8&0.6&0.0&0.0&& \\*
&2&1.6&5.7&0.1&0.1&&&& \\*
&3&0.1&1.0&0.0&0.0&&&& \\*
&4&0.0&0.1&&&&&& \\*
&5&&0.0&&&&&& \\*
&6&&0.0&&&&&& \\*
&$\Sigma$&73.8&21.9&3.5&0.7&0.0&0.0&& \\
 \midrule
 \multirow{2}{*}{$\Delta_{798}^\circ$}
&0&54.0&0.0&2.9&0.0&0.0&&& \\*
&1&19.5&11.4&1.3&0.5&0.0&0.0&& \\*
\multirow{2}{*}{$\Delta_{808}^\circ$}&2&3.1&5.2&0.2&0.2&0.0&0.0&& \\*
&3&0.3&1.1&0.0&0.0&&&& \\*
\multirow{2}{*}{$\Delta_{810}^\circ$}&4&0.0&0.1&&&&&& \\*
&5&0.0&0.0&&&&&& \\*
\multirow{2}{*}{$\Delta_{812}^\circ$}&6&&0.0&&&&&& \\*
&$\Sigma$&77.0&17.9&4.4&0.7&0.0&0.0&& \\
 \midrule
 \multirow{9}{*}{$\Delta_{254}^\circ$}
&0&54.7&&2.2&&0.0&&& \\*
&1&31.8&&1.1&&0.0&&& \\*
&2&8.2&0.3&0.2&0.0&0.0&0.0&& \\*
&3&1.2&0.2&0.0&0.0&&&& \\*
&4&0.1&0.0&0.0&0.0&&&& \\*
&5&0.0&0.0&&0.0&&&& \\*
&6&&0.0&&&&&& \\*
&7&&0.0&&&&&& \\*
&$\Sigma$&95.9&0.5&3.5&0.0&0.0&0.0&& \\
 \midrule
 \multirow{9}{*}{$\Delta_{52}^\circ$}
&0&54.7&&2.2&&0.0&&& \\*
&1&31.5&&1.3&&0.0&&& \\*
&2&8.0&0.5&0.3&0.0&&0.0&& \\*
&3&1.1&0.2&0.0&0.0&&&& \\*
&4&0.1&0.1&&0.0&&&& \\*
&5&0.0&0.0&&0.0&&&& \\*
&6&&0.0&&&&&& \\*
&7&&0.0&&&&&& \\*
&$\Sigma$&95.3&0.7&3.9&0.0&0.0&0.0&& \\
 \midrule
 \multirow{9}{*}{$\Delta_{302}^\circ$}
&0&54.5&&2.2&&0.0&&& \\*
&1&32.2&&1.1&&0.0&&& \\*
&2&7.9&0.3&0.2&0.0&0.0&&& \\*
&3&1.2&0.2&0.0&0.0&&&& \\*
&4&0.1&0.0&0.0&0.0&&&& \\*
&5&0.0&0.0&&0.0&&&& \\*
&6&&0.0&&&&&& \\*
&7&&0.0&&&&&& \\*
&$\Sigma$&95.9&0.5&3.5&0.0&0.0&&& \\
 \midrule
 \multirow{9}{*}{$\Delta_{786}^\circ$}
&0&51.3&&2.7&&0.0&&& \\*
&1&32.4&&1.7&&0.0&&& \\*
&2&9.3&0.1&0.4&0.0&0.0&0.0&& \\*
&3&1.6&0.1&0.0&0.0&0.0&0.0&& \\*
&4&0.2&0.0&0.0&0.0&&&& \\*
&5&0.0&0.0&0.0&0.0&&&& \\*
&6&0.0&0.0&&&&&& \\*
&7&&0.0&&&&&& \\*
&$\Sigma$&94.8&0.3&4.8&0.0&0.0&0.0&& \\
 \midrule
 \multirow{9}{*}{$\Delta_{762}^\circ$}
&0&51.3&&2.7&&0.0&&& \\*
&1&32.3&&1.7&&0.0&&& \\*
&2&9.3&0.1&0.4&0.0&0.0&0.0&& \\*
&3&1.6&0.1&0.1&0.0&&&& \\*
&4&0.2&0.0&0.0&0.0&&&& \\*
&5&0.0&0.0&&0.0&&&& \\*
&6&0.0&0.0&&&&&& \\*
&7&&0.0&&&&&& \\*
&$\Sigma$&94.8&0.3&4.9&0.0&0.0&0.0&& \\
 \midrule
 \multirow{9}{*}{$\Delta_{417}^\circ$}
&0&51.3&&2.7&&0.0&&0.0& \\*
&1&32.4&&1.7&&0.0&&& \\*
&2&9.3&0.1&0.4&0.0&0.0&0.0&& \\*
&3&1.6&0.1&0.0&0.0&0.0&0.0&& \\*
&4&0.2&0.0&0.0&0.0&&&& \\*
&5&0.0&0.0&0.0&0.0&&&& \\*
&6&0.0&0.0&&&&&& \\*
&7&&0.0&&&&&& \\*
&$\Sigma$&94.8&0.3&4.8&0.0&0.0&0.0&0.0& \\
 \midrule
 \multirow{9}{*}{$\Delta_{838}^\circ$}
&0&51.3&&2.7&&0.0&&& \\*
&1&32.3&&1.8&&0.0&&& \\*
&2&9.3&0.1&0.5&0.0&0.0&0.0&& \\*
&3&1.6&0.1&0.1&0.0&&0.0&& \\*
&4&0.2&0.0&0.0&0.0&&&& \\*
&5&0.0&0.0&&0.0&&&& \\*
&6&0.0&0.0&&&&&& \\*
&7&&0.0&&&&&& \\*
&$\Sigma$&94.7&0.3&5.0&0.0&0.0&0.0&& \\
 \midrule
 \multirow{9}{*}{$\Delta_{782}^\circ$}
&0&51.3&&2.7&&0.0&&& \\*
&1&32.3&&1.8&&0.0&&& \\*
&2&9.3&0.1&0.5&0.0&0.0&0.0&& \\*
&3&1.6&0.1&0.1&0.0&&0.0&& \\*
&4&0.2&0.0&0.0&0.0&&&& \\*
&5&0.0&0.0&&0.0&&&& \\*
&6&0.0&0.0&&&&&& \\*
&7&&0.0&&&&&& \\*
&$\Sigma$&94.6&0.3&5.0&0.0&0.0&0.0&& \\
 \midrule
 \multirow{3}{*}{$\Delta_{377}^\circ$}
&0&48.2&&3.1&&0.1&&& \\*
&1&32.7&&2.3&&0.0&&& \\*
&2&10.3&0.1&0.7&0.0&0.0&0.0&& \\*
\multirow{3}{*}{$\Delta_{499}^\circ$} &3&2.0&0.1&0.1&0.0&0.0&0.0&& \\*
&4&0.2&0.0&0.0&0.0&&&& \\*
&5&0.0&0.0&0.0&0.0&&&& \\*
\multirow{3}{*}{$\Delta_{503}^\circ$} &6&0.0&0.0&&&&&& \\*
&7&0.0&0.0&&&&&& \\*
&$\Sigma$&93.4&0.2&6.2&0.0&0.1&0.0&& \\
 \midrule
 \multirow{9}{*}{$\Delta_{1348}^\circ$}
&0&48.2&&3.1&&0.1&&0.0& \\*
&1&32.7&&2.3&&0.0&&& \\*
&2&10.4&&0.7&&0.0&&& \\*
&3&2.0&0.0&0.1&0.0&0.0&&& \\*
&4&0.3&0.0&0.0&0.0&&&& \\*
&5&0.0&0.0&0.0&0.0&&&& \\*
&6&0.0&0.0&&&&&& \\*
&7&0.0&0.0&&&&&& \\*
&$\Sigma$&93.7&0.0&6.2&0.0&0.1&&0.0& \\
 \midrule
 \multirow{4}{*}{$\Delta_{882}^\circ$}
&0&48.2&&3.1&&0.0&&0.0& \\*
&1&32.7&&2.3&&0.0&&& \\*
&2&10.3&0.1&0.7&0.0&0.0&0.0&& \\*
&3&2.0&0.1&0.1&0.0&0.0&0.0&& \\*
\multirow{5}{*}{\textcolor{red}{$\Delta_{856}^\circ$}}&4&0.2&0.0&0.0&0.0&&&& \\*
&5&0.0&0.0&0.0&0.0&&&& \\*
&6&0.0&0.0&&&&&& \\*
&7&0.0&0.0&&&&&& \\*
&$\Sigma$&93.4&0.3&6.2&0.0&0.1&0.0&0.0& \\
 \midrule
 \multirow{9}{*}{$\Delta_{1340}^\circ$}
&0&45.2&&3.5&&0.1&&0.0& \\*
&1&32.9&&2.8&&0.1&&0.0& \\*
&2&11.3&&1.0&&0.0&&& \\*
&3&2.4&0.0&0.2&0.0&0.0&&& \\*
&4&0.3&0.0&0.0&0.0&&&& \\*
&5&0.0&0.0&0.0&0.0&&&& \\*
&6&0.0&0.0&&&&&& \\*
&7&0.0&0.0&&&&&& \\*
&$\Sigma$&92.3&0.0&7.6&0.0&0.1&&0.0& \\
 \midrule
 \multirow{9}{*}{$\Delta_{1879}^\circ$}
&0&45.2&&3.5&&0.1&&0.0& \\*
&1&32.9&&2.8&&0.1&&0.0& \\*
&2&11.3&&1.0&&0.0&&& \\*
&3&2.4&0.0&0.2&0.0&0.0&&& \\*
&4&0.3&0.0&0.0&0.0&&&& \\*
&5&0.0&0.0&0.0&0.0&&&& \\*
&6&0.0&0.0&&&&&& \\*
&7&0.0&0.0&&&&&& \\*
&$\Sigma$&92.3&0.0&7.5&0.0&0.1&&0.0& \\
 \midrule
 \multirow{9}{*}{$\Delta_{1384}^\circ$}
&0&42.5&&3.8&&0.1&&0.0& \\*
&1&33.0&&3.4&&0.1&&0.0& \\*
&2&12.1&&1.4&&0.0&&& \\*
&3&2.8&&0.3&&0.0&&& \\*
&4&0.4&0.0&0.0&0.0&&&& \\*
&5&0.0&0.0&0.0&0.0&&&& \\*
&6&0.0&0.0&&&&&& \\*
&7&0.0&0.0&&&&&& \\*
&$\Sigma$&90.9&0.0&8.9&0.0&0.2&&0.0& \\*
\bottomrule
\end{longtable}
\end{footnotesize}

\begin{footnotesize}
\begin{longtable}{ccc|cc|cc|cc|cc}
\caption{Absolute numbers}\\
\toprule
& & $N$ & $h^0 = 3$ & $h^0 \geq 3$ & $h^0 = 4$ & $h^0 \geq 4$ & $h^0 = 5$ & $h^0 \geq 5$ & $h^0 = 6$ & $h^0 \geq 6$ \\
\midrule
\endfirsthead
\multicolumn{8}{c}
{\tablename\ \thetable\ -- \textit{Absolute numbers -- continued from previous page}} \\
\midrule
& & $N$ & $h^0 = 3$ & $h^0 \geq 3$ & $h^0 = 4$ & $h^0 \geq 4$ & $h^0 = 5$ & $h^0 \geq 5$ & $h^0 = 6$ & $h^0 \geq 6$ \\
\midrule
\endhead
\midrule \multicolumn{8}{r}{\textit{Absolute numbers -- continued on next page}} \\
\endfoot
\endlastfoot
\multirow{8}{*}{$\Delta_{88}^\circ$} & \multirow{8}{*}{$\mu = 20^5$}
& 0 & 781680888 & 62712 & 25196800 & & 106800 & \\*
& & 1 & 163221088 & 206886912 & 5967200 & 5200000 & & 7200 \\*
& & 2 & 13270504 & 66489896 & 399200 & 880400 & & \\*
& & 3 & 504800 & 9361600 & & 45600 & & \\*
& & 4 & & 692000 & & & & \\*
& & 5 & & 24800 & & & & \\*
& & 6 & & 1600 & & & & \\*
& & $\Sigma$ & 958677280 & 283519520 & 31563200 & 6126000 & 106800 & 7200 \\
\midrule
\multirow{8}{*}{$\Delta_{110}^\circ$} & \multirow{8}{*}{$\mu = 20^5$}
& 0 & 738662983 & 301017 & 27732238 & 5762 & 168000 & \\*
& & 1 & 270495495 & 116866505 & 10971166 & 3856834 & & \\*
& & 2 & 41953954 & 52251246 & 979978 & 958822 & & \\*
& & 3 & 3212904 & 10434296 & & 76800 & & \\*
& & 4 & 83188 & 930812 & & & & \\*
& & 5 & & 54000 & & & & \\*
& & 6 & & 4000 & & & & \\*
& & $\Sigma$ & 1054408524 & 180841876 & 39683382 & 4898218 & 168000 & \\
\midrule
\multirow{4}{*}{$\Delta_{272}^\circ$} & \multirow{8}{*}{$\mu = 20^5$}
& 0 & 736011640 & 131160 & 30428570 & 2230 & 196800 & \\*
& & 1 & 229717974 & 156098026 & 11338671 & 5229729 & 2000 & 31200 \\*
& & 2 & 31225880 & 62229320 & 1224000 & 1388800 & & \\*
& & 3 & 2052498 & 11425902 & & 105600 & & \\*
\multirow{4}{*}{$\Delta_{274}^\circ$} & & 4 & 43200 & 1064800 & & & & \\*
& & 5 & & 49600 & & & & \\*
& & 6 & & 2400 & & & & \\*
& & $\Sigma$ & 999051192 & 231001208 & 42991241 & 6726359 & 198800 & 31200  \\
\midrule
\multirow{8}{*}{$\Delta_{387}^\circ$} & \multirow{8}{*}{$\mu = 20^5$}
& 0 & \textcolor{green}{733798300} & 59300 & 32742000 & & 306000 & \\*
& & 1 & 189788940 & 194187060 & 10830000 & 7116000 & 72000 & 36000 \\*
& & 2 & 20364840 & 73241160 & 1224000 & 1596000 & & \\*
& & 3 & 1134000 & 12170000 & 60000 & 120000 & & \\*
& & 4 & 34000 & 1074000 & & & & \\*
& & 5 & & 44400 & & & & \\*
& & 6 & & 2000 & & & & \\*
& & $\Sigma$ & 945120080 & 280777920 & 44856000 & 8832000 & 378000 & 36000 \\
\midrule
\multirow{2}{*}{$\Delta_{798}^\circ$} & \multirow{8}{*}{$\mu = 20^5$}
& 0 & 690950608 & 123792 & 37074570 & 2230 & 469600 & \\*
& & 1 & 250171050 & 146465350 & 16720271 & 6645729 & 128800 & 54400 \\*
\multirow{2}{*}{$\Delta_{808}^\circ$} & & 2 & 40113524 & 66594076 & 2840800 & 2065600 & 400 & 1600 \\*
& & 3 & 3586377 & 13878823 & 147600 & 211200 & & \\*
\multirow{2}{*}{$\Delta_{810}^\circ$} & & 4 & 157600 & 1506400 & & & & \\*
& & 5 & 2400 & 80800 & & & & \\*
\multirow{2}{*}{$\Delta_{812}^\circ$} & & 6 & & 6400 & & & & \\*
& & $\Sigma$ & 984981559 & 228655641 & 56783241 & 8924759 & 598800 & 56000 \\
\midrule
\multirow{9}{*}{$\Delta_{254}^\circ$} & \multirow{9}{*}{$\mu = 20^6$}
& 0 & 35004914 & & 1396842 & & 9258 & \\*
& & 1 & 20350604 & & 706580 & & 1822 & \\*
& & 2 & 5220580 & 194384 & 129078 & 6010 & 94 & 4 \\*
& & 3 & 759982 & 111668 & 11322 & 2098 & & \\*
& & 4 & 60688 & 27286 & 360 & 276 & & \\*
& & 5 & 2616 & 3238 & & 12 & & \\*
& & 6 & & 274 & & & & \\*
& & 7 & & 10 & & & & \\*
& & $\Sigma$ & 61399384 & 336860 & 2244182 & 8396 & 11174 & 4 \\
\midrule
\multirow{9}{*}{$\Delta_{52}^\circ$} & \multirow{9}{*}{$\mu = 20^6$}
& 0 & 34980351 & & 1438026 & & 9794 & \\*
& & 1 & 20149232 & & 862358 & & 1408 & \\*
& & 2 & 5110469 & 292371 & 173179 & 10368 & & 54 \\*
& & 3 & 715533 & 149594 & 10409 & 5304 & & \\*
& & 4 & 52346 & 33342 & & 597 & & \\*
& & 5 & 1308 & 3817 & & 9 & & \\*
& & 6 & & 127 & & & & \\*
& & 7 & & 4 & & & & \\*
& & $\Sigma$ & 61009239 & 479255 & 2483972 & 16278 & 11202 & 54 \\
\midrule
\multirow{9}{*}{$\Delta_{302}^\circ$} & \multirow{9}{*}{$\mu = 20^6$}
& 0 & 34908682 & & 1396458 & & 6388 & \\*
& & 1 & 20592434 & & 715731 & & 4874 & \\*
& & 2 & 5055454 & 194096 & 121628 & 5960 & 320 & \\*
& & 3 & 741181 & 118536 & 11182 & 2346 & & \\*
& & 4 & 85844 & 25418 & 744 & 174 & & \\*
& & 5 & 6776 & 4922 & & 39 & & \\*
& & 6 & & 786 & & & & \\*
& & 7 & & 27 & & & & \\*
& & $\Sigma$ & 61390371 & 343785 & 2245743 & 8519 & 11582 & \\
\midrule
\multirow{9}{*}{$\Delta_{786}^\circ$} & \multirow{9}{*}{$\mu = 20^6$}
& 0 & 32860461 & & 1719897 & & 19661 & \\*
& & 1 & 20719735 & & 1076991 & & 6835 & \\*
& & 2 & 5976644 & 91108 & 265865 & 3978 & 667 & 13 \\*
& & 3 & 1029259 & 60313 & 31227 & 2121 & 14 & 11 \\*
& & 4 & 106187 & 17559 & 1575 & 264 & & \\*
& & 5 & 6217 & 2758 & 18 & 18 & & \\*
& & 6 & 238 & 331 & & & & \\*
& & 7 & & 35 & & & & \\*
& & $\Sigma$ & 60698741 & 172104 & 3095573 & 6381 & 27177 & 24 \\
\midrule
\multirow{9}{*}{$\Delta_{762}^\circ$} & \multirow{9}{*}{$\mu = 20^6$}
& 0 & 32858151 & & 1722871 & & 18984 & \\*
& & 1 & 20703000 & & 1093360 & & 6702 & \\*
& & 2 & 5962907 & 91385 & 281675 & 3935 & 262 & 18 \\*
& & 3 & 1020651 & 63981 & 33392 & 2526 & & \\*
& & 4 & 105082 & 19543 & 1338 & 567 & & \\*
& & 5 & 6182 & 3022 & & 30 & & \\*
& & 6 & 160 & 258 & & & & \\*
& & 7 & & 18 & & & & \\*
& & $\Sigma$ & 60656133 & 178207 & 3132636 & 7058 & 25948 & 18 \\
\midrule
\multirow{9}{*}{$\Delta_{417}^\circ$} & \multirow{9}{*}{$\mu = 20^6$}
& 0 & 32857596 & & 1723530 & & 18545 & & 77 & \\*
& & 1 & 20722495 & & 1074066 & & 7982 & & & \\*
& & 2 & 5980745 & 90711 & 260581 & 4102 & 880 & 15 & & \\*
& & 3 & 1024325 & 66272 & 30773 & 2116 & 14 & 5 & & \\*
& & 4 & 104665 & 19054 & 1548 & 333 & & & & \\*
& & 5 & 6204 & 2749 & 18 & 6 & & & & \\*
& & 6 & 242 & 320 & & & & & & \\*
& & 7 & & 31 & & & & & & \\*
& & $\Sigma$ & 60696272 & 179137 & 3090516 & 6557 & 27421 & 20 & 77 & \\
\midrule
\multirow{9}{*}{$\Delta_{838}^\circ$} & \multirow{9}{*}{$\mu = 20^6$}
& 0 & 32845047 & & 1739918 & & 20615 & \\*
& & 1 & 20641615 & & 1137880 & & 9155 & \\*
& & 2 & 5963926 & 90449 & 290774 & 4354 & 692 & 25 \\*
& & 3 & 1015475 & 66705 & 36569 & 2288 & & 13 \\*
& & 4 & 104650 & 18640 & 1686 & 324 & & \\*
& & 5 & 5772 & 2871 & & 27 & & \\*
& & 6 & 146 & 354 & & & & \\*
& & 7 & & 30 & & & & \\*
& & $\Sigma$ & 60576631 & 179049 & 3206827 & 6993 & 30462 & 38 \\
\midrule
\multirow{9}{*}{$\Delta_{782}^\circ$} & \multirow{9}{*}{$\mu = 20^6$}
& 0 & 32844379 & & 1740840 & & 20654 & \\*
& & 1 & 20644971 & & 1134088 & & 8228 & \\*
& & 2 & 5953958 & 91168 & 302751 & 4189 & 563 & 16 \\*
& & 3 & 1015304 & 64607 & 36518 & 2664 & & 22 \\*
& & 4 & 105016 & 18859 & 1431 & 489 & & \\*
& & 5 & 5831 & 2976 & & 30 & & \\*
& & 6 & 120 & 303 & & & & \\*
& & 7 & & 25 & & & & \\*
& & $\Sigma$ & 60569579 & 177938 & 3215628 & 7372 & 29445 & 38 \\
\midrule
\multirow{3}{*}{$\Delta_{377}^\circ$} & \multirow{9}{*}{$\mu = 20^6$}
& 0 & 30846440 & & 1997602 & & 34270 & \\*
& & 1 & 20949495 & & 1457977 & & 17032 & \\*
& & 2 & 6576718 & 85540 & 450756 & 4812 & 2094 & 24 \\*
\multirow{3}{*}{$\Delta_{499}^\circ$} & & 3 & 1265969 & 50759 & 69181 & 3007 & 50 & 34 \\*
& & 4 & 153456 & 15220 & 4308 & 576 & & \\*
& & 5 & 10846 & 2811 & 48 & 63 & & \\*
\multirow{3}{*}{$\Delta_{503}^\circ$} & & 6 & 616 & 240 & & & & \\*
& & 7 & 16 & 40 & & & & \\*
& & $\Sigma$ & 59803556 & 154610 & 3979872 & 8458 & 53446 & 58 \\
\midrule
\multirow{9}{*}{$\Delta_{1348}^\circ$} & \multirow{9}{*}{$\mu = 20^6$}
& 0 & 30845702 & & 2000040 & & 33043 & & 54 & \\*
& & 1 & 20949103 & & 1455999 & & 17509 & & & \\*
& & 2 & 6682151 & & 437511 & & 2633 & & & \\*
& & 3 & 1304785 & 14941 & 67651 & 478 & 140 & & & \\*
& & 4 & 160502 & 8102 & 4788 & 93 & & & & \\*
& & 5 & 12234 & 1508 & 96 & 3 & & & & \\*
& & 6 & 646 & 223 & & & & & & \\*
& & 7 & 28 & 37 & & & & & & \\*
& & $\Sigma$ & 59955151 & 24811 & 3966085 & 574 & 53325 & & 54 & \\
\midrule
\multirow{4}{*}{$\Delta_{882}^\circ$} & \multirow{9}{*}{$\mu = 20^6$}
& 0 & 30840098 & & 2005760 & & 31362 & & 308 & \\*
& & 1 & 20954897 & & 1454627 & & 18060 & & & \\*
& & 2 & 6582752 & 84392 & 439884 & 5174 & 3280 & 26 & & \\*
& & 3 & 1259073 & 62243 & 66781 & 3463 & 50 & 10 & & \\*
\multirow{5}{*}{\textcolor{red}{$\Delta_{856}^\circ$}} & & 4 & 150706 & 17636 & 4170 & 768 & & & & \\*
& & 5 & 10700 & 2707 & 48 & 57 & & & & \\*
& & 6 & 544 & 372 & & & & & & \\*
& & 7 & 16 & 36 & & & & & & \\*
& & $\Sigma$ & 59798786 & 167386 & 3971270 & 9462 & 52752 & 36 & 308 & \\
\midrule
\multirow{9}{*}{$\Delta_{1340}^\circ$} & \multirow{9}{*}{$\mu = 20^6$}
& 0 & 28954543 & & 2237908 & & 49877 & & 216 & \\*
& & 1 & 21074384 & & 1820659 & & 33865 & & 46 & \\*
& & 2 & 7243577 & & 642014 & & 6961 & & & \\*
& & 3 & 1554653 & 4776 & 121250 & 186 & 625 & & & \\*
& & 4 & 217875 & 3851 & 10569 & 105 & & & & \\*
& & 5 & 19271 & 1008 & 321 & 6 & & & & \\*
& & 6 & 1232 & 104 & & & & & & \\*
& & 7 & 88 & 30 & & & & & & \\*
& & $\Sigma$ & 59065623 & 9769 & 4832721 & 297 & 91328 & & 262 & \\
\midrule
\multirow{9}{*}{$\Delta_{1879}^\circ$} & \multirow{9}{*}{$\mu = 20^6$}
& 0 & 28950852 & & 2242073 & & 48482 & & 370 & \\*
& & 1 & 21078338 & & 1819768 & & 33768 & & 92 & \\*
& & 2 & 7243403 & & 636926 & & 7878 & & & \\*
& & 3 & 1554736 & 9494 & 118700 & 396 & 754 & & & \\*
& & 4 & 215750 & 5782 & 10320 & 243 & & & & \\*
& & 5 & 18880 & 1142 & 312 & 24 & & & & \\*
& & 6 & 1266 & 143 & & & & & & \\*
& & 7 & 84 & 24 & & & & & & \\*
& & $\Sigma$ & 59063309 & 16585 & 4828099 & 663 & 90882 & & 462 & \\
\midrule
\multirow{9}{*}{$\Delta_{1384}^\circ$} & \multirow{9}{*}{$\mu = 20^6$}
& 0 & 27178020 & & 2439920 & & 68270 & & 540 & \\*
& & 1 & 21092430 & & 2184225 & & 56190 & & 230 & \\*
& & 2 & 7757630 & & 870520 & & 15010 & & & \\*
& & 3 & 1806325 & & 193330 & & 2170 & & & \\*
& & 4 & 280930 & 1330 & 21060 & 30 & & & & \\*
& & 5 & 27970 & 610 & 930 & 15 & & & & \\*
& & 6 & 2060 & 40 & & & & & & \\*
& & 7 & 200 & 15 & & & & & & \\*
& & $\Sigma$ & 58145565 & 1995 & 5709985 & 45 & 141640 & & 770 & \\
\bottomrule
\end{longtable}
\end{footnotesize}

\section{Stationary circuits in \texorpdfstring{$\mathcal{O}(10^{11})$}{10 to 11} F-theory QSMs} \label{sec:StationaryCircuits}

In this section, we count the global sections on the stationary circuits encountered when enumerating the limit roots for the nodal quark-doublet curve in the family of QSM base spaces $B_3( \Delta_4^\circ )$. For this we always encounter nodal curves whose reducible components are smooth rational curves. In the following, we display their dual graphs. The line bundle in question is listed implicitly, by stating the number of sections on each irreducible component. The goal is to argue that for all these configuration, the number of global sections is exactly three.

\paragraph{Three non-resolved nodes}

Up to symmetry, we are looking at the following setup:
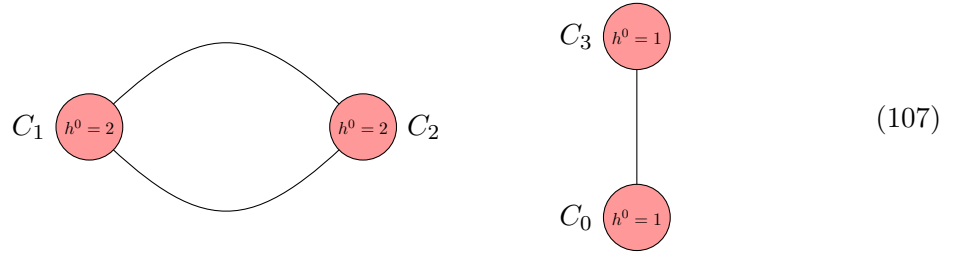
\begin{equation}
\begin{tikzpicture}[scale=0.6, baseline=(current  bounding  box.center)]
      
      \def\s{3.0};
      \def\h{2};
      
      \path[-,out = 45, in = 135, looseness = 1.5] (0,0) edge (2*\s,0);
      \path[-,out = -45, in = -135, looseness = 1.5] (0,0) edge (2*\s,0);
      \path[-] (4*\s,\h) edge (4*\s,-\h);
      
      \node at (0,0) [stuff_fill_red, scale=0.6, label=left:$C_1$]{$h^0 = 2$};
      \node at (2*\s,0) [stuff_fill_red, scale=0.6, label=right:$C_2$]{$h^0 = 2$};
      \node at (4*\s,\h) [stuff_fill_red, scale=0.6, label=left:$C_3$]{$h^0 = 1$};
      \node at (4*\s,-\h) [stuff_fill_red, scale=0.6, label=left:$C_0$]{$h^0 = 1$};
      
\end{tikzpicture}
\end{equation}
As explained in \oref{subsec:StationaryCircuits}, there are exactly two global sections on the circuit $C_1 \cup C_2$. On the tree-like curve $C_3 \cup C_4$, there exists exactly one global section by the technology described in \oref{sec:LineBundleCohomologyOnTreelikeCurves}. Hence, we conclude $h^0( C^\bullet, P^\bullet) = 3$.

\paragraph{Four non-resolved nodes}

Up to symmetry, there are two configurations. The first one is:
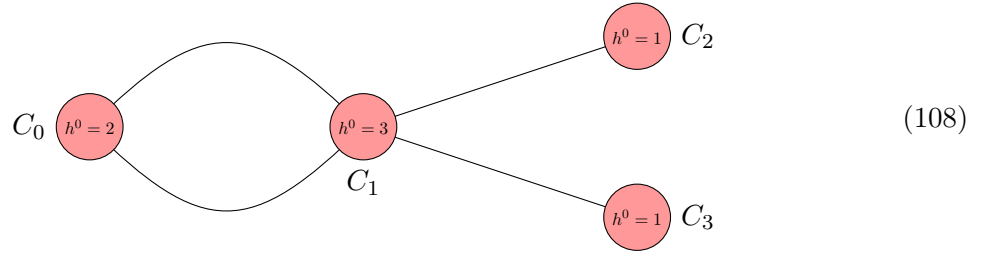
\begin{equation}
\begin{tikzpicture}[scale=0.6, baseline=(current  bounding  box.center)]
      
      \def\s{3.0};
      \def\h{2};
      
      \path[-,out = 45, in = 135, looseness = 1.5] (0,0) edge (2*\s,0);
      \path[-,out = -45, in = -135, looseness = 1.5] (0,0) edge (2*\s,0);
      \path[-] (2*\s,0) edge (4*\s,\h);
      \path[-] (2*\s,0) edge (4*\s,-\h);
      
      \node at (0,0) [stuff_fill_red, scale=0.6, label=left:$C_0$]{$h^0 = 2$};
      \node at (2*\s,0) [stuff_fill_red, scale=0.6, label=below:$C_1$]{$h^0 = 3$};
      \node at (4*\s,\h) [stuff_fill_red, scale=0.6, label=right:$C_2$]{$h^0 = 1$};
      \node at (4*\s,-\h) [stuff_fill_red, scale=0.6, label=right:$C_3$]{$h^0 = 1$};
      
\end{tikzpicture}
\end{equation}
To determine the number of global sections, we recall that a line bundle on a nodal curve whose irreducible components are all isomorphic to $\mathbb{P}^1$ is uniquely specified by two pieces of information \cite{harris2006moduli}:
\begin{itemize}
 \item The degree of the line bundle on each irreducible component.
 \item The descent data, which corresponds to a number $\lambda \in \mathbb{C}^\ast$ for each node. In particular, the values of the global sections agree at each node up to this factor $\lambda$.
\end{itemize}
Hence, the information displayed above does not specify the descent data. Rather, this diagram encodes an infinite family of line bundles. Our claim is that every line bundle in this family has exactly three global sections.

To see this, we first notice that the (constant) sections on $C_2$ and $C_3$ are uniquely fixed by the gluing condition at the nodes $C_1 \cap C_2$ and $C_1 \cap C_3$, respectively. For this reason, we can focus on the following simpler configuration:
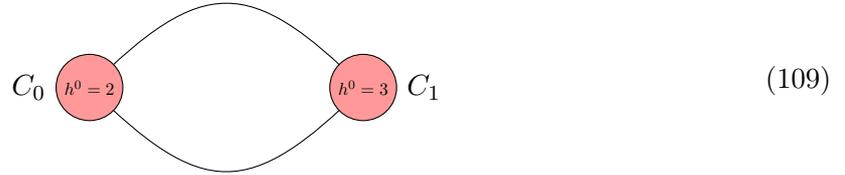
\begin{equation}
\begin{tikzpicture}[scale=0.6, baseline=(current  bounding  box.center)]
      
      \def\s{3.0};
      \def\h{2};
      
      \path[-,out = 45, in = 135, looseness = 1.5] (0,0) edge (2*\s,0);
      \path[-,out = -45, in = -135, looseness = 1.5] (0,0) edge (2*\s,0);
      
      \node at (0,0) [stuff_fill_red, scale=0.6, label=left:$C_0$]{$h^0 = 2$};
      \node at (2*\s,0) [stuff_fill_red, scale=0.6, label=right:$C_1$]{$h^0 = 3$};
      
\end{tikzpicture}
\end{equation}
We pick local coordinates and parametrize the local sections by $\left( \alpha_1, \dots, \alpha_5 \right) \in \mathbb{C}^5$:
\begin{align}
\begin{tabular}{ccc}
\toprule
Curve & Coordinates & Sections \\
\midrule
$C_0$ & $[a:b]$ & $\alpha_1 a + \alpha_2 b$ \\
$C_1$ & $[c:d]$ & $\alpha_3 c^2 + \alpha_4 c d + \alpha_5 d^2$ \\
\bottomrule
\end{tabular}
\end{align}
By use of a M{\"o}bius transformation, the nodes are located at the following positions:
\begin{align}
\begin{tabular}{c|cc|c}
\toprule
Label & $C_0$ & $C_1$ & Descent data \\
\midrule
$n_1$ & $[a:b] = [1:0]$ & $[c:d] = [1:0]$ & $1$ \\
$n_2$ & $[a:b] = [0:1]$ & $[c:d] = [0:1]$ & $\lambda$ \\
\bottomrule
\end{tabular}
\end{align}
Upon rescaling of the local sections, we can assume that the gluing conditions are
\begin{align}
\left( \alpha_1 a + \alpha_2 b \right)\left( [1:0] \right) &= \left( \alpha_3 c^2 + \alpha_4 c d + \alpha_5 d^2 \right)\left( [1:0] \right) \, , \\
\left( \alpha_1 a + \alpha_2 b \right)\left( [0:1] \right) &= \lambda \cdot \left( \alpha_3 c^2 + \alpha_4 c d + \alpha_5 d^2 \right)\left( [0:1] \right) \, ,
\end{align}
where $\lambda \in \mathbb{C}^\ast$. This is equivalent to
\begin{align}
\alpha_1 = \alpha_3 \, , \qquad \alpha_2 = \lambda \alpha_5 \, .
\end{align}
Hence, the global sections are paramerized by $(\alpha_3, \alpha_4, \alpha_5) \in \mathbb{C}^3$:
\begin{align}
\begin{tabular}{ccc}
\toprule
Curve & Coordinates & Sections \\
\midrule
$C_0$ & $[a:b]$ & $\alpha_3 a + \lambda \alpha_5 b$ \\
$C_1$ & $[c:d]$ & $\alpha_3 c^2 + \alpha_4 cd + \alpha_5 d^2$ \\
\bottomrule
\end{tabular}
\end{align}
In particular, $h^0( C^\bullet, P^\bullet) = 3$. We now turn to the second configuration:
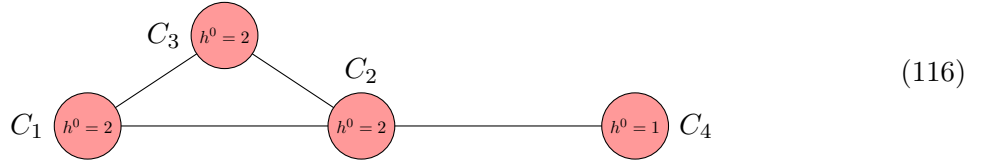
\begin{equation}
\begin{tikzpicture}[scale=0.6, baseline=(current  bounding  box.center)]
      
      \def\s{3.0};
      \def\h{2};

      \path[-] (0,0) edge (2*\s,0);
      \path[-] (2*\s,0) edge (\s,\h);
      \path[-] (\s,\h) edge (0,0);
      \path[-] (2*\s,0) edge (4*\s,0);
      
      \node at (0,0) [stuff_fill_red, scale=0.6, label=left:$C_1$]{$h^0 = 2$};
      \node at (2*\s,0) [stuff_fill_red, scale=0.6, label=above:$C_2$]{$h^0 = 2$};
      \node at (\s,\h) [stuff_fill_red, scale=0.6, label=left:$C_3$]{$h^0 = 2$};
      \node at (4*\s,0) [stuff_fill_red, scale=0.6, label=right:$C_4$]{$h^0 = 1$};
      
\end{tikzpicture}
\end{equation}
Just as above, we can ignore $C_4$ and instead focus on the following simpler configuration:
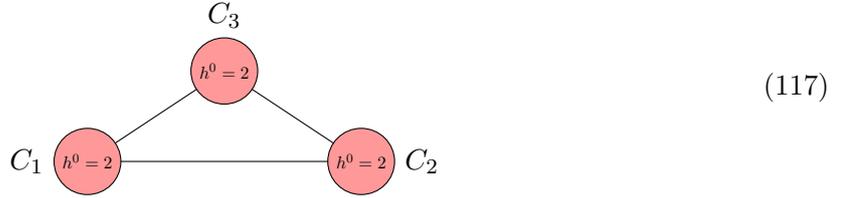
\begin{equation}
\begin{tikzpicture}[scale=0.6, baseline=(current  bounding  box.center)]
      
      \def\s{3.0};
      \def\h{2};

      \path[-] (0,0) edge (2*\s,0);
      \path[-] (2*\s,0) edge (\s,\h);
      \path[-] (\s,\h) edge (0,0);
      
      \node at (0,0) [stuff_fill_red, scale=0.6, label=left:$C_1$]{$h^0 = 2$};
      \node at (2*\s,0) [stuff_fill_red, scale=0.6, label=right:$C_2$]{$h^0 = 2$};
      \node at (\s,\h) [stuff_fill_red, scale=0.6, label=above:$C_3$]{$h^0 = 2$};
      
\end{tikzpicture}
\end{equation}
We pick homogeneous coordinates and parametrize the local sections by $\left( \alpha_1, \dots, \alpha_6 \right) \in \mathbb{C}^6$:
\begin{align}
\begin{tabular}{ccc}
\toprule
Curve & Coordinates & Sections \\
\midrule
$C_1$ & $[a:b]$ & $\alpha_1 a + \alpha_2 b$ \\
$C_2$ & $[c:d]$ & $\alpha_3 c + \alpha_4 d$ \\
$C_3$ & $[e:f]$ & $\alpha_5 e + \alpha_6 f$ \\
\bottomrule
\end{tabular}
\end{align}
The three nodes are placed, by use of a M\"obius transformation, at the following positions:
\begin{align}
\begin{tabular}{c|ccc|c}
\toprule
Label & $C_1$ & $C_2$ & $C_3$ & Descent data \\
\midrule
$n_1$ & $[a:b] = [1:0]$ & $[c:d] = [1:0]$ & & $1$ \\
$n_2$ & $[a:b] = [0:1]$ & & $[e:f] = [1:0]$ & $1$ \\
$n_3$ & & $[c:d] = [0:1]$ & $[e:f] = [0:1]$ & $\lambda \in \mathbb{C}^\ast$ \\
\bottomrule
\end{tabular}
\end{align}
Note that upon a suitable rescaling of the sections on $C_2$ and $C_3$, we can ensure gluing conditions as listed above. Hence, the three gluing conditions are:
\begin{align}
\alpha_1 = \alpha_3 \, \qquad \alpha_2 = \alpha_5 \, , \qquad \alpha_4 = \lambda \alpha_6 \, .
\end{align}
Thus, the global sections are always parametrized by $(\alpha_3, \alpha_5, \alpha_6) \in \mathbb{C}^3$ via:
\begin{align}
\begin{tabular}{ccc}
\toprule
Curve & Coordinates & Sections \\
\midrule
$C_1$ & $[a:b]$ & $\alpha_3 a + \alpha_5 b$ \\
$C_2$ & $[c:d]$ & $\alpha_3 c + \lambda \alpha_6 d$ \\
$C_3$ & $[e:f]$ & $\alpha_5 e + \alpha_6 f$ \\
\bottomrule
\end{tabular}
\end{align}
In particular, we find $h^0( C^\bullet, P^\bullet) = 3$.

\paragraph{Five non-resolved nodes}

Up to symmetry, there is only one configuration:
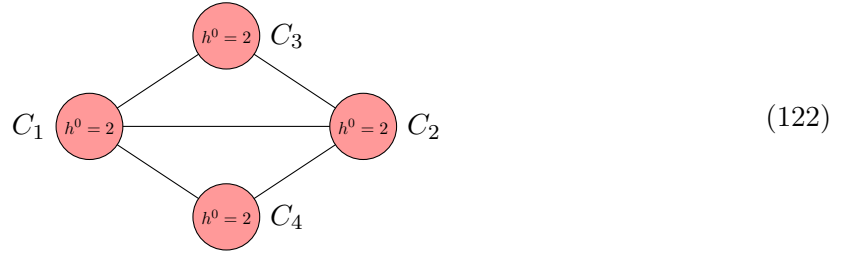
\begin{equation}
\begin{tikzpicture}[scale=0.6, baseline=(current  bounding  box.center)]
      
      \def\s{3.0};
      \def\h{2};

      \path[-] (0,0) edge (2*\s,0);
      \path[-] (2*\s,0) edge (\s,\h);
      \path[-] (\s,\h) edge (0,0);
      \path[-] (2*\s,0) edge (\s,-\h);
      \path[-] (0,0) edge (\s,-\h);
      
      \node at (0,0) [stuff_fill_red, scale=0.6, label=left:$C_1$]{$h^0 = 2$};
      \node at (2*\s,0) [stuff_fill_red, scale=0.6, label=right:$C_2$]{$h^0 = 2$};
      \node at (\s,\h) [stuff_fill_red, scale=0.6, label=right:$C_3$]{$h^0 = 2$};
      \node at (\s,-\h) [stuff_fill_red, scale=0.6, label=right:$C_4$]{$h^0 = 2$};
      
\end{tikzpicture}
\end{equation}
From the previous case study, we know that the triangle $C_1 - C_2 - C_3$ has exactly three global sections. The values of these sections at $C_1 \cap C_4$ and $C_2 \cap C_4$ uniquely fix the sections on $C_4$ (cf. lemma 7 in \cite{Bies:2021nje}) and $h^0( C^\bullet, P^\bullet) = 3$.

\section{Jumping circuit in \texorpdfstring{$\mathcal{O}(10^{11})$}{10 to 11} F-theory QSMs} \label{sec:TriaIndependence}

In this section we study the location of the nodes of the nodal quark-doublet curve $C^\bullet_{\left( \mathbf{3}, \mathbf{2} \right)_{1/6}}$ which was originally introduced in \cite{Bies:2021nje}. Specifically, we will argue that, in terms of suitable local homogeneous coordinates, the location of these nodes is the same for all QSM geometries associated to the 3-dimensional, reflexive polytope $\Delta_4^\circ$ in the Kreuzer-Skarke database \cite{Kreuzer:1998vb}. Explicitly, this polytope is given by
\begin{align}
\Delta_4^\circ = \mathrm{Conv} \left\{ \left[ \begin{array}{c} -1 \\ -1 \\ -1 \end{array} \right], \left[ \begin{array}{c} 2 \\ -1 \\ -1 \end{array} \right], \left[ \begin{array}{c} -1 \\ 2 \\ -1 \end{array} \right], \left[ \begin{array}{c} -1 \\ -1 \\ 5 \end{array} \right] \right\} \subseteq \mathbb{R}^3 \, .
\end{align}
More background information on the constructions can be found e.g. in \cite{trias, Halverson:2016tve, Cvetic:2019gnh, Bies:2021xfh}. This leads to a family $B_3( \Delta_4^\circ )$ of smooth, projective toric varieties.  It can be estimated that $B_3( \Delta_4^\circ )$ consists of $3 \times 10^{11}$ toric spaces \cite{Halverson:2016tve}.

Recall that the nodal quark-doublet curve $C^\bullet_{\left( \mathbf{3}, \mathbf{2} \right)_{1/6}}$ is given by \cite{Bies:2021nje}:
\begin{align}
C^\bullet_{\left( \mathbf{3}, \mathbf{2} \right)_{1/6}} = V \left( \prod_{i = 0}^{28}{x_i}, s_9 \right) \subseteq B_3 \, .
\end{align}
In this expression $s_9 \in H^0( X_\Sigma, \overline{K}_{X_\Sigma} ) \cong \mathbb{C}^6$ and will argue momentarily that
\begin{align}
\begin{split}
s_9 & = \alpha_1 \cdot x_{0}^6 x_{4}^5 x_{5}^4 x_{6}^3 x_{7}^2 x_{8} x_{9}^4 x_{10}^3 x_{11}^2 x_{12} x_{14}^2 x_{15} x_{17}^4 x_{18}^3 x_{19}^2 x_{20} x_{22}^2 x_{25}^2 x_{26} \\
   &\qquad + \alpha_2 \cdot x_{0}^3 x_{3}^3 x_{4}^3 x_{5}^3 x_{6}^3 x_{7}^3 x_{8}^3 x_{9}^2 x_{10}^2 x_{11}^2 x_{12}^2 x_{13}^2 x_{14} x_{15} x_{16} x_{17}^2 x_{18}^2 x_{19}^2 x_{20}^2 x_{21}^2 x_{22} x_{23} x_{25} x_{26} x_{27} \\
   &\qquad + \alpha_3 \cdot x_{0} x_{1} x_{2} x_{3} x_{4} x_{5} x_{6} x_{7} x_{8} x_{9} x_{10} x_{11} x_{12} x_{13} x_{14} x_{15} x_{16} x_{17} x_{18} \\
   & \hspace{20em} \times x_{19} x_{20} x_{21} x_{22} x_{23} x_{24} x_{25} x_{26} x_{27} x_{28} \\
   &\qquad + \alpha_4 \cdot x_{3}^6 x_{4} x_{5}^2 x_{6}^3 x_{7}^4 x_{8}^5 x_{10} x_{11}^2 x_{12}^3 x_{13}^4 x_{15} x_{16}^2 x_{18} x_{19}^2 x_{20}^3 x_{21}^4 x_{23}^2 x_{26} x_{27}^2 \\
   &\qquad + \alpha_5 \cdot x_{2}^3 x_{9} x_{10} x_{11} x_{12} x_{13} x_{14}^2 x_{15}^2 x_{16}^2 x_{22} x_{23} x_{24}^2 x_{28}\\
   &\qquad + \alpha_6 \cdot x_{1}^3 x_{17} x_{18} x_{19} x_{20} x_{21} x_{22} x_{23} x_{24} x_{25}^2 x_{26}^2 x_{27}^2 x_{28}^2 \, , \label{equ:Section}
\end{split}
\end{align}
where $x_i$ are the homogeneous coordinates of $X_\Sigma \in B_3( \Delta_4^\circ )$. The parameters $\alpha_i \in \mathbb{C}$ provide a (possibly redundant) parametrization of the complex structure moduli of 
$C^\bullet_{\left( \mathbf{3}, \mathbf{2} \right)_{1/6}}$. As explained in \oref{subsec:SimplificationForPartialBlowups}, to count limit roots we can always pass to a simplified nodal curve by removing circuits. The dual graph of this simplified curve is for all spaces $X_\Sigma \in B_3( \Delta_4^\circ )$ given by:
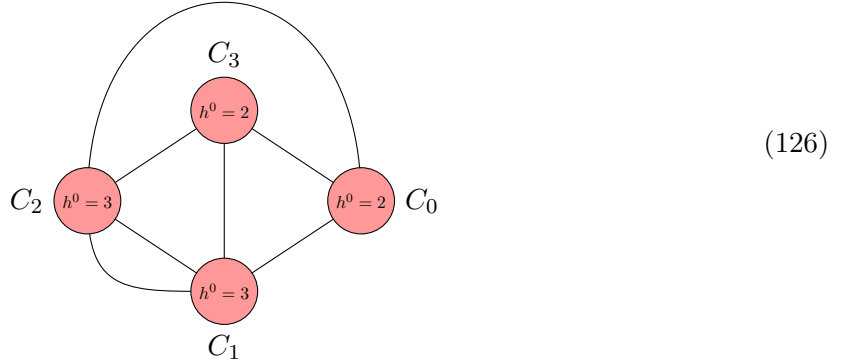
\begin{equation}
\begin{tikzpicture}[scale=0.6, baseline=(current  bounding  box.center)]
      
      \def\s{3.0};
      \def\h{2.0};
      
      \path[-] (-\s,0) edge (0,\h);
      \path[-] (-\s,0) edge (0,-\h);
      \path[-, out = -90, in = 180, looseness = 1.5] (-\s,0) edge (0,-\h);
      \path[-] (\s,0) edge (0,\h);
      \path[-] (\s,0) edge (0,-\h);
      \path[-] (0,\h) edge (0,-\h);
      \path[-, out = 90, in = 90, looseness = 2.5] (-\s,0) edge (\s,0);
      
      \node at (1*\s,0) [stuff_fill_red, scale=0.6, label=right:$C_0$]{$h^0 = 2$};
      \node at (0,-\h) [stuff_fill_red, scale=0.6, label=below:$C_1$]{$h^0 = 3$};
      \node at (-1*\s,0) [stuff_fill_red, scale=0.6, label=left:$C_2$]{$h^0 = 3$};
      \node at (0,\h) [stuff_fill_red, scale=0.6, label=above:$C_3$]{$h^0 = 2$};
      
\end{tikzpicture} \label{equ:JumpingCircuit}
\end{equation}
The irreducible components $C_i$ displayed in \oref{equ:JumpingCircuit} are given by $C_i = V(x_i, s_9)$. We conduct a detailed study of this curve under the simplifying assumption
\begin{align}
\alpha_4 \neq 0 \, , \qquad \alpha_2^2 \neq 4 \alpha_1 \alpha_4 \, ,
\end{align}
We will identify homogeneous coordinates $[u_i \colon v_i]$ of $C_i$ and show that the location of the nodes of the circuit \oref{equ:JumpingCircuit} are independent of the FRSTs of $\Delta_4^\circ$. More explicitly, we will argue that for all $X_\Sigma \in B_3( \Delta_4^\circ )$, the location of these nodes is then given as follows:
\begin{align}
\begin{tabular}{cc|cc}
\toprule
Label & Superset & Coordinates in first curve & Coordinates in second curve \\
\midrule
$n_1$ & $C_0 \cap C_1$ & $[ u_0 \colon v_0 ] = [\alpha_5:-\alpha_4]$ & $[u_1 \colon v_1] = [0:1]$\\
$n_2$ & $C_0 \cap C_2$ & $[ u_0 \colon v_0 ] = [1:0]$ & $[u_2 \colon v_2] = [0:1]$\\
$n_3$ & $C_0 \cap C_3$ & $[ u_0 \colon v_0 ] = [0:1]$ & $[u_3 \colon v_3] = [0:1]$\\
\midrule
$n_4$ & $C_1 \cap C_2$ & $\left[ 1 : \frac{- \alpha_2 -  \sqrt{\alpha_2^2 - 4 \alpha_1 \alpha_4}}{2 \alpha_4} \right]$ & $\left[ 1 : \frac{- \alpha_2 -  \sqrt{\alpha_2^2 - 4 \alpha_1 \alpha_4}}{2 \alpha_4} \right]$ \\
$n_5$ & $C_1 \cap C_2$ & $\left[ 1 : \frac{\alpha_2 - \sqrt{\alpha_2^2 - 4 \alpha_1 \alpha_4}}{2 \alpha_4} \right]$ & $\left[ 1 : \frac{\alpha_2 - \sqrt{\alpha_2^2 - 4 \alpha_1 \alpha_4}}{2 \alpha_4} \right]$ \\
$n_6$ & $C_1 \cap C_3$ & $[u_1 \colon v_1] = [1:0]$ & $[u_3 \colon v_3] = [\alpha_5:- \alpha_1]$\\
\midrule
$n_7$ & $C_2 \cap C_3$ & $[u_2 \colon v_2] = [1:0]$ & $[u_3 \colon v_3] = [1:0]$\\
\bottomrule
\end{tabular}
\end{align}

\subsection{Ray generators} \label{sec:Rays}

For every $X_\Sigma \in B_3( \Delta_4^\circ )$, the ray generators/toric divisors are one-to-one to the non-trivial lattice points of the polytope
\begin{align}
\Delta_4^\circ = \mathrm{Conv} \left\{ \left[ \begin{array}{c} -1 \\ -1 \\ -1 \end{array} \right], \left[ \begin{array}{c} 2 \\ -1 \\ -1 \end{array} \right], \left[ \begin{array}{c} -1 \\ 2 \\ -1 \end{array} \right], \left[ \begin{array}{c} -1 \\ -1 \\ 5 \end{array} \right] \right\} \, .
\end{align}
This polytope has exactly 29 non-trivial lattice points and we make the following assignment of homogeneous variables:
\begin{align}
\begin{tabular}{ccc|ccc}
\toprule
Ray & Lattice points & Variable & Ray & Lattice points & Variable \\
\midrule
$r_0$ & $[-1, -1, -1]$ & $x_0$ & $r_{15}$ & $[-1, 1, 0]$ & $x_{15}$ \\
$r_1$ & $[2, -1, -1]$ & $x_1$ & $r_{16}$ & $[-1, 1, 1]$ & $x_{16}$ \\
$r_2$ & $[-1, 2, -1]$ & $x_2$ & $r_{17}$ & $[0, -1, -1]$ & $x_{17}$ \\
$r_3$ & $[-1, -1, 5]$ & $x_3$ & $r_{18}$ & $[0, -1, 0]$ & $x_{18}$ \\
$r_4$ & $[-1, -1, 0]$ & $x_4$ & $r_{19}$ & $[0, -1, 1]$ & $x_{19}$ \\
\midrule
$r_5$ & $[-1, -1, 1]$ & $x_5$ & $r_{20}$ & $[0, -1, 2]$ & $x_{20}$ \\
$r_6$ & $[-1, -1, 2]$ & $x_6$ & $r_{21}$ & $[0, -1, 3]$ & $x_{21}$ \\
$r_7$ & $[-1, -1, 3]$ & $x_7$ & $r_{22}$ & $[0, 0, -1]$ & $x_{22}$ \\
$r_8$ & $[-1, -1, 4]$ & $x_8$ & $r_{23}$ & $[0, 0, 1]$ & $x_{23}$ \\
$r_9$ & $[-1, 0, -1]$ & $x_9$ & $r_{24}$ & $[0, 1, -1]$ & $x_{24}$ \\
\midrule
$r_{10}$ & $[-1, 0, 0]$ & $x_{10}$ & $r_{25}$ & $[1, -1, -1]$ & $x_{25}$ \\
$r_{11}$ & $[-1, 0, 1]$ & $x_{11}$ & $r_{26}$ & $[1, -1, 0]$ & $x_{26}$ \\
$r_{12}$ & $[-1, 0, 2]$ & $x_{12}$ & $r_{27}$ & $[1, -1, 1]$ & $x_{27}$ \\
$r_{13}$ & $[-1, 0, 3]$ & $x_{13}$ & $r_{28}$ & $[1, 0, -1]$ & $x_{28}$ \\
$r_{14}$ & $[-1, 1, -1]$ & $x_{14}$ \\
\bottomrule
\end{tabular} \label{equ:RayGenerators}
\end{align}
These rays are same for every FRST of $\Delta_4^\circ$. Furthermore, since we construct projective toric varieties $X_\Sigma$, they have no torus-factor. Hence, we have the following short-exact sequence:
\begin{align}
0 \to M \to \mathrm{Div}_T( X_\Sigma ) \to \mathrm{Cl}( X_\Sigma ) \, .
\end{align}
The first map in this sequence is given by $A \in \mathbb{M}( 29 \times 3, \mathbb{Z} )$ with rows given by the rays in \oref{equ:RayGenerators}. Hence, $\mathrm{Div}_T( X_\Sigma ) \to \mathrm{Cl}( X_\Sigma )$ is the cokernel of $A$. The latter defines the grading of the Cox ring. Hence, for all FRSTs of $\Delta_4^\circ$, the grading of the Cox ring is the same. Specifically, we can take the Cox ring $S = \mathbb{C}[ x_0, \dots, x_{28} ]$ to be $\mathbb{Z}^{26}$-graded via
\begin{align}
\begin{adjustbox}{max width=0.9\textwidth}
\begin{tabular}{ccccc|ccccc|ccccc|ccccc|ccccc|ccccc}
\toprule
$x_{0}$ & $x_{1}$ & $x_{2}$ & $x_{3}$ & $x_{4}$ & $x_{5}$ & $x_{6}$ & $x_{7}$ & $x_{8}$ & $x_{9}$ & $x_{10}$ & $x_{11}$ & $x_{12}$ & $x_{13}$ & $x_{14}$ & $x_{15}$ & $x_{16}$ & $x_{17}$ & $x_{18}$ & $x_{19}$ & $x_{20}$ & $x_{21}$ & $x_{22}$ & $x_{23}$ & $x_{24}$ & $x_{25}$ & $x_{26}$ & $x_{27}$ & $x_{28}$ \\
\midrule
26 & -21 & -14 & -6 & -27 & & & & & -21 & & & & & & & & & & & & & & & & & & & \\
3 & -3 & -2 & -1 & -3 & & & & & -3 & & & & & & & & & & & & & & & & & & & \\
7 & -6 & -4 & -2 & -8 & 1 & & & & -6 & & & & & & & & & & & & & & & & & & & \\
11 & -9 & -6 & -3 & -12 & & 1 & & & -9 & & & & & & & & & & & & & & & & & & & \\
15 & -12 & -8 & -4 & -16 & & & 1 & & -12 & & & & & & & & & & & & & & & & & & & \\
19 & -15 & -10 & -5 & -20 & & & & 1 & -15 & & & & & & & & & & & & & & & & & & & \\
8 & -7 & -5 & -2 & -9 & & & & & -6 & & & & & & & & & & & & & & & & & & & \\
12 & -10 & -7 & -3 & -13 & & & & & -10 & 1 & & & & & & & & & & & & & & & & & & \\
16 & -13 & -9 & -4 & -17 & & & & & -13 & & 1 & & & & & & & & & & & & & & & & & \\
20 & -16 & -11 & -5 & -21 & & & & & -16 & & & 1 & & & & & & & & & & & & & & & & \\
24 & -19 & -13 & -6 & -25 & & & & & -19 & & & & 1 & & & & & & & & & & & & & & & \\
17 & -14 & -10 & -4 & -18 & & & & & -14 & & & & & 1 & & & & & & & & & & & & & & \\
21 & -17 & -12 & -5 & -22 & & & & & -17 & & & & & & 1 & & & & & & & & & & & & & \\
25 & -20 & -14 & -6 & -26 & & & & & -20 & & & & & & & 1 & & & & & & & & & & & & \\
-13 & 10 & 7 & 3 & 13 & & & & & 10 & & & & & & & & 1 & & & & & & & & & & & \\
-9 & 7 & 5 & 2 & 9 & & & & & 7 & & & & & & & & & 1 & & & & & & & & & & \\
-5 & 4 & 3 & 1 & 5 & & & & & 4 & & & & & & & & & & 1 & & & & & & & & & \\
-1 & 1 & 1 & & 1 & & & & & 1 & & & & & & & & & & & 1 & & & & & & & & \\
3 & -2 & -1 & -1 & -3 & & & & & -2 & & & & & & & & & & & & 1 & & & & & & & \\
-4 & 3 & 2 & 1 & 4 & & & & & 3 & & & & & & & & & & & & & 1 & & & & & & \\
4 & -3 & -2 & -1 & -4 & & & & & -3 & & & & & & & & & & & & & & 1 & & & & & \\
5 & -4 & -3 & -1 & -5 & & & & & -4 & & & & & & & & & & & & & & & 1 & & & & \\
-25 & 20 & 14 & 6 & 26 & & & & & 20 & & & & & & & & & & & & & & & & 1 & & & \\
-21 & 17 & 12 & 5 & 22 & & & & & 17 & & & & & & & & & & & & & & & & & 1 & & \\
-17 & 14 & 10 & 4 & 18 & & & & & 14 & & & & & & & & & & & & & & & & & & 1 & \\
-16 & 13 & 9 & 4 & 17 & & & & & 13 & & & & & & & & & & & & & & & & & & & 1 \\
\bottomrule
\end{tabular}
\end{adjustbox}
\end{align}
From this it follows that for every FRST of $\Delta_4^\circ$, we can express $s \in H^0( B_3, \overline{K}_{B_3} )$ as in \oref{equ:Section}. The coefficients $\alpha_i \in \mathbb{C}$ used in the expansion of $s$ provide a (possibly redundant) description of the complex structure moduli of the nodal curve in question. Therefore, these parameters will play a crucial role in the following analysis.

\subsection{Vanishing intersections and approximation of Stanley-Reisner ideal} \label{sec:SRs}

There are many FRSTs of the lattice points in $\Delta_4^\circ$. To approximate them, we follow \cite{Halverson:2016tve} and compute FRTs of the four facets of $\Delta_4^\circ$:
\begin{align}
\begin{split}
f_1 &= \mathrm{Conv} \left\{ r_0, r_1, r_2 \right\} \, , \qquad f_2 = \mathrm{Conv} \left\{ r_0, r_1, r_3 \right\} \, , \\
f_3 &= \mathrm{Conv} \left\{ r_0, r_2, r_2 \right\} \, , \qquad f_4 = \mathrm{Conv} \left\{ r_1, r_2, r_3 \right\} \, .
\end{split}
\end{align}
These have 10, 16, 16 and 10 lattice points respectively. Hence, for the facets the FRTs can be computed relatively quickly. It then follows that every FRST of $\Delta_4^\circ$ is the ``union'' of an FRT of each of its facets. Crucially though, such a ``union'' need not yield a regular triangulation. In this sense, the FRTs of the facets approximate the FRSTs of $\Delta_4^\circ$.

\subsubsection*{Triangulations of the facets}

The software \texttt{TOPCOM} \cite{Rambau:TOPCOM:2002} can enumerate the FRTs of $f_1$.\footnote{Recently, parallel enumerations of triangulations were investigated in \cite{JordanJowigKastner}. This can lead to significant performance improvements. A computer implementation is available at \href{https://polymake.org/doku.php/mptopcom}{https://polymake.org/doku.php/mptopcom}.} By adding the origin, we obtain fanes which define toric varieties. By computing the intersection of the Stanley-Reisner ideals of these tori varieties, we find the ideal $I_1$ of those elements common to all Stanley-Reisner ideals. All of these tasks can be achieved very efficiently with the computer algebra system \texttt{OSCAR} \cite{OSCAR, OSCAR-book}. Explicitly, we run the following code in \texttt{Julia} \cite{Julia-2017}:
\jlinputlisting{sr_intersect.jl}
This identifies 79 FRTs for $f_1$ and leads to
\begin{align}
\begin{split}
I_1 &= \langle x_{24} x_{1}, x_{17} x_{1}, x_{2} x_{1}, x_{0} x_{1}, x_{2} x_{28}, x_{9} x_{28}, x_{14} x_{25}, x_{0} x_{25}, x_{17} x_{24}, x_{9} x_{2}, x_{0} x_{2}, x_{0} x_{14},\\
& \qquad x_{9} x_{14} x_{1}, x_{0} x_{24} x_{28}, x_{14} x_{17} x_{28}, x_{9} x_{24} x_{25}, x_{2} x_{17} x_{25}, \\
& \qquad x_{22} x_{25} x_{28} x_{1}, x_{14} x_{22} x_{28} x_{1}, x_{9} x_{22} x_{25} x_{1}, x_{22} x_{24} x_{25} x_{28}, x_{17} x_{22} x_{25} x_{28}, x_{14} x_{22} x_{24} x_{28}, \\
& \qquad x_{0} x_{17} x_{22} x_{28}, x_{2} x_{22} x_{24} x_{25}, x_{9} x_{17} x_{22} x_{25}, x_{14} x_{2} x_{22} x_{24}, x_{9} x_{14} x_{22} x_{24}, x_{0} x_{9} x_{22} x_{24}, \\
& \qquad x_{14} x_{2} x_{17} x_{22}, x_{9} x_{14} x_{17} x_{22}, x_{0} x_{9} x_{17} x_{22} \rangle \, .
\end{split}
\end{align}
Similarly, we find 14295, 14295 and 79 FRTs for $f_2$, $f_3$, $f_4$. Hence, an upper bound for the FRSTs of $\Delta_4^\circ$ is $79^2 \times 14295^2 \sim 1.275 \times 10^{12}$ which matches the result in \cite{Halverson:2016tve}. By repeating the above ideal intersection code for $f_2$ we find

\begin{align}
\begin{split}
I_2 &= \langle x_{21} x_{1}, x_{19} x_{1}, x_{17} x_{1}, x_{3} x_{1}, x_{6} x_{1}, x_{0} x_{1}, x_{25} x_{27}, x_{3} x_{27}, x_{7} x_{27}, x_{5} x_{27}, x_{0} x_{27}, x_{8} x_{26},\\
& \qquad x_{6} x_{26}, x_{4} x_{26}, x_{3} x_{25}, x_{7} x_{25}, x_{5} x_{25}, x_{0} x_{25}, x_{19} x_{21}, x_{18} x_{21}, x_{17} x_{21}, x_{18} x_{20}, x_{17} x_{20},\\
& \qquad x_{17} x_{19}, x_{7} x_{3}, x_{6} x_{3}, x_{5} x_{3}, x_{4} x_{3}, x_{0} x_{3}, x_{6} x_{8}, x_{5} x_{8}, x_{4} x_{8}, x_{0} x_{8}, x_{5} x_{7}, x_{4} x_{7}, x_{0} x_{7},\\
& \qquad x_{4} x_{6}, x_{0} x_{6}, x_{0} x_{5}, x_{18} x_{27} x_{1}, x_{4} x_{27} x_{1}, x_{20} x_{25} x_{1}, x_{8} x_{25} x_{1}, x_{5} x_{20} x_{1}, x_{4} x_{20} x_{1},\\
& \qquad x_{8} x_{18} x_{1}, x_{7} x_{18} x_{1}, x_{7} x_{8} x_{1}, x_{4} x_{5} x_{1}, x_{6} x_{21} x_{27}, x_{4} x_{21} x_{27}, x_{4} x_{20} x_{27}, x_{8} x_{19} x_{27},\\
& \qquad x_{8} x_{18} x_{27}, x_{6} x_{18} x_{27}, x_{8} x_{17} x_{27}, x_{6} x_{17} x_{27}, x_{4} x_{17} x_{27}, x_{7} x_{21} x_{26}, x_{5} x_{21} x_{26}, x_{0} x_{21} x_{26},\\
& \qquad x_{5} x_{20} x_{26}, x_{0} x_{20} x_{26}, x_{3} x_{19} x_{26}, x_{0} x_{19} x_{26}, x_{3} x_{18} x_{26}, x_{7} x_{18} x_{26}, x_{3} x_{17} x_{26}, x_{7} x_{17} x_{26},\\
& \qquad x_{5} x_{17} x_{26}, x_{8} x_{21} x_{25}, x_{6} x_{21} x_{25}, x_{4} x_{21} x_{25}, x_{6} x_{20} x_{25}, x_{4} x_{20} x_{25}, x_{4} x_{19} x_{25}, x_{8} x_{18} x_{25},\\
& \qquad x_{8} x_{17} x_{25}, x_{6} x_{17} x_{25}, x_{20} x_{26} x_{27} x_{1}, x_{8} x_{20} x_{27} x_{1}, x_{18} x_{25} x_{26} x_{1}, x_{7} x_{20} x_{26} x_{1}, \\
& \qquad x_{5} x_{18} x_{26} x_{1}, x_{4} x_{18} x_{25} x_{1}, x_{20} x_{21} x_{26} x_{27}, x_{19} x_{20} x_{26} x_{27}, x_{18} x_{19} x_{26} x_{27}, x_{17} x_{18} x_{26} x_{27},\\
& \qquad x_{8} x_{20} x_{21} x_{27}, x_{6} x_{19} x_{20} x_{27}, x_{4} x_{18} x_{19} x_{27}, x_{20} x_{21} x_{25} x_{26}, x_{19} x_{20} x_{25} x_{26}, x_{18} x_{19} x_{25} x_{26},\\
& \qquad x_{17} x_{18} x_{25} x_{26}, x_{3} x_{20} x_{21} x_{26}, x_{7} x_{19} x_{20} x_{26}, x_{5} x_{18} x_{19} x_{26}, x_{0} x_{17} x_{18} x_{26}, x_{8} x_{19} x_{20} x_{25},\\
& \qquad x_{6} x_{18} x_{19} x_{25}, x_{4} x_{17} x_{18} x_{25}, x_{8} x_{3} x_{20} x_{21}, x_{7} x_{8} x_{20} x_{21}, x_{6} x_{7} x_{20} x_{21}, x_{5} x_{6} x_{20} x_{21},\\
& \qquad x_{4} x_{5} x_{20} x_{21}, x_{0} x_{4} x_{20} x_{21}, x_{8} x_{3} x_{19} x_{20}, x_{7} x_{8} x_{19} x_{20}, x_{6} x_{7} x_{19} x_{20}, x_{5} x_{6} x_{19} x_{20},\\
& \qquad x_{4} x_{5} x_{19} x_{20}, x_{0} x_{4} x_{19} x_{20}, x_{8} x_{3} x_{18} x_{19}, x_{7} x_{8} x_{18} x_{19}, x_{6} x_{7} x_{18} x_{19}, x_{5} x_{6} x_{18} x_{19},\\
& \qquad x_{4} x_{5} x_{18} x_{19}, x_{0} x_{4} x_{18} x_{19}, x_{8} x_{3} x_{17} x_{18}, x_{7} x_{8} x_{17} x_{18}, x_{6} x_{7} x_{17} x_{18}, x_{5} x_{6} x_{17} x_{18},\\
& \qquad x_{4} x_{5} x_{17} x_{18}, x_{0} x_{4} x_{17} x_{18} \rangle \, .
\end{split}
\end{align}
We would now like to conclude that $I_1 + I_2 \subseteq I_{\text{SR}}( X_\Sigma )$. There is a minor caveat to this. Suppose that $r_i, r_j \in f_1 \cap f_2$. If $x_i x_j \in I_1$, this means that there is never a triangle in the FRTs of $f_1$ with $r_i$ and $r_j$ as vertices. However, this is not necessarily the case in $f_2$, in which case $x_i x_j \notin I_2$. So, we have to look for such discrepancies and then correct for them. Note that $f_1 \cap f_2 = \left\{ r_{0}, r_{1}, r_{17}, r_{25} \right\} \leftrightarrow \left\{ x_{0}, x_{1}, x_{17}, x_{25} \right\}$. The relevant generators of $I_1$ are $x_{0} x_{1}$, $x_{0} x_{25}$, $x_1 x_{17}$. All are contained in $I_2$, so no adjustment is necessary.

By the same logic we find from $f_3$:
\begin{align}
\begin{split}
I_3 &= \langle x_{13} x_{2}, x_{11} x_{2}, x_{9} x_{2}, x_{3} x_{2}, x_{6} x_{2}, x_{0} x_{2}, x_{14} x_{16}, x_{3} x_{16}, x_{7} x_{16}, x_{5} x_{16}, x_{0} x_{16}, x_{8} x_{15},\\
& \qquad x_{6} x_{15}, x_{4} x_{15}, x_{3} x_{14}, x_{7} x_{14}, x_{5} x_{14}, x_{0} x_{14}, x_{11} x_{13}, x_{10} x_{13}, x_{9} x_{13}, x_{10} x_{12}, x_{9} x_{12},\\
& \qquad x_{9} x_{11}, x_{7} x_{3}, x_{6} x_{3}, x_{5} x_{3}, x_{4} x_{3}, x_{0} x_{3}, x_{6} x_{8}, x_{5} x_{8}, x_{4} x_{8}, x_{0} x_{8}, x_{5} x_{7}, x_{4} x_{7}, x_{0} x_{7},\\
& \qquad x_{4} x_{6}, x_{0} x_{6}, x_{0} x_{5}, x_{10} x_{16} x_{2}, x_{4} x_{16} x_{2}, x_{12} x_{14} x_{2}, x_{8} x_{14} x_{2}, x_{5} x_{12} x_{2}, x_{4} x_{12} x_{2},\\
& \qquad x_{8} x_{10} x_{2}, x_{7} x_{10} x_{2}, x_{7} x_{8} x_{2}, x_{4} x_{5} x_{2}, x_{6} x_{13} x_{16}, x_{4} x_{13} x_{16}, x_{4} x_{12} x_{16}, x_{8} x_{11} x_{16},\\
& \qquad x_{8} x_{10} x_{16}, x_{6} x_{10} x_{16}, x_{8} x_{9} x_{16}, x_{6} x_{9} x_{16}, x_{4} x_{9} x_{16}, x_{7} x_{13} x_{15}, x_{5} x_{13} x_{15}, x_{0} x_{13} x_{15},\\
& \qquad x_{5} x_{12} x_{15}, x_{0} x_{12} x_{15}, x_{3} x_{11} x_{15}, x_{0} x_{11} x_{15}, x_{3} x_{10} x_{15}, x_{7} x_{10} x_{15}, x_{3} x_{9} x_{15}, x_{7} x_{9} x_{15},\\
& \qquad x_{5} x_{9} x_{15}, x_{8} x_{13} x_{14}, x_{6} x_{13} x_{14}, x_{4} x_{13} x_{14}, x_{6} x_{12} x_{14}, x_{4} x_{12} x_{14}, x_{4} x_{11} x_{14}, x_{8} x_{10} x_{14},\\
& \qquad x_{8} x_{9} x_{14}, x_{6} x_{9} x_{14}, x_{12} x_{15} x_{16} x_{2}, x_{8} x_{12} x_{16} x_{2}, x_{10} x_{14} x_{15} x_{2}, x_{7} x_{12} x_{15} x_{2},\\
& \qquad x_{5} x_{10} x_{15} x_{2}, x_{4} x_{10} x_{14} x_{2}, x_{12} x_{13} x_{15} x_{16}, x_{11} x_{12} x_{15} x_{16}, x_{10} x_{11} x_{15} x_{16},\\
& \qquad x_{9} x_{10} x_{15} x_{16}, x_{8} x_{12} x_{13} x_{16}, x_{6} x_{11} x_{12} x_{16}, x_{4} x_{10} x_{11} x_{16}, x_{12} x_{13} x_{14} x_{15},\\
& \qquad x_{11} x_{12} x_{14} x_{15}, x_{10} x_{11} x_{14} x_{15}, x_{9} x_{10} x_{14} x_{15}, x_{3} x_{12} x_{13} x_{15}, x_{7} x_{11} x_{12} x_{15},\\
& \qquad x_{5} x_{10} x_{11} x_{15}, x_{0} x_{9} x_{10} x_{15}, x_{8} x_{11} x_{12} x_{14}, x_{6} x_{10} x_{11} x_{14}, x_{4} x_{9} x_{10} x_{14},\\
& \qquad x_{8} x_{3} x_{12} x_{13}, x_{7} x_{8} x_{12} x_{13}, x_{6} x_{7} x_{12} x_{13}, x_{5} x_{6} x_{12} x_{13}, x_{4} x_{5} x_{12} x_{13}, x_{0} x_{4} x_{12} x_{13},\\
& \qquad x_{8} x_{3} x_{11} x_{12}, x_{7} x_{8} x_{11} x_{12}, x_{6} x_{7} x_{11} x_{12}, x_{5} x_{6} x_{11} x_{12}, x_{4} x_{5} x_{11} x_{12}, x_{0} x_{4} x_{11} x_{12},\\
& \qquad x_{8} x_{3} x_{10} x_{11}, x_{7} x_{8} x_{10} x_{11}, x_{6} x_{7} x_{10} x_{11}, x_{5} x_{6} x_{10} x_{11}, x_{4} x_{5} x_{10} x_{11}, x_{0} x_{4} x_{10} x_{11},\\
& \qquad x_{8} x_{3} x_{9} x_{10}, x_{7} x_{8} x_{9} x_{10}, x_{6} x_{7} x_{9} x_{10}, x_{5} x_{6} x_{9} x_{10}, x_{4} x_{5} x_{9} x_{10}, x_{0} x_{4} x_{9} x_{10} \rangle \, .
\end{split}
\end{align}
Finally, $f_4$ leads to
\begin{align}
\begin{split}
I_4 &= \langle x_{24} x_{1}, x_{21} x_{1}, x_{2} x_{1}, x_{3} x_{1}, x_{2} x_{28}, x_{13} x_{28}, x_{16} x_{27}, x_{3} x_{27}, x_{21} x_{24}, x_{13} x_{2}, x_{3} x_{2}, x_{3} x_{16},\\
& \qquad x_{13} x_{16} x_{1}, x_{3} x_{24} x_{28}, x_{16} x_{21} x_{28}, x_{13} x_{24} x_{27}, x_{2} x_{21} x_{27}, x_{23} x_{27} x_{28} x_{1}, x_{16} x_{23} x_{28} x_{1},\\
& \qquad x_{13} x_{23} x_{27} x_{1}, x_{23} x_{24} x_{27} x_{28}, x_{21} x_{23} x_{27} x_{28}, x_{16} x_{23} x_{24} x_{28}, x_{3} x_{21} x_{23} x_{28}, x_{2} x_{23} x_{24} x_{27},\\
& \qquad x_{13} x_{21} x_{23} x_{27}, x_{16} x_{2} x_{23} x_{24}, x_{13} x_{16} x_{23} x_{24}, x_{3} x_{13} x_{23} x_{24}, x_{16} x_{2} x_{21} x_{23}, x_{13} x_{16} x_{21} x_{23},\\
& \qquad x_{3} x_{13} x_{21} x_{23} \rangle \, .
\end{split}
\end{align}
There are no discrepancies as mentioned above. Therefore, it follows that for every FRST of $\Delta_4^\circ$ it holds $I_1 + I_2 + I_3 + I_4 \subseteq I_{\text{SR}}( X_\Sigma )$.

\subsubsection*{Interplay of different facets}

We can improve this statement further. Namely, $r_i \in f_1$ and $r_j \in f_2$ cannot appear as vertices of a cone of $\Sigma$ unless $r_i, r_j \in f_1 \cap f_2$. Put differently, if $r_i \in f_1 \setminus f_2$ and $r_j \in f_2 \setminus f_1$, then $x_i x_j \in I_{\text{SR}}( X_\Sigma )$. Hence we notice:
\begin{align}
\begin{split}
\left\{ r_{0}, r_{1}, r_{2}, r_{9}, r_{14}, r_{17}, r_{22}, r_{24}, r_{25}, r_{28} \right\} &\subseteq f_1 \, , \\
\left\{ r_{0}, r_{1}, r_{3}, r_{4}, r_{5}, r_{6}, r_{7}, r_{8}, r_{17}, r_{18}, r_{19}, r_{20}, r_{21}, r_{25}, r_{26}, r_{27} \right\} &\subseteq f_2 \, , \\
\left\{ r_{0}, r_{2}, r_{3}, r_{4}, r_{5}, r_{6}, r_{7}, r_{8}, r_{9}, r_{10}, r_{11}, r_{12}, r_{13}, r_{14}, r_{15}, r_{16} \right\} &\subseteq f_3 \, , \\
\left\{ r_{1}, r_{2}, r_{3}, r_{13}, r_{16}, r_{21}, r_{23}, r_{24}, r_{27}, r_{28} \right\} &\subseteq f_4 \, .
\end{split}
\end{align}
Therefore, the above argument implies that for each FRST of $\Delta_4^\circ$, we have $I_5 \subseteq I_{\text{SR}}( X_\Sigma )$ where
\begin{align}
\begin{split}
I_5 &= \langle x_{0} x_{23}, x_{4} x_{22}, x_{4} x_{23}, x_{4} x_{24}, x_{4} x_{28}, x_{5} x_{22}, x_{5} x_{23}, x_{5} x_{24}, x_{5} x_{28}, x_{6} x_{22}, x_{6} x_{23},\\
& \qquad x_{6} x_{24}, x_{6} x_{28}, x_{7} x_{22}, x_{7} x_{23}, x_{7} x_{24}, x_{7} x_{28}, x_{8} x_{22}, x_{8} x_{23}, x_{8} x_{24}, x_{8} x_{28}, x_{3} x_{22},\\
& \qquad x_{9} x_{18}, x_{9} x_{19}, x_{9} x_{20}, x_{9} x_{21}, x_{9} x_{23}, x_{9} x_{26}, x_{9} x_{27}, x_{10} x_{17}, x_{10} x_{18}, x_{10} x_{19}, x_{10} x_{20},\\
& \qquad x_{10} x_{21}, x_{10} x_{22}, x_{10} x_{23}, x_{10} x_{24}, x_{10} x_{25}, x_{10} x_{26}, x_{10} x_{27}, x_{10} x_{28}, x_{10} x_{1}, x_{11} x_{17},\\
& \qquad x_{11} x_{18}, x_{11} x_{19}, x_{11} x_{20}, x_{11} x_{21}, x_{11} x_{22}, x_{11} x_{23}, x_{11} x_{24}, x_{11} x_{25}, x_{11} x_{26}, x_{11} x_{27},\\
& \qquad x_{11} x_{28}, x_{11} x_{1}, x_{12} x_{17}, x_{12} x_{18}, x_{12} x_{19}, x_{12} x_{20}, x_{12} x_{21}, x_{12} x_{22}, x_{12} x_{23}, x_{12} x_{24},\\
& \qquad x_{12} x_{25}, x_{12} x_{26}, x_{12} x_{27}, x_{12} x_{28}, x_{12} x_{1}, x_{13} x_{17}, x_{13} x_{18}, x_{13} x_{19}, x_{13} x_{20}, x_{13} x_{22},\\
& \qquad x_{13} x_{25}, x_{13} x_{26}, x_{14} x_{18}, x_{14} x_{19}, x_{14} x_{20}, x_{14} x_{21}, x_{14} x_{23}, x_{14} x_{26}, x_{14} x_{27}, x_{15} x_{17},\\
& \qquad x_{15} x_{18}, x_{15} x_{19}, x_{15} x_{20}, x_{15} x_{21}, x_{15} x_{22}, x_{15} x_{23}, x_{15} x_{24}, x_{15} x_{25}, x_{15} x_{26}, x_{15} x_{27},\\
& \qquad x_{15} x_{28}, x_{15} x_{1}, x_{16} x_{17}, x_{16} x_{18}, x_{16} x_{19}, x_{16} x_{20}, x_{16} x_{22}, x_{16} x_{25}, x_{16} x_{26}, x_{2} x_{18},\\
& \qquad x_{2} x_{19}, x_{2} x_{20}, x_{2} x_{26}, x_{17} x_{23}, x_{18} x_{22}, x_{18} x_{23}, x_{18} x_{24}, x_{18} x_{28}, x_{19} x_{22}, x_{19} x_{23},\\
& \qquad x_{19} x_{24}, x_{19} x_{28}, x_{20} x_{22}, x_{20} x_{23}, x_{20} x_{24}, x_{20} x_{28}, x_{21} x_{22}, x_{22} x_{23}, x_{22} x_{26}, x_{22} x_{27},\\
& \qquad x_{23} x_{25}, x_{23} x_{26}, x_{24} x_{26}, x_{26} x_{28} \rangle \, .
\end{split}
\end{align}
Consequently, for every FRST, it follows that
\begin{align}
I_1 + I_2 + I_3 + I_4 + I_5 \subseteq I_{\text{SR}}( X_\Sigma ) \, . 
\end{align}
This argument can of course be extended to three and four rays. For our purposes, it will be sufficient to notice that this includes $I_6 \subseteq I_{\text{SR}}( X_\Sigma )$ where
\begin{align}
\begin{split}
I_6 &= \langle x_0 x_{13} x_{24}, x_3 x_9 x_{24}, x_9 x_{13} x_{24}, x_{13} x_{14} x_{24}, x_3 x_{17} x_{28}, x_0 x_{21} x_{28},  x_{17} x_{27} x_{28},x_{21}x_{25}x_{28},   \\
& \qquad x_{21}x_0x_9, x_{21}x_0x_{28}, x_4x_9x_{17}, x_1x_9x_4, x_1x_9x_5, x_1x_9x_6, x_1x_9x_7, x_1x_9x_8, x_1x_9x_{16},  \\
& \qquad  x_1x_{13}x_4, x_1x_{13}x_5, x_1x_{13}x_6, x_1x_{13}x_7, x_1x_{13}x_8, x_1x_{13}x_{14}, x_1x_{14}x_4, x_1x_{14}x_6,    \\
& \qquad x_1x_{14}x_8, x_1x_{16}x_4,, x_1x_{16}x_6, x_1x_{16}x_8, x_1x_{28}x_4, x_1x_{28}x_5, x_1x_{28}x_6, x_1x_{28}x_7, x_1x_{28}x_8,     \\
& \qquad x_2x_{25}x_4, x_2x_{25}x_8, x_2x_{25}x_{21}, x_2x_{21}x_4, x_2x_{21}x_5, x_2x_{21}x_7, x_2x_{21}x_8, x_2x_{17}x_4, x_2x_{17}x_5,   \\
& \qquad  x_2x_{17}x_7, x_2x_{17}x_8, x_2x_{17}x_{27}, x_2x_{27}x_4, x_2x_{27}x_8, x_2x_{24}x_4, x_2x_{24}x_5, x_2x_{24}x_7,  \\
&\qquad x_2x_{24}x_8, x_3x_9x_{17}, x_3x_9x_{24}, x_3x_{17}x_{13}, x_3x_{17}x_{28}, x_3x_8x_{24}, x_3x_8x_{28}, x_8x_{13}x_{21},  \\
&\qquad x_0x_{24}x_{28}x_3, x_0x_{24}x_{28}x_{16}, x_0x_{24}x_{28}x_{27}, x_3x_{24}x_{28}x_{14}, x_3x_{24}x_{28}x_{25}, x_{14}x_{24}x_{28}x_{16},\\
&\qquad x_{14}x_{24}x_{28}x_{27} \rangle \, .
\end{split}
\end{align}

\subsubsection*{Vanishing intersections from facet interiors}

The following homogeneous coordinates correspond to facet interior points of $\Delta_4^\circ$:
\begin{align}
I_7 = \left\{ x_{10}, x_{11}, x_{12}, x_{15}, x_{18}, x_{19}, x_{20}, x_{22}, x_{23}, x_{26} \right\} \, .
\end{align}
Therefore, $V( z, s ) = \emptyset$ for all $z \in I_7$ \cite{Bies:2021xfh}.

\subsection{The nodal curve}

\subsubsection*{The components}

In \cite{Bies:2021xfh} we argued that the structure of the canonical nodal quark-doublet curve is independent of the FRST. Specifically, we have
\begin{align}
C^\bullet_{(\mathbf{3}, \mathbf{2})_{1/6}} = V \left( \prod_{i = 0}^{28}{x_i}, s_9 \right) \subseteq B_3 \, .
\end{align}
By taking $I_7$ into account, it follows that $C^\bullet$ has the following 17 irreducible components (each a smooth $\mathbb{P}^1$ for generic $s \in H^0( B_3, \overline{K}_{B_3} )$):
\begin{align}
\begin{tabular}{ccccc}
$C^\bullet_0 = V( x_0, s_9 )$ & $C^\bullet_1 = V( x_1, s_9 )$ & $C^\bullet_2 = V( x_2, s_9 )$ & $C^\bullet_3 = V( x_3, s_9 )$ \\
$C^\bullet_4 = V( x_4, s_9 )$ & $C^\bullet_5 = V( x_5, s_9 )$ & $C^\bullet_6 = V( x_6, s_9 )$ & $C^\bullet_7 = V( x_7, s_9 )$ \\
$C^\bullet_8 = V( x_8, s_9 )$ & $C^\bullet_9 = V( x_9, s_9 )$ & $C^\bullet_{13} = V( x_{13}, s_9 )$ & $C^\bullet_{14} = V( x_{14}, s_9 )$ \\
$C^\bullet_{16} = V( x_{16}, s_9 )$ & $C^\bullet_{17} = V( x_{17}, s_9 )$ & $C^\bullet_{21} = V( x_{21}, s_9 )$ & $C^\bullet_{25} = V( x_{25}, s_9 )$ \\
$C^\bullet_{27} = V( x_{27}, s_9 )$ \\
\end{tabular}
\end{align}
In addition, $C^\bullet_{24} = V( x_{24}, s_9 )$ and $C^\bullet_{28} = V( x_{28}, s_9 )$ are reducible (cf. \cite{Bies:2021xfh} and references therein):
\begin{align}
C^\bullet_{24} &= V( x_{24}, s_9 ) = C^\bullet_{24,0} \cup C^\bullet_{24,1} \, , \quad C^\bullet_{28} = V( x_{28}, s_9 ) = C^\bullet_{28,0} \cup C^\bullet_{28,1} \, .
\end{align}
We now wish to identify the dual graph and the location of the intersection points of the components of $C^\bullet$. In order to analyze $C_{24}^\bullet$, $C_{28}^\bullet$ efficiently, we employ the simplifying assumption 
\begin{align}
\alpha_4 \neq 0 \, .
\end{align}
We will see momentarily, that this allows to express $C^\bullet_{24,0}$, $C^\bullet_{24,1}$, $C^\bullet_{28,0}$, $C^\bullet_{28,1}$ in terms of
\begin{align}
\widetilde{\alpha}_1 &= \frac{\alpha_2 + \sqrt{\alpha_2^2 - 4 \alpha_1 \alpha_4}}{2 \alpha_4} \, , \qquad \widetilde{\alpha}_3 = \frac{\alpha_2 - \sqrt{\alpha_2^2 - 4 \alpha_1 \alpha_4}}{2 \alpha_4} \, . \label{equ:DefinitionOfTildeCoefficients}
\end{align}
With that said, we will now discuss various vanishing sets and their intersections. Eventually, we wish to identify homogeneous coordinates of $C_0^\bullet$, $C_1^\bullet$, $C_2^\bullet$, $C_3^\bullet$.

\subsubsection*{Details of \texorpdfstring{$C^\bullet_{24}$}{C24} and \texorpdfstring{$C^\bullet_{28}$}{C28}}

Recall that $C^\bullet_{24} = V( x_{24}, s_9 )$. By recalling $I_1 + I_2 + I_3 + I_4 + I_5 + I_6 \subseteq I_{\text{SR}}( X_\Sigma )$ , we note that $x_{1}, x_{4}, x_{5}, x_{6}, x_{7}, x_{8}, x_{10}, x_{11}, x_{12}, x_{15}, x_{17}, x_{18},  x_{19}, x_{20}, x_{21}, x_{26}$ cannot vanish on $C^\bullet_{24}$. Let us set these variables temporarily to $1$. Then we see
\begin{align}
C^\bullet_{24} = V( x_{24}, \alpha_1 z_1^2 + \alpha_2 z_1 z_2 + \alpha_4 z_2^2 ) \, , \quad z_1 = x_0^3 x_9^2 x_{14} x_{22} x_{25} \, , \quad z_2 = x_3^3 x_{13}^2 x_{16} x_{23} x_{27} \, .
\end{align}
Provided that $\alpha_4 \neq 0$ -- this is where we use our simplifying assumption -- we can write
\begin{align}
\begin{split}
\alpha_1 z_1^2 + \alpha_2 z_1 z_2 + \alpha_4 z_2^2 &= \alpha_4 \cdot \left( z_2 + \frac{\alpha_2 + \sqrt{\alpha_2^2 - 4 \alpha_1 \alpha_4}}{2 \alpha_4} \cdot z_1 \right) \\
& \qquad \qquad \qquad \qquad \times \left( z_2 + \frac{\alpha_2 - \sqrt{\alpha_2^2 - 4 \alpha_1 \alpha_4}}{2 \alpha_4} \cdot z_1 \right) \\
&= \left(z_2 + \widetilde{\alpha_1} z_1 \right) \left( z_2 + \widetilde{\alpha_3} z_1 \right) \, .
\end{split}
\end{align}
We use this result to identify the two irreducible components of $C_{24}^\bullet$. To express these components economically and homogeneously, we define
\begin{align}
d_1 = x_0^3 x_4^2 x_5  x_9^2 x_{10} x_{17}^2 x_{18} x_{22} x_{25}\, , \quad d_2 = x_3^3 x_7 x_8^2 x_{12} x_{13}^2 x_{20} x_{21}^2 x_{23} x_{27} \, .
\end{align}
By taking into account that $I_1 + I_2 + I_3 + I_4 + I_5 + I_6 \subseteq I_{\text{SR}}( X_\Sigma )$ and that $V(z,s) = \emptyset$ for all $z \in I_7$ it then follows that $d_1, d_2 \neq 0$ on $C_{24}^\bullet$. We set
\begin{align}
S_{24} = \left\{ 0, 1, 2, 3, \dots, 28 \right\} \setminus \left\{ x_2, x_{14}, x_{16}, x_{24}, x_{28} \right\} \, .
\end{align}
Then we have $C^\bullet_{24} = C^\bullet_{24,0} \cup C^\bullet_{24,1}$ and
\begin{align}
C^\bullet_{24,0} &= \left\{ x \in X_\Sigma \left| x_{24} = 0 \, , x_{16} = - \widetilde{\alpha}_1 \cdot \frac{d_1 x_{14}}{d_2} \, , x_i \neq 0 \, \forall i \in S_{24} \right. \right\} \, , \\
C^\bullet_{24,1} &= \left\{ x \in X_\Sigma \left| x_{24} = 0 \, , x_{16} = - \widetilde{\alpha}_3 \cdot \frac{d_1 x_{14}}{d_2} \, , x_i \neq 0 \, \forall i \in S_{24} \right. \right\} \, .
\end{align}
Notice that $x_2 x_{28} \in I_1$, so that $C_{24,0}^\bullet, C_{24,1}^\bullet \cong \mathbb{P}^1_{[x_2 \colon x_{28}]}$. Similarly, we set
\begin{align}
S_{28} = \left\{ 0, 1, 2, 3, \dots, 28 \right\} \setminus \left\{ x_1, x_{14}, x_{16}, x_{24}, x_{28} \right\} \, .
\end{align}
Then we have $C^\bullet_{28} = C^\bullet_{28,0} \cup C^\bullet_{28,1}$ and 
\begin{align}
C^\bullet_{28,0} &= \left\{ x \in X_\Sigma \left| x_{28} = 0 \, , x_{16} = - \widetilde{\alpha}_1 \cdot \frac{d_1 x_{14}}{d_2} \, , x_i \neq 0 \, \forall i \in S_{28} \right. \right\} \, , \\
C^\bullet_{28,1} &= \left\{ x \in X_\Sigma \left| x_{28} = 0 \, , x_{16} = - \widetilde{\alpha}_3 \cdot \frac{d_1 x_{14}}{d_2} \, , x_i \neq 0 \, \forall i \in S_{28} \right. \right\} \, .
\end{align}
Notice that $x_1 x_{24} \in I_1$, so that $C_{28,0}^\bullet, C_{28,1}^\bullet \cong \mathbb{P}^1_{[x_1 \colon x_{24}]}$. We can now look at the intersection of $C^\bullet_{24,0}$ and $C^\bullet_{28,1}$. If we write this non-homogeneous, then
\begin{align}
C^\bullet_{24,0} \cap C^\bullet_{28,1} &= V \left( x_{24}, x_{28}, x_{16} = -\widetilde{\alpha}_1 = -\widetilde{\alpha}_3 \right) \, .
\end{align}
This has a solution iff $\widetilde{\alpha}_1 = \widetilde{\alpha}_3$. By use of \oref{equ:DefinitionOfTildeCoefficients}, we can see that this is equivalent to
\begin{align}
0 = \sqrt{\alpha_2^2 - 4 \alpha_1 \alpha_4} \, .
\end{align}
Hence, as long as we ensure $\alpha_2^2 \neq 4 \alpha_1 \alpha_4$, we have $C^\bullet_{24,0} \cap C^\bullet_{28,1} = \emptyset$. Similarly, this condition ensures $C^\bullet_{24,1} \cap C^\bullet_{28,0} = \emptyset$. At the same time, it then follows that $C^\bullet_{24,0} \cap C^\bullet_{28,0}$ and $C^\bullet_{24,1} \cap C^\bullet_{28,1}$ are both non-empty and consist of exactly one point.

\subsubsection*{Simplifying the nodal curve by removing chains}

By repeating this exercise, it is not too hard to verify that the dual graph of $C^\bullet$ is given in \oref{equ:FullDualGraph}. Here, $L_{i_1,...,i_k}$ represents a chain of $k$ genus zero curves $C_{i_j}$ such that $C_{i_j}$ and $C_{i_{j+1}}$ intersect each other at a single node. As a dual graph, $L_{i_1, ..., i_k}$ is a string of red vertices. The chain is ordered so that $C_{i_1}$ intersects the $\mathbb{P}^1$ on the left and $C_{i_k}$ intersects the $\mathbb{P}^1$ on the right. For $L_{21, 27}$, we have that $C_{21}$ intersects $C_3$ while $C_{27}$ intersects $C_1$. 
\begin{equation}
\begin{tikzpicture}[scale=0.6, baseline=(current  bounding  box.center)]
      
      \def\s{5.0};
      \def\h{4.0};
      
      \path[-] (-\s,0) edge node[fill = white] {$L_{16,13}$} (0,\h);
      \path[-] (-\s,0) edge node[fill = white] {$L^0_{24,28}$} (0,-\h);
      \path[-, out = -90, in = 180, looseness = 1.5] (-\s,0) edge node[fill = white] {$L^1_{24,28}$} (0,-\h);
      \path[-] (\s,0) edge node[fill = white] {$L_{8,7,6,5,4}$} (0,\h);
      \path[-] (\s,0) edge node[fill = white] {$L_{25,17}$} (0,-\h);
      \path[-] (0,\h) edge node[fill = white] {$L_{21,27}$} (0,-\h);
      \path[-, out = 90, in = 90, looseness = 2.5] (-\s,0) edge node[fill = white] {$L_{14,9}$} (\s,0);
      
      \node at (1*\s,0) [stuff_fill_red, scale=0.6, label=right:$C_0$]{$h^0 = 2$};
      \node at (0,-\h) [stuff_fill_red, scale=0.6, label=below:$C_1$]{$h^0 = 3$};
      \node at (-1*\s,0) [stuff_fill_red, scale=0.6, label=left:$C_2$]{$h^0 = 3$};
      \node at (0,\h) [stuff_fill_red, scale=0.6, label=above:$C_3$]{$h^0 = 2$};
      
\end{tikzpicture} \label{equ:FullDualGraph}
\end{equation}

As argued in \cite{Bies:2021xfh} and \oref{subsec:SimplificationForPartialBlowups}, we can simplify the counting task by looking at a simplified nodal curve. The dual graph of this simplified curve is obtained by replacing each of the chains $T_{ij}$ in \oref{equ:FullDualGraph} by a single edge. This leads to a nodal curve with components $C_0^\bullet$, $C_1^\bullet$, $C_2^\bullet$, $C_3^\bullet$ whose dual graph is exactly the jumping circuit in \oref{equ:JumpingCircuit}. The position of the nodes on this simplified nodal curve are as follows:
\begin{align}
\begin{adjustbox}{max width=0.9\textwidth}
\begin{tabular}{cc|cc}
\toprule
Label & Superset & Coordinates in first curve & Coordinates in second curve \\
\midrule
$n_1$ & $C_0^\bullet \cap C_1^\bullet$ & $V(x_0, x_{17}, s)$ & $V(x_1, x_{25}, s)$\\
$n_2$ & $C_0^\bullet \cap C_2^\bullet$ & $V(x_0, x_{9}, s)$ & $V(x_2, x_{14}, s)$\\
$n_3$ & $C_0^\bullet \cap C_3^\bullet$ & $V(x_0, x_{4}, s)$ & $V(x_3, x_{8}, s)$\\
\midrule
$n_4$ & $C_1^\bullet \cap C_2^\bullet$ & $C_{28,0}^\bullet \cap C_1^\bullet$ & $C_{24,0}^\bullet \cap C_2^\bullet$ \\
$n_5$ & $C_1^\bullet \cap C_2^\bullet$ & $C_{28,1}^\bullet \cap C_1^\bullet$ & $C_{24,1}^\bullet \cap C_2^\bullet$ \\
$n_6$ & $C_1^\bullet \cap C_3^\bullet$ & $V(x_1, x_{27}, s)$ & $V(x_3, x_{21}, s)$\\
\midrule
$n_7$ & $C_2^\bullet \cap C_3^\bullet$ & $V(x_2, x_{16}, s)$ & $V(x_3, x_{13}, s)$\\
\bottomrule
\end{tabular}
\end{adjustbox}
\end{align}

\subsection{The location of the nodes}

\subsubsection*{Nodes on \texorpdfstring{$C_0^\bullet$}{C0}}

Recall that $C_0^\bullet = V( x_0, s )$. We define
\begin{align}
\begin{split}
g_1 &= x_{5}^2 x_{6}^3 x_{7}^4 x_{8}^5 x_{3}^6 x_{10} x_{11}^2 x_{12}^3 x_{13}^4 x_{15} x_{16}^2 x_{18} x_{19}^2 x_{20}^3 x_{21}^4 x_{23} x_{26} x_{27}^2 \, ,\\
g_2 &= x_{10} x_{11} x_{12} x_{13} x_{14}^2 x_{15}^2 x_{16}^2 x_{2}^3 x_{22} x_{24}^2 x_{28} \, , \\
g_3 &= x_{1}^3 x_{18} x_{19} x_{20} x_{21} x_{22} x_{24} x_{25}^2 x_{26}^2 x_{27}^2 x_{28}^2 \, .
\end{split}
\end{align}
By use of $I_1 + I_2 + I_3 + I_4 + I_5 + I_6 \subseteq I_{\text{SR}}( X_\Sigma )$ and that $V(z,s) = \emptyset$ for all $z \in I_7$, it follows that $g_1, g_2, g_3 \neq 0$ on $C_{0}^\bullet$ and that we can write
\begin{align}
C_0^\bullet &= V \left( x_0, \alpha_4 \left( \frac{g_1 x_{4}}{g_3} \right) + \alpha_5 \left( \frac{g_2 x_9}{g_3} \right) + \alpha_6 x_{17} \right) \, ,\\
1 &\cong \left( 0, \dots, 0, 1, 0, \dots, 0 \right) = w \left( \frac{g_1 x_4}{g_3} \right) = w \left( \frac{g_2 x_9}{g_3} \right) = w( x_{17} ) \, .
\end{align}
Since $x_4 x_9 x_{17} \in I_6 \subseteq I_{\text{SR}}( X_\Sigma )$, we can consider $C_0^\bullet$ as hypersurface in $\mathbb{P}^2_{[x_4 : x_9 : x_{17}]}$. In particular, the following map is well-defined
\begin{align}
\varphi \colon \mathbb{P}^1_{[u:v]} \to C_0^\bullet \, , \, \left[ u:v \right] \mapsto \left[ \frac{g_1 x_4}{g_3} = u : \frac{g_2 x_9}{g_3} = v : x_{17} = - \frac{\alpha_4 u + \alpha_5 v}{\alpha_6} \right] \, .
\end{align}
This establishes $C_0^\bullet \cong \mathbb{P}^1_{[u:v]}$ with $\left[ \frac{g_1 x_4}{g_3} \colon \frac{g_2 x_9}{g_3} \right] = \left[ u_0 \colon v_0 \right]$ as homogeneous coordinates of $C_0^\bullet$. In terms of $\left[ u_0 \colon v_0 \right]$, the locations of the nodes on $C$ adjacent to $C_0^\bullet$ are as follows:
\begin{align}
n_1 = \left[ 1 \colon - \frac{\alpha_4}{\alpha_5} \right] \, , \qquad n_2 = \left[ 1 \colon 0 \right] \, , \qquad n_3 = \left[ 0 \colon 1 \right] \, .
\end{align}
We want to emphasize that it is important to represent the node $n_1$ as $\left[ 1 \colon - \frac{\alpha_4}{\alpha_5} \right]$ for the computation of the global sections. To see this, recall that we are looking at the nodal curve
\begin{align}
C^\bullet = V \left( \prod_{i = 1}^{28}{x_i}, s_9 \right) \, .
\end{align}
Apparently, this curve remains the same if we replace $s_9$ by $\lambda \cdot s_9$, where $\lambda \in \mathbb{C}^\ast$ is arbitrary but fixed. Consequently, the location of the nodes must remain unchanged upon rescaling the coefficients $\alpha_i$ of $s_9$, i.e. upon
\begin{align}
\alpha_i \to \lambda \cdot \alpha_i \, , \qquad \lambda \in \mathbb{C}^\ast \, .
\end{align}
This scale invariance must also be true for the number of global sections on $C^\bullet$, which we will eventually compute from the right nullpace of a $\mathbb{C}$-valued matrix. To ``inform`` this matrix about the scale invariance, it is imperative to use the complex numbers $1$ and $-\frac{\alpha_4}{\alpha_5}$ (or $1$ and $-\frac{\alpha_5}{\alpha_4}$) as coordinates for the node $n_1$.

\subsubsection*{Nodes on \texorpdfstring{$C_1^\bullet$}{C1}}

We define
\begin{align}
\begin{split}
g_4 &= x_0^6 x_4^5 x_5^4 x_6^3 x_7^2 x_8 x_9^4 x_{10}^2 x_{11} x_{14}^2 x_{17}^4 x_{18}^3 x_{19}^2 x_{20} x_{22}^2 \, , \\
g_5 &= x_0^3 x_4^3 x_5^3 x_6^3 x_7^3 x_8^3 x_9^2 x_{10} x_{11} x_{12} x_{13}^2 x_{14} x_{16} x_{17}^2 x_{18}^2 x_{19}^2 x_{20}^2 x_{21}^2 x_{22} x_{23} x_3^3 \, , \\
g_6 &= x_4 x_5^2 x_6^3 x_7^4 x_8^5 x_{11} x_{12}^2 x_{13}^4 x_{16}^2 x_{18} x_{19}^2 x_{20}^3 x_{21}^4 x_{23}^2 x_3^6 \, , \\
g_7 &= x_2^3 x_9 x_{13} x_{14}^2 x_{15} x_{16}^2 x_{22} x_{23} x_{24}^2 \, .
\end{split}
\end{align}
By use of $I_1 + I_2 + I_3 + I_4 + I_5 + I_6 \subseteq I_{\text{SR}}( X_\Sigma )$ and that $V(z,s) = \emptyset$ for all $z \in I_7$, it follows that $g_4, g_5, g_6, g_7 \neq 0$ on $C_{1}^\bullet$ and that we can write
\begin{align}
C_1^\bullet &= V \left( x_1, \alpha_1 x_{25}^2 + \alpha_2 x_{25} \cdot \frac{g_5 \cdot x_{27}}{g_4} + \alpha_4 \cdot \frac{g_6 x_{27}^2}{g_4} + \alpha_5 \cdot \frac{g_7 x_{28}}{g_4 x_{26}} \right) \, .
\end{align}
An explicit computation shows that $g_4 g_6 - g_5^2 = 0$. Consequently,
\begin{align}
C_1^\bullet = V \left( x_1, \alpha_1 x_{25}^2 + \alpha_2 x_{25} \cdot \frac{g_5 \cdot x_{27}}{g_4} + \alpha_4 \cdot \left( \frac{g_5 \cdot x_{27}}{g_4} \right)^2 + \alpha_5 \cdot \frac{g_7 x_{28}}{g_4 x_{26}} \right) \, .
\end{align}
We notice that
\begin{align}
w \left( x_{25} \right) = w \left( \frac{g_5 x_{27}}{g_4} \right) = w \left( \frac{g_7 x_{28}}{g_4 x_{26}} \right) = \left( 0, \dots, 0, 1, 0, \dots, 0 \right) \cong 1 \, .
\end{align}
Also $x_{25} x_{27} \in I_2$. Therefore, the following map is well-defined:
\begin{align}
\varphi \colon \mathbb{P}^1_{[u:v]} \to C_1^\bullet \, , \, \left[ u:v \right] \mapsto \left[ x_{25} = u : \frac{g_5 x_{27}}{g_4} = v : \frac{g_7 x_{28}}{g_4 x_{26}} = - \frac{\alpha_1 u^2 + \alpha_2 u v + \alpha_4 v^2}{\alpha_5} \right] \, .
\end{align}
This establishes $C_1^\bullet \cong \mathbb{P}^1_{[u:v]}$ and allows us to pick $[ x_{25} : \frac{g_5 x_{27}}{g_4} ] \equiv \left[ u_1 \colon v_1 \right]$ as its homogeneous coordinates. In terms of these coordinates, the location of the nodes $n_1$, $n_4$, $n_5$, $n_6$ are:
\begin{align}
\begin{tabular}{cc}
\toprule
Node & Coordinates on $C_1^\bullet$ \\
\midrule
$n_1$ & $[0 : 1]$ \\
$n_4$ & $\left[ 1 : - \widetilde{\alpha_1} \right] = \left[ 1 : \frac{- \alpha_2 -  \sqrt{\alpha_2^2 - 4 \alpha_1 \alpha_4}}{2 \alpha_4} \right]$ \\
$n_5$ & $\left[ 1 : - \widetilde{\alpha_3} \right] = \left[ 1 : \frac{\alpha_2 - \sqrt{\alpha_2^2 - 4 \alpha_1 \alpha_4}}{2 \alpha_4} \right]$ \\
$n_6$ & $[1 : 0]$ \\
\bottomrule
\end{tabular}
\end{align}

\subsubsection*{Nodes on \texorpdfstring{$C_2^\bullet$}{C2}}

We define
\begin{align}
\begin{split}
g_8 &= x_0^6 x_{4}^5 x_{5}^4 x_{6}^3 x_{7}^2 x_{8} x_{9}^4 x_{10}^3 x_{11}^2 x_{12} x_{17}^4 x_{18}^2 x_{19} x_{22}^2 x_{25}^2 \, , \\
g_9 &= x_0^3 x_{3}^3 x_{4}^3 x_{5}^3 x_{6}^3 x_{7}^3 x_{8}^3 x_{9}^2 x_{10}^2 x_{11}^2 x_{12}^2 x_{13}^2 x_{17}^2 x_{18} x_{19} x_{20} x_{21}^2 x_{22} x_{23} x_{25} x_{27} \, , \\
g_{10} &= x_{3}^6 x_{4} x_{5}^2 x_{6}^3 x_{7}^4 x_{8}^5 x_{10} x_{11}^2 x_{12}^3 x_{13}^4 x_{19} x_{20}^2 x_{21}^4 x_{23}^2 x_{27}^2 \, , \\
g_{11} &= x_{1}^3 x_{17} x_{21} x_{22} x_{23} x_{25}^2 x_{26} x_{27}^2 x_{28}^2 \, .
\end{split}
\end{align}
By use of $I_1 + I_2 + I_3 + I_4 + I_5 + I_6 \subseteq I_{\text{SR}}( X_\Sigma )$ and that $V(z,s) = \emptyset$ for all $z \in I_7$, it follows that $g_8, g_9, g_{10}, g_{11} \neq 0$ on $C_{2}^\bullet$ and that we can write
\begin{align}
C_2^\bullet = V \left( x_2, \alpha_1 x_{14}^2 + \alpha_2 \cdot x_{14} \cdot \frac{g_9 x_{16}}{g_8 } + \alpha_4 \cdot \frac{g_{10} x_{16}^2}{g_8} + \alpha_6 \cdot \frac{g_{11} x_{24}}{g_8 x_{15}} \right) \, .
\end{align}
An explicit computation shows that $g_8 g_{10} - g_9^2 = 0$. Hence, we have
\begin{align}
C_2^\bullet = V \left( x_2, \alpha_1 x_{14}^2 + \alpha_2 \cdot x_{14} \cdot \frac{g_9 x_{16}}{g_8} + \alpha_4 \cdot \left( \frac{g_9 x_{16}}{g_8} \right)^2 + \alpha_6 \cdot \frac{g_{11} x_{24}}{g_8 x_{15}} \right) \, .
\end{align}
We notice that
\begin{align}
w( x_{14} ) = w \left( \frac{g_9 x_{16}}{g_8} \right) = w \left( \frac{g_{11} x_{24}}{g_8 x_{15}} \right) = \left( 0, \dots, 0, 1, 0, \dots, 0 \right) \cong 1 \, .
\end{align}
Notice that $x_{14} x_{16} \in I_3$. Therefore, the following map is well-defined:
\begin{align}
\varphi \colon \mathbb{P}^1_{[u:v]} \to C_2^\bullet \, , \, \left[ u:v \right] \mapsto \left[ x_{14} = u : \frac{g_9 x_{16}}{g_8} = v : \frac{x_{24}}{x_{15}} = - \frac{\alpha_1 u^2 + \alpha_2 u v + \alpha_4 v^2}{\alpha_6} \right] \, .
\end{align}
This establishes $C_2^\bullet \cong \mathbb{P}^1_{[u:v]}$ and allows us to pick $[ x_{14} : \frac{g_9 x_{16}}{g_8} ] \equiv \left[ u_2 \colon v_2 \right]$ as its homogeneous coordinates. In terms of these coordinates, the location of the nodes $n_2$, $n_4$, $n_5$ and $n_7$ are:
\begin{align}
\begin{tabular}{cc}
\toprule
Node & Coordinates on $C_2^\bullet$ \\
\midrule
$n_2$ & $[0 : 1]$ \\
$n_4$ & $\left[ 1 : - \widetilde{\alpha_1} \right] = \left[ 1 : \frac{- \alpha_2 -  \sqrt{\alpha_2^2 - 4 \alpha_1 \alpha_4}}{2 \alpha_4} \right]$ \\
$n_5$ & $\left[ 1 : - \widetilde{\alpha_3} \right] = \left[ 1 : \frac{\alpha_2 - \sqrt{\alpha_2^2 - 4 \alpha_1 \alpha_4}}{2 \alpha_4} \right]$ \\
$n_7$ & $[1 : 0]$ \\
\bottomrule
\end{tabular}
\end{align}

\subsubsection*{Nodes on \texorpdfstring{$C_3^\bullet$}{C3}}

We frist define
\begin{align}
\begin{split}
g_{12} &= x_{0}^6 x_{4}^5 x_{5}^4 x_{6}^3 x_{7}^2 x_{9}^4 x_{10}^3 x_{11}^2 x_{12} x_{14}^2 x_{15} x_{17}^4 x_{18}^3 x_{19}^2 x_{20} x_{22}^2 x_{25}^2 x_{26} \, ,\\
g_{13} &= x_{9} x_{10} x_{11} x_{12} x_{14}^2 x_{15}^2 x_{16}^2 x_{2}^3 x_{22} x_{23} x_{24}^2 x_{28} \, ,\\
g_{14} &= x_{1}^3  x_{17} x_{18} x_{19} x_{20} x_{22} x_{23} x_{24} x_{25}^2 x_{26}^2 x_{27}^2 x_{28}^2 \, .
\end{split}
\end{align}
By use of $I_1 + I_2 + I_3 + I_4 + I_5 + I_6 \subseteq I_{\text{SR}}( X_\Sigma )$ and that $V(z,s) = \emptyset$ for all $z \in I_7$, it follows that $g_{12}, g_{13}, g_{14} \neq 0$ on $C_{3}^\bullet$ and that we can write
\begin{align}
C_3^\bullet &= V \left( x_3, \alpha_1 \cdot \frac{g_{12} x_8}{g_{14}} + \alpha_5 \cdot \frac{g_{13} x_{13}}{g_{14}} + \alpha_6 \cdot x_{21} \right) \, , \\
w \left( \frac{g_{12} x_8}{g_{14}} \right) &= w \left( \frac{g_{13} x_{13}}{g_{14}} \right) = w \left( x_{21} \right) = \left( 0, \dots, 0, 1, 0, \dots, 0 \right) \cong 1 \, .
\end{align}
Since $x_8 x_{13} x_{21} \in I_6$, we can think of $C_3^\bullet$ as hypersurface in $\mathbb{P}^2_{[x_8 : x_{13} : x_{21}]}$. Alternatively, the following map is well-defined:
\begin{align}
\varphi \colon \mathbb{P}^1_{[u:v]} \to C_3^\bullet \, , \, \left[ u:v \right] \mapsto \left[ \frac{g_{12} x_8}{g_{14}} = u : \frac{g_{13} x_{13}}{g_{14}} = v : x_{21} = - \frac{\alpha_1 u + \alpha_5 v}{\alpha_6} \right] \, .
\end{align}
This establishes $C^\bullet_3 \cong \mathbb{P}^1_{[u: v]}$ and identifies $[\frac{g_{12} x_{8}}{g_{14}}: \frac{g_{13} x_{13}}{g_{14}}] = \left[ u_3 \colon v_3 \right]$ as its homogeneous coordinates. The locations of the nodes are
\begin{align}
n_3 = [0:1] \, , \qquad n_6 = \left[ 1 \colon - \frac{\alpha_1}{\alpha_5} \right] \, , \qquad n_7 = [1 : 0] \, .
\end{align}

\bibliography{references}{}
\bibliographystyle{JHEP}

\end{document}